%% file: ivashchenko.sergijenko.torbaniuk.tex
\documentclass[useAMS,usenatbib]{mn2e}
\usepackage{amssymb,amsmath}
\usepackage{graphicx,epsfig,rotating,euscript}
\title[Composite spectra of quasars with different UV spectral index]{Composite spectra of quasars with\\ different UV spectral index}
\author[G.~Ivashchenko, O.~Sergijenko, O.~Torbaniuk]{Ivashchenko~G.$^{1}$\thanks{E-mail: g.ivashchenko@gmail.com}, Sergijenko~O.$^{2}$\thanks{E-mail: olka@astro.franko.lviv.ua}, Torbaniuk~O.$^{3}$\thanks{E-mail: el.torbaniuk@gmail.com}\\
$^1$Astronomical Observatory of the Taras Shevchenko National University of Kyiv, Observatorna str., 3, 04058, Kyiv, Ukraine\\
$^2$Astronomical Observatory of the Ivan Franko National University of Lviv, Kyryla i Methodia str., 8, 79005, Lviv, Ukraine\\
$^3$Faculty of Physics of the Taras Shevchenko National University of Kyiv, Glushkova ave., 4, 03127, Kyiv, Ukraine\\
}
\begin{document}
\date{Accepted \underline{\hspace*{2cm}}. Received \underline{\hspace*{2cm}}; in original form \underline{\hspace*{2cm}}}
\pagerange{\pageref{firstpage}--\pageref{lastpage}} \pubyear{0000}
\maketitle

\label{firstpage}

\begin{abstract}
The composite spectra of quasars are widely used as templates for redshift determination, as well as for measurements of the mean transmission in Ly$\alpha$-forest studies, and for investigation of general spectral properties of quasars. Possibility of composite spectra utilisation in these fields is related to remarkable similarity of quasar spectra in UV-optical range. But despite of general similarity in spectral shapes, they differ in several parameters, one of which is the spectral index. In the present paper we study the possible effects, related to neglect of this difference. We compiled 16 composite spectra from subsamples of individual SDSS DR7 quasar spectra with different spectral indices $\alpha_{\lambda}$ within the wavelength range  1270--1480\,\AA, and show that (i) the redshifts measured for a test sample of high signal-to-noise ratio quasar spectra using these composites as templates appear to be systematically higher than those calculated with a traditional template, compiled from spectra with different $\alpha_{\lambda}$, with 1.5 times smaller errors in the former case; (ii) the difference in $\alpha_{\lambda}$ in individual spectra used for compilation of composites can yield the mean transmission uncertainty up to 20\%; (iii) a number of emission lines indistinguishable in ordinary composites, but seen in individual high-resolution spectra, can be detected in such composites. It is also shown, that there is no dependence of $\alpha_{\lambda}$ on quasar luminosity in SDSS $u$, $g$, $r$ and $i$ bands, and monochromatic luminosity at $1450$\,\AA. 
\end{abstract}


\section{Introduction}\label{sec:1}
\indent\indent The spectral properties of quasars in ultraviolet-optical range have been a subject of a number of studies. The smooth part (continuum) of this UV bump, called the ``Big Blue Bump'', is traditionally considered to be thermal emission from an accretion disk, while the emission lines are attributed to this emission reprocessed in surrounding clumped gas in a form of broad and narrow line regions (BLR and NLR) as well as in a dust obscured torus, but many details of this paradigm are still purely understood, as well as the nature of broad absorption lines (BAL), possessed by about 10\% of quasars. Thus, on the one hand, the detailed study of the quasar spectra plays an important role in understanding of the quasar nature. On the other hand, the accurate quasar spectrum shape, which does not vary significantly from one object to another, is needed as a template for quasar redshift measurements in large redshift surveys. Finally, the precise knowledge of the intrinsic quasar spectrum shape, prior to absorption in intergalactic medium, is required for studying the intergalactic gas, manifested as absorption features in the quasar spectra, first of all as the Hydrogen Ly$\alpha$-forest.

In all of the fields, mentioned above the composite (mean) spectra are widely used. The possibility of their utilisation is related to the fact that quasar spectra are remarkably similar from one object to another. Due to their high signal-to-noise ratio composite spectra reveal weak features that are rarely detectable in individual quasar spectra. The composite spectra were compiled for a wide set of quasar samples, e.\,g. the Large Bright Quasar Survey (LBQS) \citep{francis+91}, the First Bright Quasar Survey (FBQS) \citep{brotherton+01}, the Sloan Digital Sky Survey (SDSS) \citep{vandenberk+01,pieri+10}, quasar spectra from the Hubble Space Telescope (HST) \citep{zheng+97,telfer+02}, and Far Ultraviolet Spectroscopic Explorer (FUSE) \citep{scott2+04}.

The part of UV-optical quasar spectrum redward of the Ly$\alpha$ (1215.67\,\AA) emission line (free of hydrogen Lyman-series forests of absorption lines) is studied quite well. As it was shown e.\,g. by \citet{vandenberk+01} this region up to $\approx5000$\AA\ is described with a smooth continuum, well approximated with a power law $\sim\lambda^{\alpha_{\lambda}}$ over some limited wavelength range, broad emission and, in some cases, absorption lines. \citet{vandenberk+01} reported on $\alpha_{\lambda}$ to be $-1.54$ within 1350--4230\,\AA, that agrees well with the values $-1.54$ and $-1.68$ which were obtained by \citet{brotherton+01} within 1450--5050\,\AA\ for spectra from LBQS and FBQS respectively, and also with the results of \citet{carballo+99}, who obtained the values $-1.34\pm0.15$ and $-2.11\pm0.16$ within 1300--3000\,\AA\ and 3000--4500\,\AA\  ranges respectively for a sample of radio-loud quasars. Throughout this paper the indicated wavelengths are rest-frame ones, unless other is specified. The most complete to-date list of emission lines redward of the Ly$\alpha$ emission line found in quasar composite can be found in \citealt{vandenberk+01}. Some statistical studies were also conducted with samples of individual spectra, e.\,g. \citet{tang+12} studied the strong UV and optical emission line properties using a sample of 85 bright quasars.

The part of the spectrum blueward of the Ly$\alpha$ emission line, used for the Ly$\alpha$-forest studies, is much worse investigated due to the presence of the Lyman series forests. Direct reconstruction of intrinsic quasar spectral shape in this region is possible only for high resolution spectra, the number of which is not large, that does not allow to conduct statistical studies of the subsamples with different characteristics, e.\,g. luminosity. Several-parametric modelling of the continuum level within the Ly$\alpha$ forest region suffers from the degeneration between the mean transmission and parameters of lines and continuum, that is why extrapolation of the continuum from the red part of spectrum seems to be more reliable in some cases (see e.\,g., \citealt{press+93,bernardi+03,desjacques_07,greg+10,songaila-04}). But in practice it is only an approximation of the real continuum, because the centre of the ``Big Blue Bump'' is located at $\approx1000-1300$\,\AA. This is confirmed by the values of $\alpha_{\lambda}$ obtained from HST and FUSE spectra, $-1.01\pm0.05$ within 1050--2020\,\AA\ \citep{zheng+97}, $-0.24\pm0.12$ within 500--1200\,\AA\ \citep{telfer+02}, $-1.44^{+0.38}_{-0.28}$ within 630--1155\,\AA\ \citep{scott2+04}, when comparing with the same values for longer wavelengths mentioned above. Due to lower signal-to-noise ratio in the Lyman series forests region detection of emission lines is more complicated here. For example, \citet{vandenberk+01} listed only three emission features between Ly$\alpha$ and Ly$\beta$ (1025.72\,\AA) with large uncertainties in the central wavelength. The other studies of composites and individual spectra have not enlarged significantly their number. In Table~3 we tried to summarise the known emission features, found in optical \citep{francis+91,brotherton+01,vandenberk+01,tytler+04} and UV \citep{zheng+97,telfer+02,scott2+04} composites, and in some individual spectra \citep{brotherton+94,Laor+94,Laor+1995,laor+97,Vestergaard+01,leighly+07,binette+08}, within the range of $\sim1025-1450$\,\AA\ (between the Ly$\beta$ and C{\sc iv} lines), used for studies in the present work.

The spectral properties of quasars described above can be considered only as some generalised ones, because despite of general similarity in spectral shapes, they differ in spectral index (also called spectral slope), intensity of continuum and emission lines, lines' equivalent widths, etc. Some of these parameters are found to have some dependence on each other, e.\,g., the inverse correlation of equivalent width of some emission lines on monochromatic flux in UV region, known as the Baldwin effect \citep{baldwin}. Therefore, compilation of composite spectra from subsamples with similar properties (e.\,g., with similar luminosity) is of interest for studying these properties and there relation to other parameters. E.\,g., \citet{richards+11} studied the Baldwin effect constructing composite spectra of quasar subsamples, chosen on the base of C{\sc iv} (1549\,\AA) emission-line properties (equivalent width and blueshift). They also found, that the mean blueshift of C{\sc iv} line is approximately twice 
larger for radio-quite quasars, than that of radio-loud ones. \citet{reichard+03} studied the spectral properties of Broad Absorption Line (BAL) quasars, compiling composites from subsamples with different ionization levels of BALs, and found that there are substantial differences in the emission-line and continuum properties of high- and low-ionization BAL quasars, which, in their opinion, can be related to intrinsic quasar properties such as the continuum spectral index. Also using the composite spectra, \citet{yip+04,vandenberk+04} studied the spectral properties of quasars depending on luminosity and redshift, and concluded that aside from the Baldwin effect the average spectral properties are similar, e.\,g. the quasars with different luminosity have similar spectral index.

In the present paper we use the composite spectra compiled for subsamples of spectra with similar spectral index and study the possible effects, related to neglect of the difference in this parameter. 

The data used in this work, the sample selection and the composite spectra compilation processes are described in Section~\ref{sec:2}. The difference in redshift measurements made with the obtained composite spectra as templates compared to that with the SDSS quasar template is studied in Section~\ref{sec:3}. Section~\ref{sec:4} gives an estimation of errors introduced into the Ly$\alpha$ forest mean transmission measurement by neglect the difference in $\alpha_{\lambda}$ when using the composite spectra for such studies. In Section~\ref{sec:5} we present the results of our search for emission lines in our composites. Conclusions are presented in Section~\ref{sec:6}.

\section[]{Data}\label{sec:2}
\subsection{The SDSS DR7 quasar sample}\label{sec:2.1}

\indent\indent  Our sample is taken from the public available release of the sky-residual subtracted spectra for the Sloan Digital Sky Survey (SDSS) Legacy Release 7 \citep{Abazadjian_2009}. This release (`WH sample' hereafter), which is described in \citet{wild_10}, contains a total of 106\,006 spectra, generated using the \citet{Wild_05} scheme from the spectra of the objects from the \citet{Schneider_2010} quasar catalogue. This scheme includes a significantly improved technique of the OH sky lines subtraction. The strong OH sky emission lines extend over almost half of the wavelength range ($>6700$\,\AA) of the SDSS spectra and is not subtracted optimally enough with the previous SDSS pipelines.

The redshifts in SDSS are measured mainly with the help of two techniques --- emission line measurements and cross-correlation, or one of them. The errors of redshift (due to intrinsic line shifts etc.) could affect the results of our study hence we tried to minimize their influence using the improved redshifts of the objects from the \citet{Schneider_2010} quasar catalogue, which was generated using the \citet{hewett_10} scheme, instead of redshifts in the headers of each fits-file. According to \citet{hewett_10}, their redshifts possess systematic biases, which are for a factor of 20 smaller compared to the SDSS redshift values.

\subsection{Spectra selection}\label{sec:2.2}

\begin{figure}
\centering
\epsfig{figure=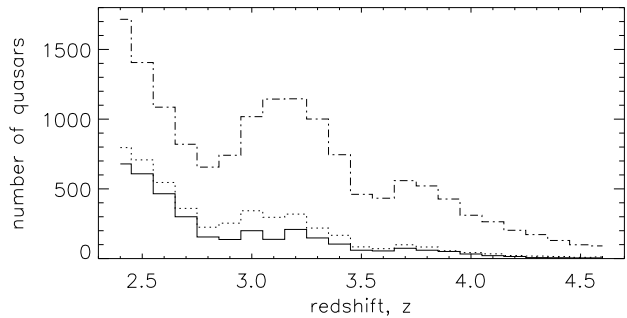,width=.98\linewidth}
\caption{Redshift distribution of all objects from the HW sample with $2.3<z<4.6$ and redshift determination confidence level $>0.9$ (\textit{dash-dot}), all visually selected (\textit{dotted}) for the present study and those with the rms of the normalization constant $A$ less than 15\% (\textit{solid}).}
\label{fig:zdistr}
\end{figure}

\indent\indent All the quasar spectra (15\,154~objects) from the HW sample with redshifts within the range 2.3$-$4.6 and redshift determination confidence level $>0.9$ were firstly selected. Because the SDSS is an automatic survey there is a possibility of pollution of the sample with `wrong' objects whose general photometry or spectroscopy properties required by the automatic selection pipelines are similar to the ones of `real' objects. Thus a visual examination was carried out. Except the non-quasar spectra the spectra of quasars with low signal-to-noise ratio, spectra of BAL quasars and those with the Damped Ly$\alpha$ (DLA) systems were also excluded during this examination. The presence of the BAL quasars in the sample could introduce additional errors into an estimation of the quasar mean continuum level. The DLA systems are a class of quasar absorbers selected for the presence of H~{\sc i} column densities $>2\cdot10^{20}$~cm$^{-2}$ (see e.\,g.~\citet{wolfe_05} for review). They are identified as 
absorption features with the rest equivalent width exceeding 5\,\AA. The nature of these systems is still not understood, however due to their high density they are usually related to the galaxy formation and therefore could not be used as representatives of the linear perturbations in the neutral intergalactic medium.

After the visual examination the sample contains 4\,779~spectra. Redshift distributions of these objects and all the objects initially selected from HW sample are presented in Figure~\ref{fig:zdistr} (\textit{dash-dot} and \textit{dotted} correspondingly).

\subsection{Subsamples and composites}\label{sec:2.3}

\indent\indent Generating composite spectra of quasars includes three steps: (i) normalization of each spectrum, (ii) setting each spectrum to the rest frame and (iii) calculation of the mean spectrum. Before we smoothed all spectra with a simple moving average over 3 points. Following \citet{mcdonald+06} we removed the following wavelength regions from our analysis because of calibration problems due to strong sky lines: $5575\mbox{\AA}<\lambda<5583\mbox{\AA}$, $5888\mbox{\AA}<\lambda<5895\mbox{\AA}$, $6296\mbox{\AA}<\lambda<6308\mbox{\AA}$, and $6862\mbox{\AA}<\lambda<6871\mbox{\AA}$, where the second region is 1\AA\ wider than that presented in \citet{mcdonald+06}, because we added also Na{\sc i} (5894.6\,\AA) interstellar line to it. Normalization of each spectrum is needed to be applied due to different apparent flux densities. Taking into account the similarity of quasar spectra and following \citet{press+93} and \citet{zheng+97} we normalize each spectrum on the (arithmetic) mean flux in all pixels within the rest wavelength range 1450--1470\,\AA. This range lies blueward from the C{\sc iv} emission line and is usually considered to be free of obvious emission and absorption. The last claim appears to be not precise because of weak emission lines, which can be seen in composite spectra, but we neglect this fact, because intensity of these lines is comparable to noise level in individual spectra. For further study to reduce possible uncertainties we used only spectra with the rms of the normalization constant $A$ less than 15\%; the number of them is 3\,493 and its redshift distribution is shown in Figure~\ref{fig:zdistr} (\textit{solid} line).

\begin{figure}
\centering
\epsfig{figure=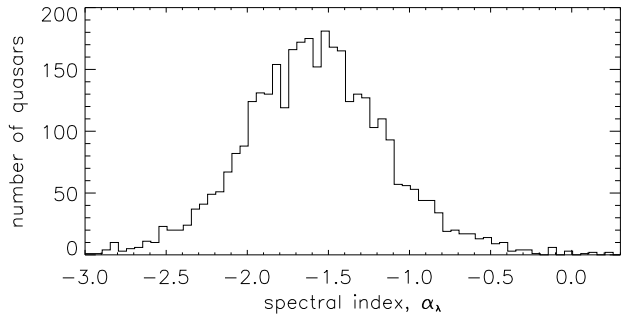,width=.98\linewidth}
\caption{The distribution of spectral indices.}
\label{fig:alpha-distr}
\end{figure}

Considering the continuum redward of Ly$\alpha$ emission line to be a power-law $\sim\lambda^{\alpha_{\lambda}}$, we calculated its index $\alpha_{\lambda}$ for each individual spectrum within the range between Ly$\alpha$ and C{\sc iv} emission lines. For this purpose we selected the following wavelength ranges from the composite spectrum compiled from all the spectra: 1278$-$1286\,\AA\AA, 1320$-$1326\,\AA\AA, 1345$-$1360\,\AA\AA\ and 1440$-$1480\,\AA\AA. The distribution of the obtained indices is shown in Figure~\ref{fig:alpha-distr}.

\begin{figure*}
\centering
\begin{minipage}{.75\linewidth}
\epsfig{figure=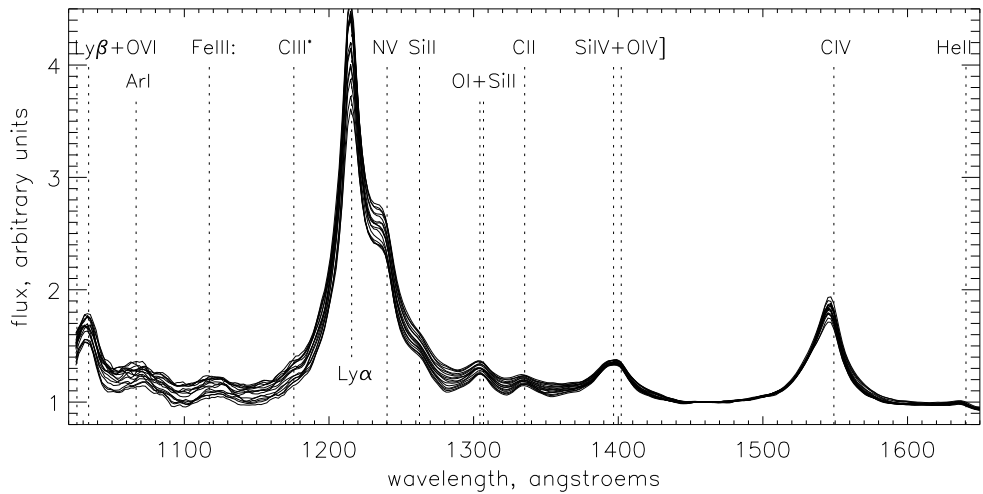,width=.99\linewidth}
\caption{Composite spectra of 16 subsamples within 1025--1650\,\AA. The highest spectrum has the steepest continuum redward of the Ly$\alpha$ emission line.}
\label{fig:all-spec}
\end{minipage}
\end{figure*}

Using the obtained values of $\alpha_{\lambda}$, we selected 16 subsamples each of 200 spectra with $\alpha_{\lambda}$ closest to $-0.7-k\cdot0.1$, $k=0..15$. Then dividing each $j$-th selected spectrum onto its normalization constant $A^{j}$, rebinning them with $\Delta\lambda_{rest}=2$\,\AA\ and stacking the spectra into the rest frame, we obtained the mean arithmetic composite spectra. The dispersion $\sigma^{2}$ of each pixel of the composite spectrum was calculated from the noises $\sigma_{i}$ of pixels of individual spectra constituting it. These spectra within 1025--1650\,\AA\ are shown in Figure~\ref{fig:all-spec}. The dashed lines indicate the laboratory wavelengths of the lines identified by \citet{vandenberk+01} in this range.

\subsection{Properties of subsamples}\label{sec:2.4}

\begin{figure}
\centering
\epsfig{figure=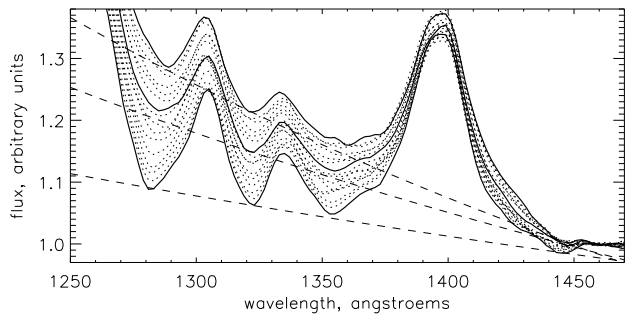,width=.98\linewidth}	
\caption{The parts of the composite spectra used for continuum approximation (\textit{short-dashed} line). The highest spectra are the steepest ones. Spectra with indices $-0.91$, $-1.62$ and $-2.14$ are shown with the \textit{solid} lines along with the fitted power-law continua (\textit{dashed} lines).}
\label{fig:lena}
\end{figure}

\indent\indent The first row of Table~\ref{tab:samples} contains the spectral indices of obtained composite spectra. In this case they were calculated in the way mentioned above, but the parts of the spectra which are the most free from emission lines were selected manually in each spectrum. The part of spectra used for $\alpha_{\lambda}$ determination is shown in Figure~\ref{fig:lena}. The regions selected for continuum fitting slightly vary from spectrum to spectrum and not all four parts used in calculation of indices of the individual spectra were taken into account in the case of composite spectra. This is clearly seen on example of the region with the shortest wavelengths: with the steepening of the spectrum its centre moves from $\sim1280$\,\AA\ to $\sim1290$\,\AA. It means that determination of the spectral index using the same regions as it was done for individual spectra is only an approximation, but it is sufficient for rough separation of spectra for compilation of composites. 

\begin{figure}
\centering
\epsfig{figure=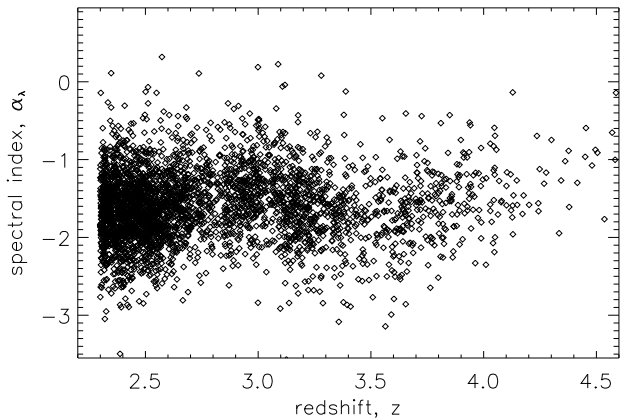,width=.98\linewidth}
\caption{Redshift distribution of spectral indices.}\label{fig:z-slo}
\end{figure}

The second row of Table~\ref{tab:samples} contains the mean redshift of each subsample. It is seen, that the mean redshift slightly increases with $\alpha_{\lambda}$, that might be an evidence for redshift evolution of the quasar spectral shape. On the other hand, the absence of such evolution in Fig.~\ref{fig:z-slo} for the whole sample of 3\,493 quasars means, that increase of the mean redshift for subsamples is probably a result of some selection effect.  

\begin{table}
 \begin{minipage}{80mm}
\centering
 \caption{The mean redshift $\bar{z}$ and the mean monochromatic luminosity $\langle\log{l_{1450}}\rangle$ of subsamples and the spectral indices $\alpha_{\lambda}$ of their composites. The quoted error bars of $\langle\log{l_{1450}}\rangle$ and $\bar{z}$ are the root mean squares for corresponding distributions.}\label{tab:samples}
\begin{tabular}{c|c|c|c}
\hline
n & $\alpha_{\lambda}$ & $\bar{z}$ & $\langle\log{l_{1450}}\rangle$ \\
\hline
1 & $-0.91\pm0.03$ & $2.90\pm0.54$ & $42.70\pm0.21$ \\
2 & $-0.97\pm0.02$ & $2.89\pm0.53$ & $42.72\pm0.21$ \\
3 & $-1.02\pm0.01$ & $2.87\pm0.51$ & $42.72\pm0.21$ \\
4 & $-1.04\pm0.05$ & $2.85\pm0.49$ & $42.72\pm0.24$ \\
5 & $-1.19\pm0.07$ & $2.82\pm0.43$ & $42.73\pm0.24$ \\
6 & $-1.35\pm0.13$ & $2.84\pm0.49$ & $42.74\pm0.21$ \\
7 & $-1.42\pm0.05$ & $2.82\pm0.44$ & $42.76\pm0.23$ \\
8 & $-1.42\pm0.02$ & $2.76\pm0.40$ & $42.78\pm0.21$ \\
9 & $-1.62\pm0.14$ & $2.77\pm0.44$ & $42.78\pm0.21$ \\
10 & $-1.64\pm0.15$ & $2.80\pm0.46$ & $42.79\pm0.27$ \\
11 & $-1.73\pm0.09$ & $2.75\pm0.45$ & $42.78\pm0.25$ \\
12 & $-1.88\pm0.07$ & $2.75\pm0.43$ & $42.78\pm0.22$ \\
13 & $-1.92\pm0.11$ & $2.74\pm0.43$ & $42.77\pm0.24$ \\
14 & $-2.02\pm0.05$ & $2.72\pm0.42$ & $42.74\pm0.23$ \\
15 & $-2.07\pm0.07$ & $2.74\pm0.42$ & $42.74\pm0.23$ \\
16 & $-2.14\pm0.05$ & $2.74\pm0.43$ & $42.75\pm0.24$ \\
\hline
\end{tabular}
\end{minipage}
\end{table}

\begin{figure*}
\centering
\epsfig{figure=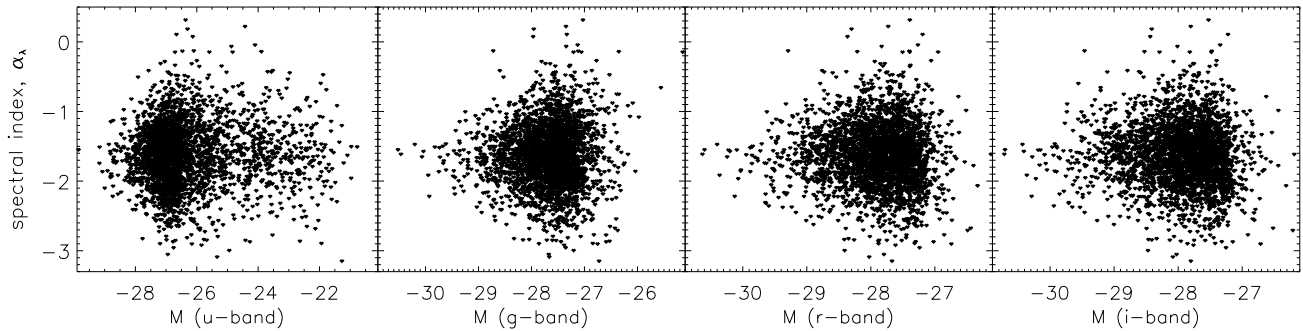,width=.98\linewidth}
\caption{Spectral index ($\alpha_{\lambda}$) -- absolute magnitude ($M$) diagrams for \textit{u,\,g,\,r,\,i}-bands.}
\label{fig:slo-mu}
\end{figure*}

\begin{figure}
\centering
\epsfig{figure=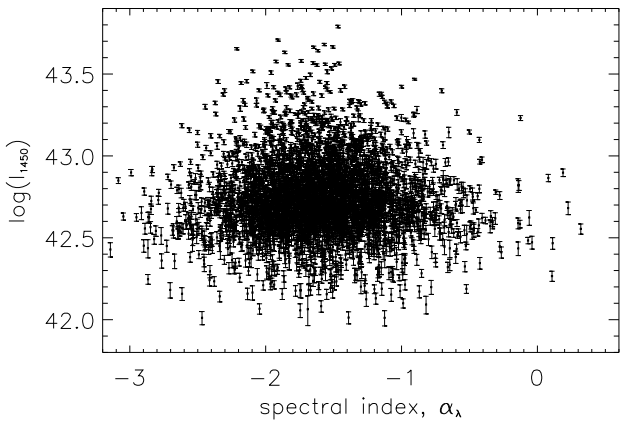,width=.98\linewidth}
\caption{Spectral index ($\alpha_{\lambda}$) -- logarithm of monochromatic luminosity ($\log{l_{1450}}$) diagram.}
\label{fig:nor-slo}
\end{figure}

To study any optical-UV luminosity dependence of the spectral index we plotted $\alpha_{\lambda}$\,--\,$M$ diagrams for absolute magnitudes $M$ in $u$ (3551\,\AA), $g$ (4686\,\AA), $r$ (6165\,\AA), $i$ ($7481$\,\AA) bands (Fig.~\ref{fig:slo-mu}). These magnitudes were calculated within the frame of the flat $\Lambda$CDM cosmological model with $H_0=70$\,km/s/Mpc, $\Omega_M=0.3$, using the apparent psf magnitudes and reddening in corresponding bands from the SDSS DR7, K-correction values from \citet{richards_06} and correction for Galactic extinction from \citet{schlegel+98}. As one can see from  Fig.~\ref{fig:slo-mu} there is no dependence between the spectral index and absolute magnitudes in these bands, however even the $u$-band central wavelength is about 2000\,\AA\ far from the bump peak, thus all these magnitudes are not a good characteristics in case of the UV-bump luminosity. Hence we also plotted the $\alpha_{\lambda}$\,--\,$\log{l_{1450}}$ diagram, where $\log{l_{1450}}$ is the monochromatic luminosity at 1450\AA, calculated from the mean flux within the wavelength range of $1449-1451$\,\AA.  Generally speaking this range is also not the best choice, but this one is the most free of obvious emission and absorption features, thus it should be the most appropriate characteristics the quasar luminosity in continuum within the bump. Any dependence of $\alpha_{\lambda}$ on monochromatic luminosity at 1450\,\AA\ can be seen from Fig.~\ref{fig:nor-slo}, as well as from its mean values listed in the third row of Table~\ref{tab:samples}.

\section{Templates for redshift measurement}\label{sec:3}
\subsection{General notes}\label{sec:3.1}
\indent\indent The precise redshift measurement for quasars is a challenging problem, because unlike normal galaxies they do not have narrow emission lines, except well-known [O{\sc iii}] (5007\,\AA), which is used for redshift calibration for quasars with $z\approx0.05-0.85$ (e.\,g. \citealt{vandenberk+01}). All other emission lines are usually broad and blended, and the rest wavelengths for high-ionization lines are known to be systematically blueshifted compared to low-ionization lines  (see e.\,g. \citealt{Gaskell+1982,Tytler+1992,Richards+2002,Shen+2007}). Thus the redshift measured with the help of some given emission line appears to be lower or higher than that measured with another line. The discrepancy is about several hundreds km/s, which, for example, is of the same order as the estimations of quasar pairwise velocities measured via redshift-space distortions (\citealt{Outram+2001,Croom+2005,daAngela+2005,daAngela+2008,Mountrichas+2009,Ivashchenko+2010}). Actually, the value, estimated from 
redshift-space distortions, namely from the `Finger of God' effect, is a superposition of the pairwise velocity and the redshift errors. Hence, the more accurate the redshifts measurement we have the more precise pairwise velocity estimation we can do.

But in addition to the blueshifting of high-ionization lines (this effect is well studied and can be taken into account somehow) the other systematic effects can contribute to the redshift errors, e.\,g. the fact that the only one template used for redshift calculation via cross-correlation technique is usually the composite spectrum stacked from a large number of single spectra without any separation by luminosity, spectral index, line equivalent width etc. But, as it was mentioned in Sec.\,\ref{sec:1}, these differences could be significant, and hence the use of only one `averaged' template for all types of quasar spectra could introduce additional uncertainties into the redshift measurements. Therefore, we tried to estimate, whether one can do better with redshift uncertainties using our composite spectra, generated with separation by spectral index, as templates.

\begin{figure*}
\centering
\epsfig{figure=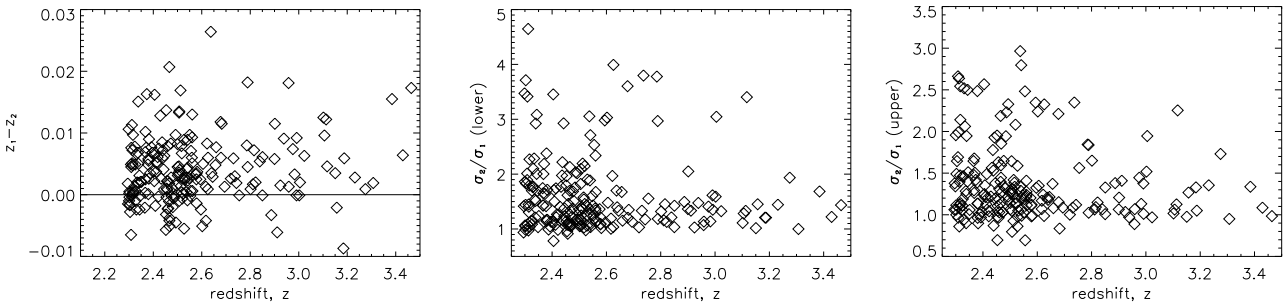,width=.98\linewidth}
\caption{Left to right: the difference between redshifts measured with our composites ($z_{1}$) and the SDSS template ($z_{2}$) as a function of $z$; ratio of the lower 1$\sigma$ marginalized uncertainties of $z_{2}$ to those of $z_{1}$; the same ratio for upper uncertainties.}
\label{fig:z-z}
\end{figure*}

\subsection{The test sample and SDSS template}\label{sec:3.2}

\indent\indent We selected by eye a test sample of spectra from our full sample of 3\,493 objects. These spectra are those of the most luminous objects, they have the highest signal-to-noise ratio and contain the smallest possible number of unwanted peculiarities,  which can influence the redshift measurement with the rough procedure described below (Sec.~\ref{sec:3.3}), like strong telluric lines  not extracted properly with SDSS pipelines. The redshifts of these objects, listed in HW catalogue, lie within the range of $2.3-3.5$, the arithmetic mean of their redshift errors from HW catalogue is 0.002, that is twice smaller than that of all objects with the same redshift range in the catalogue. We selected only the spectra with the index values within the range from $-2.2$ to $-0.8$ to match the index range of our composites. The final number of spectra in the test sample is 208. 

As a comparison template we have chosen the \citet{vandenberk+01} composite spectrum, which is used as cross-correlation template for redshift measurements in the SDSS\footnote{http://www.sdss.org/dr7/algorithms/redshift\_type.html}. It is compiled from a sample of 2204 quasar spectra from the Early Data Release of the SDSS. The sample covers a redshift range of $0.044-4.789$ and has a median redshift of $\bar{z}=1.253$. The spectral index of this composite  (within  the range of $1350-4230$\,\AA) is $\alpha_{\lambda}=-1.54$. 

It is worth to note, that our composites differ from that of \citet{vandenberk+01} not only in spectral index discretization, but also in the mean redshifts of the samples, which is about 2.8 for our ones. The reason for this difference is our intention to use selected sample for Ly$\alpha$-forest study, while \citet{vandenberk+01} utilised the whole redshift range of SDSS quasars. Here the question about redshift evolution of the quasar spectral shape and its influence on the redshift measurement accuracy arises. This problem is hard to study properly because of limitations imposed by the atmosphere on the observable spectral range. In fact, each observed quasar spectrum covers only a narrow part of the Big Blue bump, and hence each piece of the composite spectrum appears to be compiled from spectra of quasars with redshifts within some limited redshift range. Therefore, we can neglect the mean redshift difference between our samples and the sample of \citet{vandenberk+01}. 

\vspace*{-2ex}
\subsection{Algorithm}\label{sec:3.3}

\indent\indent The redshift measurement algorithm described below does not compete with specialised SDSS pipelines. It is developed only to test the difference between templates on example of `good' spectra, like those from our test sample, and cannot treat properly spectra with low signal-to-noise ratio, BALs, etc. Before utilisation we interpolated our composite spectra with spline polynomials, because they have about two-five times wider wavelength bins (2\AA), depending on the wavelength (Note, that the \citet{vandenberk+01} template has nonlinear dispersion as single SDSS spectra). 

To calculate the redshift with the help of our composites we were searching for the maximum value of the likelihood function $\EuScript{L}\sim\exp\left(-\chi^{2}/2\right)$, where
$$
\chi^{2}=\sum\left[f_{obs}(\lambda_{obs})-Af_{comp}(n,\lambda_{obs}/(1+z))\right]\sigma_{f}^{-2}.
$$
Here $f_{obs}$ and $f_{comp}$ are the fluxes in the test and template spectra, $\sigma_{f}$ is the noise of the test spectrum, and free parameters are flux normalization factor $A$, the composite number $n$, and the quasar redshift $z$. After finding the best fit values of parameters, $A$ and $n$ were fixed and 1$\sigma$ marginalized errors of $z$ were calculated. 

In the case of the SDSS template we used the same technique, but with only two free parameters: the flux normalization factor and the redshift.

\subsection{Results and discussion}\label{sec:3.4}

\indent\indent In Fig.~\ref{fig:z-z} (left panel) the difference $z_{1}-z_{2}$ is shown as a function of quasar redshift, where we denote the redshift measured with our templates as $z_{1}$ and those measured with the SDSS template as $z_{2}$. The middle and right panels of Fig.~\ref{fig:z-z} present the ratio $\sigma_{z,2}/\sigma_{z,1}$ of their lower and upper 1$\sigma$ uncertainties. One can see, that the redshifts measured with our templates are systematically higher, than those measured with the SDSS template, with the mean difference of 0.004. Meanwhile, the both lower and upper 1$\sigma$ uncertainties of the redshifts measured with our templates are systematically smaller. The mean difference of the lower and upper uncertainties are 0.022 and 0.008, correspondingly. Surely, the absolute values of the error differences cannot be compared to the redshift errors obtained with the SDSS pipeline, but their mean ratio (1.6 and 1.4 for lower and upper uncertainties, correspondingly) claims for possibility to reduce redshift errors up to 1.5 times when using a set of templates with different spectral indices instead of one mean template.   

\section[]{Application for $\bar{F}(z)$ measurement}\label{sec:4}
\subsection{General notes}\label{sec:4.1}

\indent\indent Composite spectra of quasars are also used in Ly$\alpha$-forest studies to determine intrinsic shape of quasar spectrum within the Ly$\alpha$-forest region prior to its absorption by intergalactic neutral Hydrogen. In these studies the Ly$\alpha$ forest region usually means the wavelength range between Ly$\beta$ and Ly$\alpha$ emission lines ($\sim1025-1216$\,\AA\ or more narrow), where the Ly$\alpha$-forest is not `contaminated' by other Ly-series forests. In high-resolution spectra, where it is easy to find unabsorbed parts of spectrum within the Ly$\alpha$-forest this procedure of `continuum' fitting is done directly with manual selection of such regions in each spectrum. Here, the term `continuum' is used for the whole intrinsic quasar spectrum including quasar emission lines. But for spectra with medium resolution spectra, like those in the SDSS, it is difficult to select unabsorbed regions due to blending of absorption lines. In this case another techniques are applied. They are based on similarity of quasar spectra, which allows to utilise composite spectra for estimation of the mean transmission at different redshifts. It was done, e.\,g. by \citet{bernardi+03,desjacques_07,greg+10}. More complicated, but similar to this one, method was proposed by \citet{mcdonald+06}. Recently, \citet{busca+12} used the idea of composite spectra utilisation for continuum fitting (to first approximation) in study of the Ly$\alpha$-forest with the first Baryon Oscillation Spectroscopic Survey (BOSS) data. 

The main idea of composite spectra utilisation for Ly$\alpha$ forest studies is the following. If the data (spectrum) are given in the form of pixels with (observed) wavelengths labelled $\lambda_{i}$, the flux density value $f_{i}$ and the noise $n_{i}$, one can present the observed flux density $f_{i}^{j}$ in $i$-th pixel within the Ly$\alpha$-forest region of $j$-th quasar as 
\begin{equation}\label{eq:flux_dens}
  f_{i}^{j} = A^{j}\bar{C}(\lambda_{rest})(1+\delta_{C,i}^{j})\bar{F}(z)(1+\delta_{F,i}^{j})+n_{i}^{j}, 
\end{equation}
where  $A_{j}$ is the flux normalization constant, the wavelength $\lambda_{i}$ of the absorption Ly$\alpha$ feature produced by the `cloud' of intergalactic H{\sc i} is related to its redshift $z_{i}$ as $\lambda_{i} = 1215.67(1+z_{i})$, $C(\lambda)$ is the mean `continuum' level (i.\,e. the mean intrinsic quasar spectrum),  $\delta_{C}$ are deviations of individual continuum from the mean one, $n$ is the noise, $\bar{F}$ is the mean transmission of intergalactic medium in the Ly$\alpha$ line for a given redshift, and $\delta_{F}$ are variance (or fluctuations) of transmitted flux. Different terms of expression~\eqref{eq:flux_dens} depend on different variables: namely, `continuum' $(C+\delta_{C})$ is a function of the rest wavelength $\lambda_{rest}$; mean transmission $\bar{F}$ is a function of redshift, $\delta_{F}$ depends on a line of sight (presented by a given quasar); and $n$ depends on the instrument properties and conditions of observation (thus can also be considered as a function of the observed 
wavelength $\lambda_{obs}$). Hence, the mean arithmetic composite spectrum for a given subsample with the mean redshift $\bar{z}$ can be presented as a product of the mean `continuum' and the mean transmission:
\begin{equation}\label{eq:bar_f}
f(\lambda_{rest})=\bar{C}(\lambda_{rest})\bar{F}(\bar{z}), 
\end{equation}
because the values $\langle\delta_{F,i}^{j}\rangle$, $\langle\delta_{C,i}^{j}\rangle$ and $\langle n_{i}^{j}\rangle$ are equal to zero when averaging over a large number of quasar spectra within a given redshift bin (see e.\,g.~\citet{bernardi+03,mcdonald+06} for details). Therefore, having several composite spectra stacked from subsamples of quasar spectra with different $z$ and assuming, that the mean quasar spectrum does not evolve with time, one can determine redshift dependence of $\bar{F}$ .

\subsection{Estimation of errors}\label{sec:4.2}

\begin{figure}\label{fig:norma-q}
\centering
\epsfig{figure=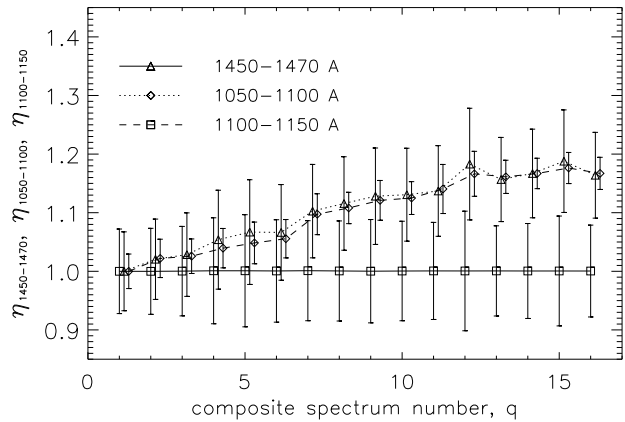,width=.99\linewidth}
\caption{Ratio of the mean flux in each composite spectrum and composite spectrum with $\alpha_{\lambda}=-0.91$ within the wavelength ranges $1450-1470$, $1050-1100$ and $1100-1150$\,\AA.}
\label{fig:spec-norma}
\end{figure}

\indent\indent To estimate the possible errors introduced to estimations of $\bar{F}(z)$ by neglect of the differences in spectrum shape, namely of the spectral index, we have calculated the following ratio:
\begin{equation*}
 \eta=\frac{\langle{f_{i}}\rangle_{q}}{\langle{f_{i}}\rangle_{1}},
\end{equation*}
where $\langle{f_{i}}\rangle$ is the mean flux within the given wavelength range in $q$-th and first composite spectra. We have chosen the following wavelength ranges: $1450-1470$, $1050-1100$ and $1100-1150$\,\AA. The first range is the one over which we normalized the spectra (see Sec.~\ref{sec:2.3}), that is why the ratio $\eta$ for it is equal to unity for all	 spectra. The latter two ranges are two parts of the Ly$\alpha$-forest region. In Fig.~\ref{fig:norma-q} the obtained values of $\eta$ are shown as function of the spectrum number. As one can see from this figure, deviation of $\eta$ from unity for spectra with different indices, which has to cause additional uncertainty in estimation of $\bar{F}$, can reach the value of 20\,\%, if the sample used for stacking composite contains spectra with indices within the range between $-2.14$ and $-0.91$. Note, that the real distribution of spectral indices is even wider (see Fig.~\ref{fig:alpha-distr}).

\section{Searching for emission lines}\label{sec:5}
\subsection{General notes}\label{sec:5.1}

\indent\indent One more opportunity provided by composite spectra is detection of weak lines, which are not resolved in single spectra due to low signal-to-noise ratio. We tried to search for emission lines within some parts of our composite spectra and to compare obtained results with previously known, firstly, with the most detailed study of quasar composite spectrum by \citet{vandenberk+01}. For this purpose we have chosen two wavelength ranges: $\approx1050-1200$\,\AA\ and $\approx1210-1450$\,\AA. The former is the so-called Ly$\alpha$-forest region containing three broad emission features identified by \citet{vandenberk+01} as Ar{\sc i}, Fe{\sc iii} and C{\sc iii}*, the latter lies redward of the Ly$\alpha$ emission line, it is free of intergalactic Ly$\alpha$ absorption and contains five broad emission features associated by \citet{vandenberk+01} mainly with N{\sc v}, Si{\sc ii}, O{\sc i}+Si{\sc ii}, C{\sc ii} and Si{\sc iv}+O{\sc iv}]. Due to blending C{\sc iii}*, N{\sc v} and Si{\sc ii} features appear to merge into one $\approx130$\,\AA-wide emission feature with the strongest Ly$\alpha$ emission line between them. 

To search for emission lines in these eight spectral ranges we modelled each one with the sum of continuum and the smallest possible number of lines in a form of the Gaussian profile. Due to line blending we modelled some features together: N{\sc v} with Si{\sc ii} and O{\sc i}+Si{\sc ii}, O{\sc iv}+Ca{\sc ii} with Si{\sc iv}+O{\sc iv}], hence the total number of separate spectrum parts, which were analysed, is five. The Ly$\alpha$ emission line has intensity several times higher then that of other features, and its profile is asymmetric (even if there was no blending by other lines) because its blue part is absorbed by intergalactic H{\sc i}. Therefore we have not tried to analyse the region around its peak, but included it separately into C{\sc iii}* and N{\sc v}+Si{\sc ii} features in a form of one additional line, despite the fact that it also can have a complex structure. 

As it was discussed above (see Sec.~\ref{sec:1}), the quasar continuum redward of the Ly$\alpha$ emission line is fitted well the power law. The wavelength range blueward from it is also considered to be a power law, but much steeper, or even close to unity \citep{zheng+97}. One should also keep in mind, that in both cases the power law is just an approximation of some smooth curve at limited ranges (indeed a tangent to a real curve determining UV continuum of quasars). That is why consideration of such narrow wavelength bins ($\approx$50-100\,\AA\ wide) as a constant, wherever it is possible, seems to be precise enough, when the main goal is detection of lines.

The main difference in modelling of the features blueward and redward of 1216\,\AA\ is the presence of some absorption described by $\bar{F}$ parameter according to \eqref{eq:bar_f}. Taking into account the similarity of the mean redshifts for all our subsamples we can assume $\bar{F}(z)=\bar{F}'$ to be the same for each subsample and renormalize \eqref{eq:bar_f} as $f(\lambda_{rest})=\bar{C}'(\lambda_{rest})$, where $\bar{C}'=\bar{C}\bar{F}'$.
 
\subsection{Spectra modelling}\label{sec:5.2}

\indent\indent We fitted each of five wavelength regions described above with the following model:
\begin{equation}\label{eq:compos}
 f(\lambda)=C+\sum\limits_{k}a_{k}\exp\left[-\frac{(\lambda-\lambda^{0}_{k})^{2}}{2w_{k}^{2}}\right].
\end{equation}
Here $\lambda$ is the rest frame wavelength, $a_{k}$, $\lambda_{k}$, $w_{k}$ are the amplitudes, the central wavelengths and FWHM (up to factor of $\sqrt{2}$) of each $k$-th emission feature, and $C$ describes continuum in a form of (i) constant $C=b$ or (ii) power law $C=d\lambda^{\alpha_{\lambda}}$ with fixed $\alpha_{\lambda}$ from Table~\ref{tab:samples}. The wavelength ranges along with the type of continuum shape used for modelling are presented in Table~\ref{tab-ranges}. For three first ranges the wavelength were selected manually for each spectrum and thus their wavelength ranges slightly differ.

\begin{table}
\centering
\caption{The spectral ranges considered in the present study and the forms of continuum used for their modelling.}\label{tab-ranges}
\vspace*{1ex}
\begin{tabular}{ccc}
\hline
n & range, \AA & continuum \\
\hline
1 & $\approx$1050--1095 & (i) \\
2 & $\approx$1095--1150 & (i) \\
3 & $\approx$1150--1185 & (i) \\
4 & 1215--1321 & (i) \\
5 & 1323--1447 & (ii) \\
\hline 
\end{tabular}
\end{table}

The fitting procedure was conducted in the following two steps: (a) using the \texttt{IDL lmfit} subroutine for each range from Table~\ref{tab-ranges} the best fit model in a form of \eqref{eq:compos} with the smallest possible number of Gaussians was found; (b) the central wavelengths $\lambda^{0}$ obtained at the first step were fixed and the best fit values of other parameters $\{b/d,a_{k},w_{k}\}$ with the 1$\sigma$ marginalized errors were calculated by the Markov Chain Monte Carlo (MCMC) method using \texttt{CosmoMC} package as a generic sampler (with the values of $\{b/d,a_{k},w_{k}\}$ obtained previously as starting values). The MCMC technique has been chosen as it is fast and accurate method of exploration of high-dimensional parameter spaces. In each case we generated 8 chains which have converged to $R-1<0.0015$.

Due to the small values of amplitudes of all lines compared to the Ly$\alpha$ emission line and larger uncertainty at Ly$\alpha$-forest range, the central wavelength of Ly$\alpha$ was determined only while fitting the 4-th range, and the same value was used as fixed while fitting the 3-rd range. Note, that we include the Ly$\alpha$ feature into both wavelength ranges only to take into account the wings of this feature, the influence of which we cannot neglect in both cases. Thus the parameters of this feature obtained from fitting of both regions cannot be considered as the parameters of the Ly$\alpha$ emission line.

Due to the small values the flux dispersion, $\sigma_{f}^{2}$, in composite spectra determined from the covariance matrix to estimate errors correctly we introduced additional intrinsic errors, $\sigma_{int}$, determined such as the total dispersion, $\sigma^{2}=\sigma_{f}^{2}+\sigma_{int}^{2}$, is such resulting the minimal $\chi^{2}/$d.o.f. to be $1-0$.

\subsection{Line identification}\label{sec:5.3}

\indent\indent The central wavelengths of all the lines found in composite spectra are presented in Tables~4 and 5 for ranges 1-3 and 4-5, correspondingly. Only the lines for which the parameters were calculated with the help of MCMC method are presented. The arrows stand for regions for which the fit failed, `tent.' means tentative identification by eye (which was not included in fit). 

Despite of variance in central wavelengths and the number of lines in different spectra we tried to systematize them. In most cases, except the the 3rd and 5th ranges, the lines seem to be the same in all spectra, but due to the difference in their FWHM in one spectrum two given lines are resolved, while in another spectrum they are blended and appear to be fitted better with one Gaussian. The values in brackets in Tables~4, 5 mean that such dublets are fitted by one Gaussian.

For the line identifications we used information from Table~3, where we listed emission lines found previously in composite UV spectra of quasars from HST and FUSE missions by \citet{zheng+97,telfer+02,scott2+04}, high-resolution by \citet{brotherton+94,Laor+94,Laor+1995,laor+97,Vestergaard+01,leighly+07,binette+08}, and also in composite spectra of quasars from optical Kast \citep{tytler+04} and SDSS \citep{vandenberk+01} surveys. The lines which were not identified with known lines from these papers are labelled with X$_{k}$.

\subsection{Parameters of lines and continuum}\label{sec:5.4}

\begin{figure}
\centering
\epsfig{figure=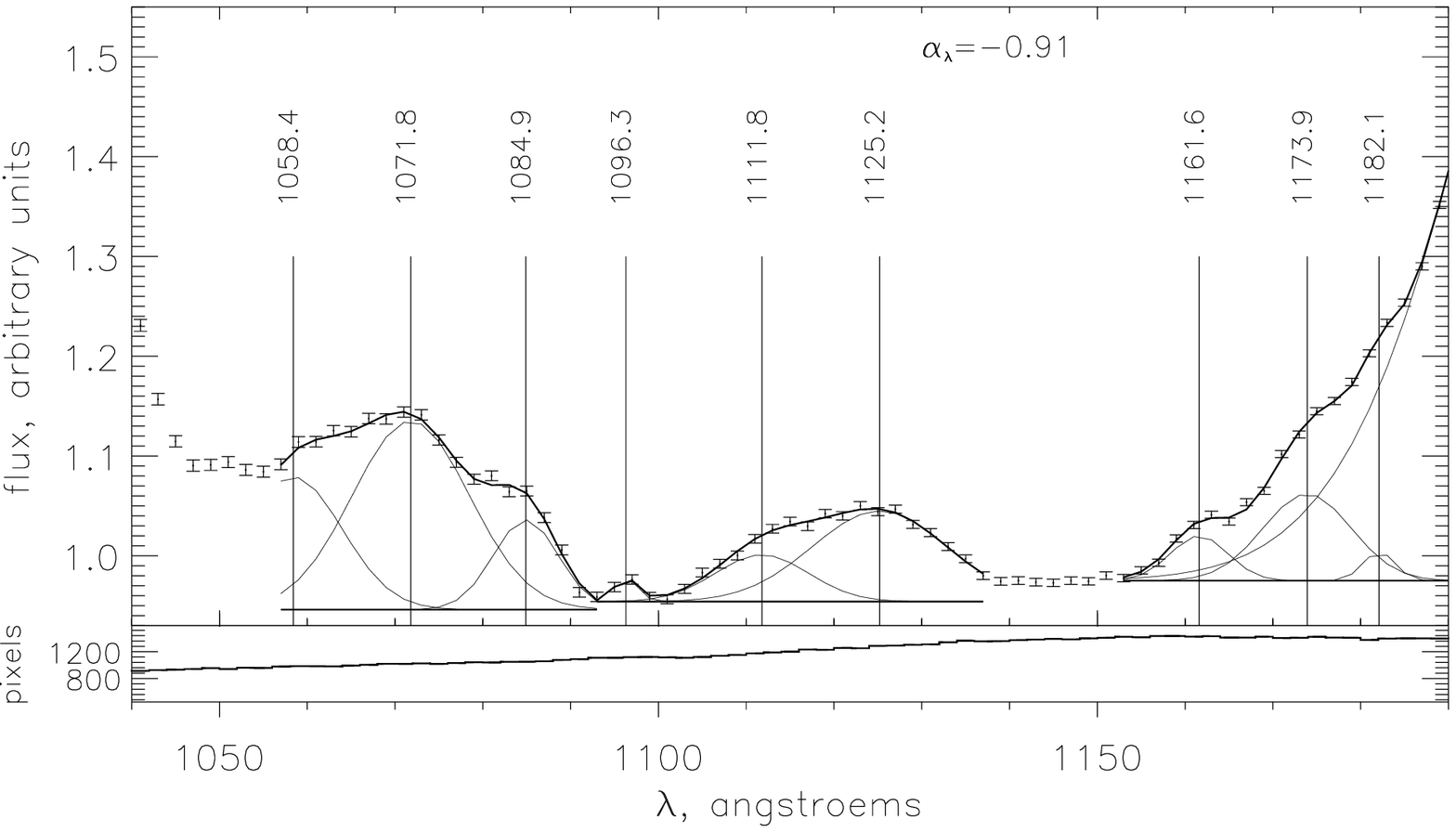,width=.99\linewidth}
\epsfig{figure=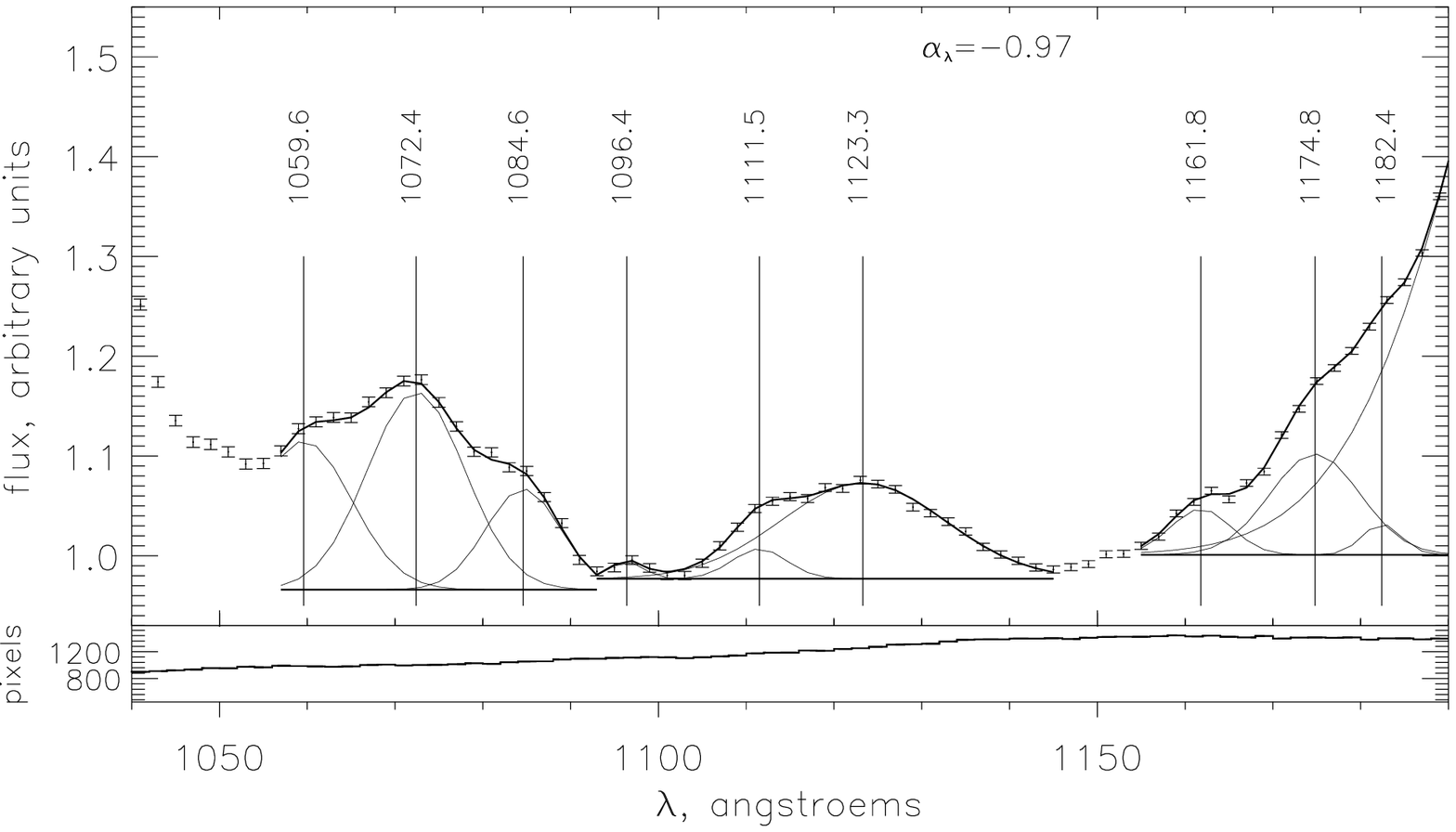,width=.99\linewidth}
\epsfig{figure=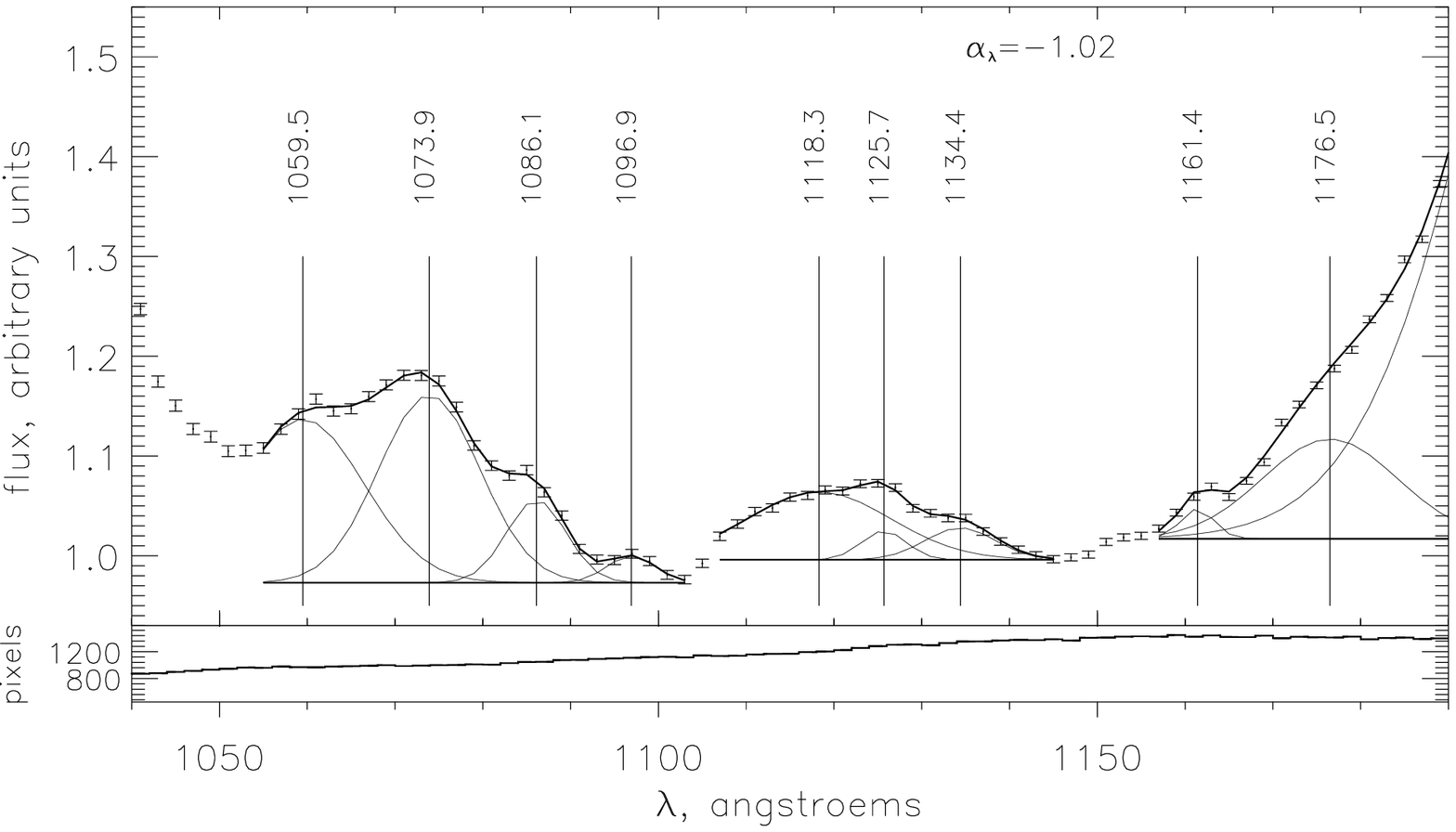,width=.99\linewidth}
\epsfig{figure=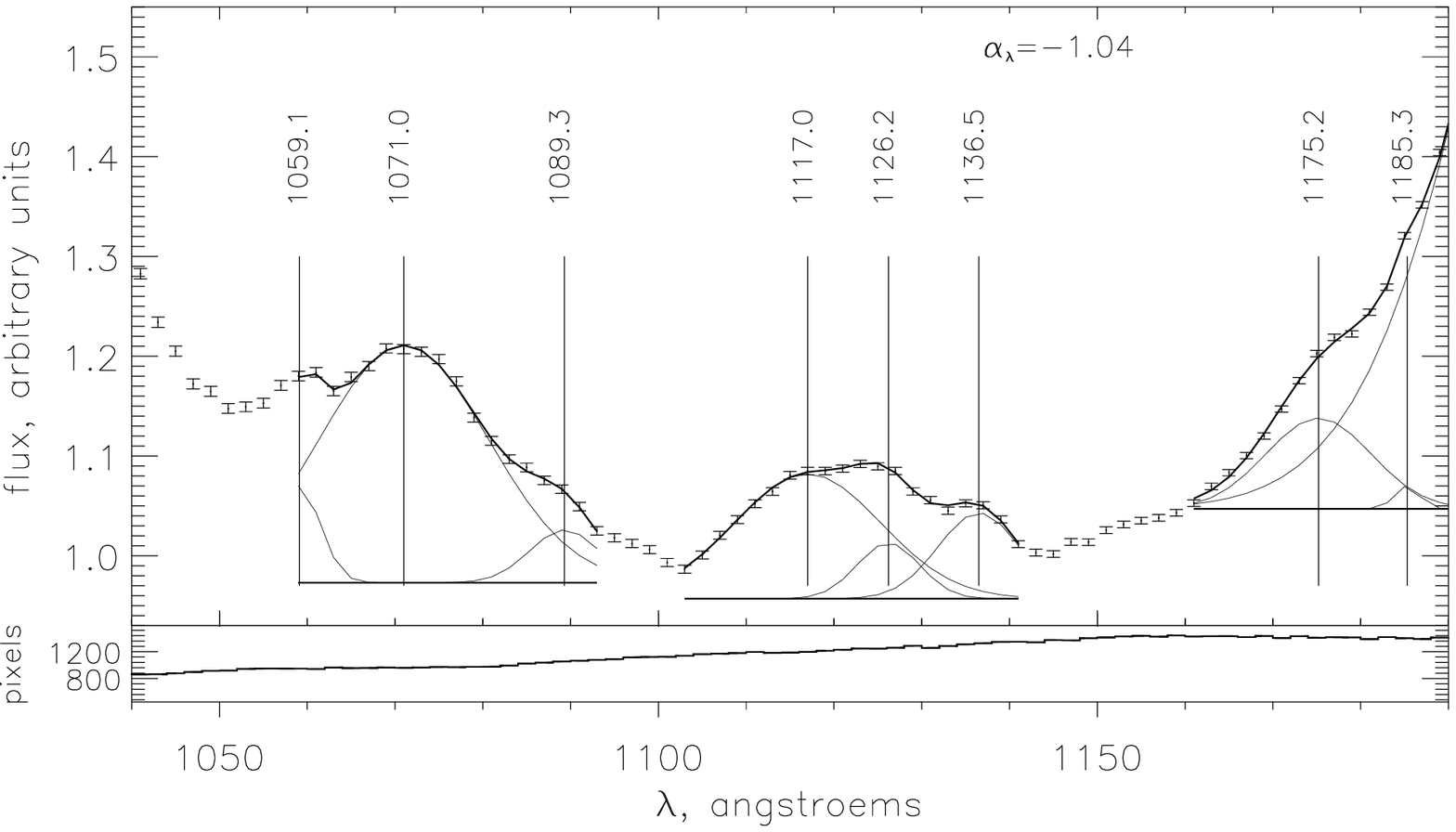,width=.99\linewidth}
\caption{Composite spectra 1$-$4 with $\alpha_{\lambda}$ (from top to bottom): $-0.91$, $-0.97$, $-1.02$, $-1.04$ (blue part).}
\label{fig:spec-1-4-a}
\end{figure}

\begin{figure}
\centering
\epsfig{figure=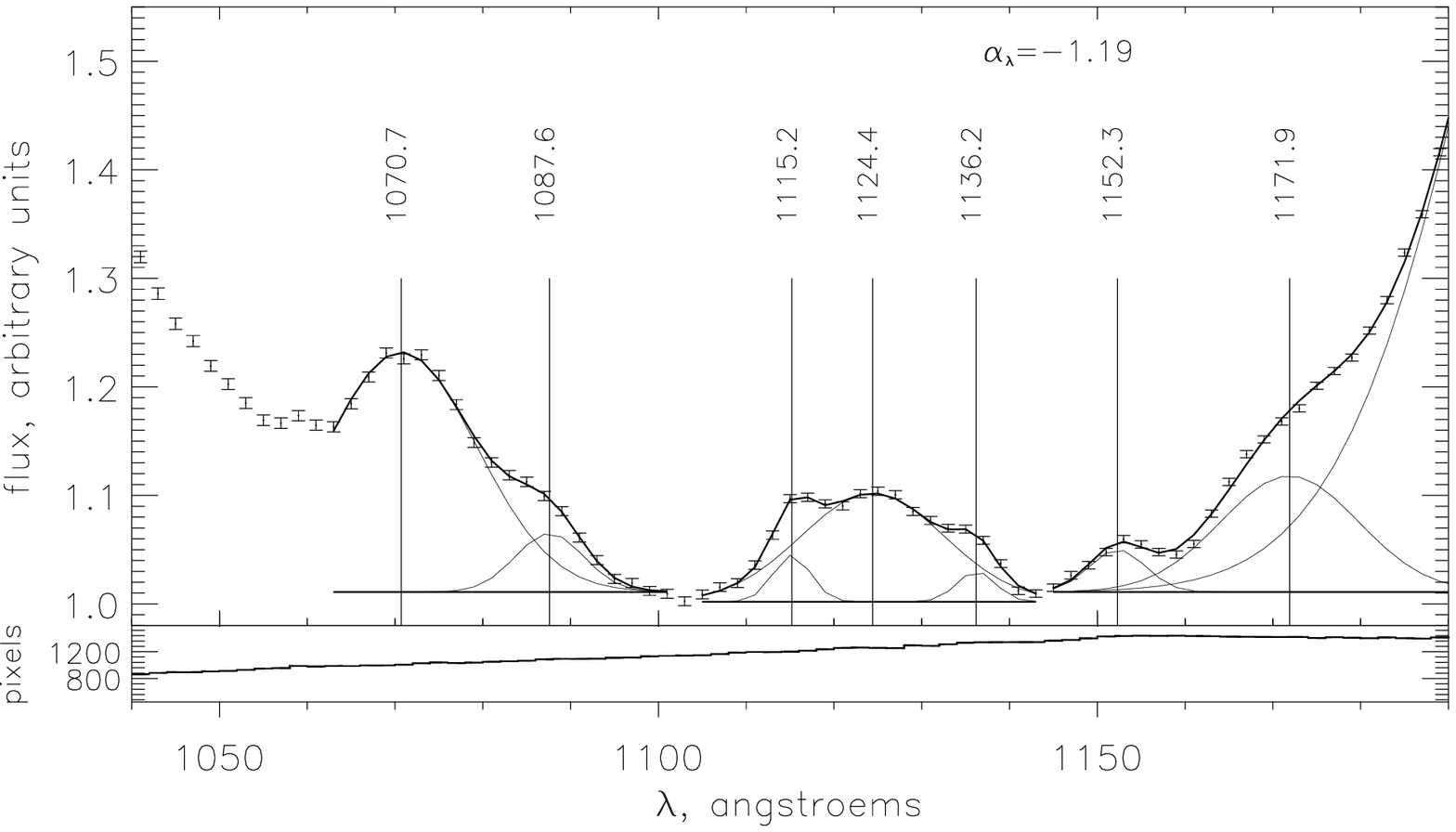,width=.99\linewidth}
 \epsfig{figure=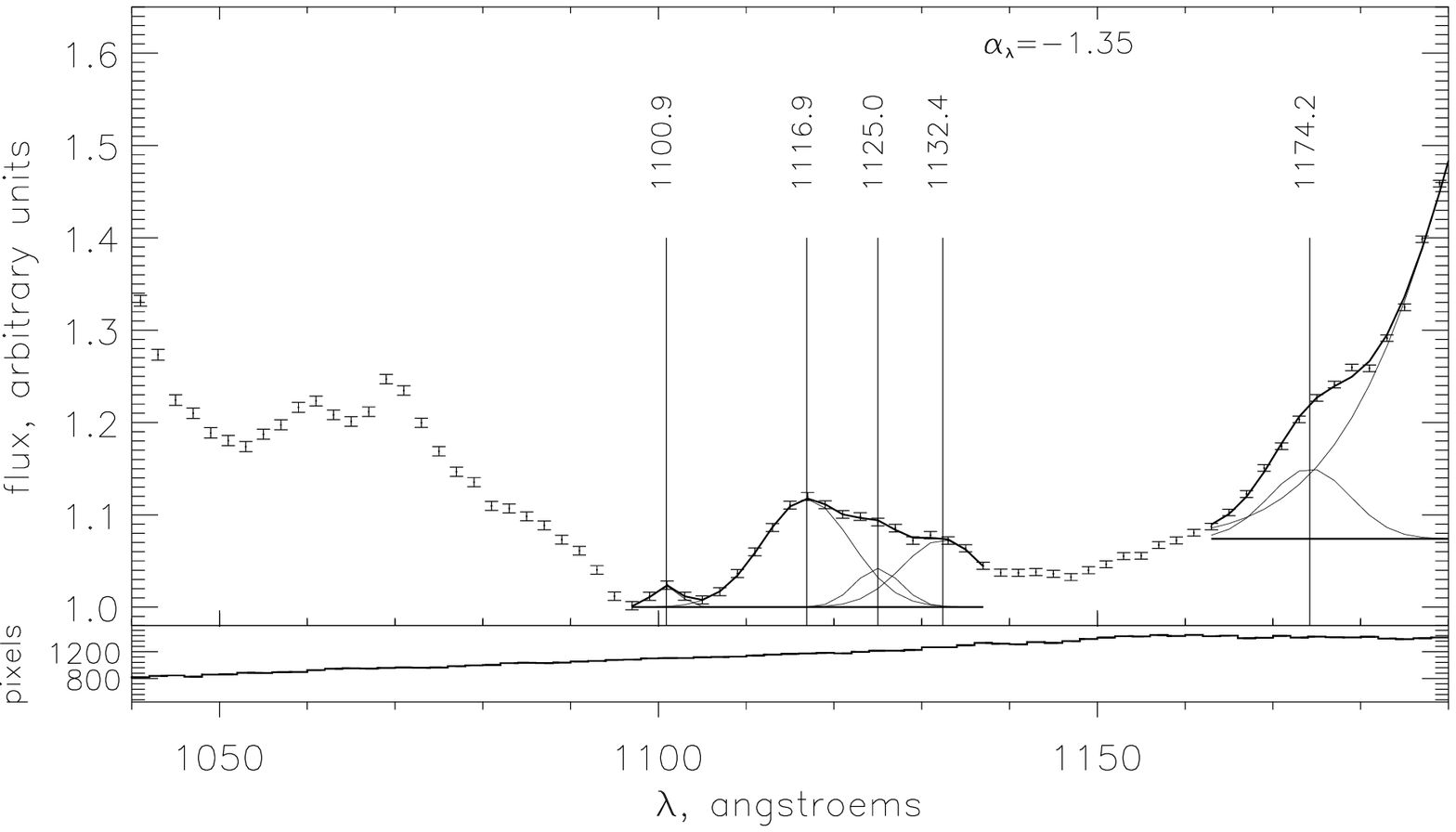,width=.99\linewidth}
 \epsfig{figure=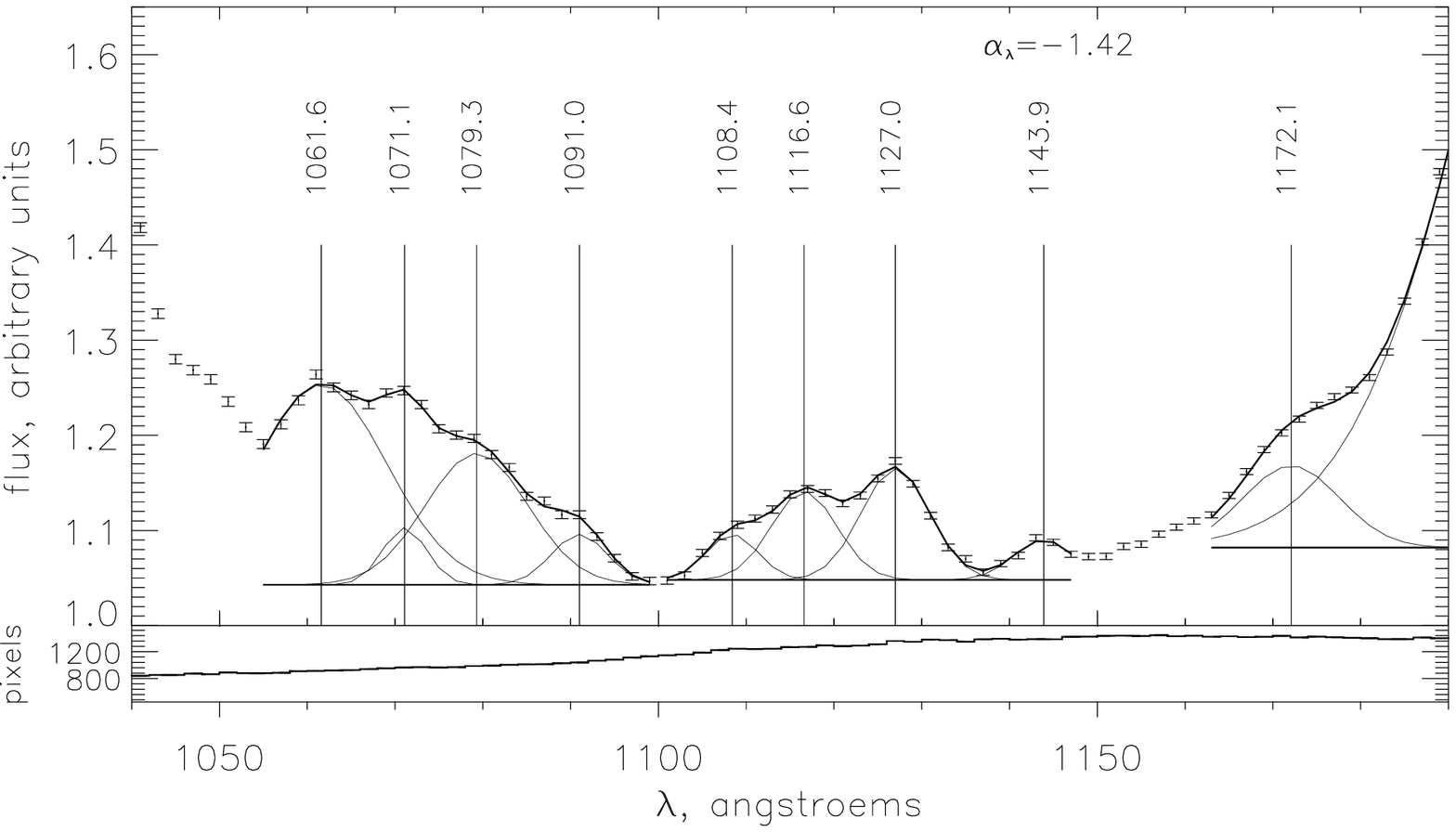,width=.99\linewidth}
 \epsfig{figure=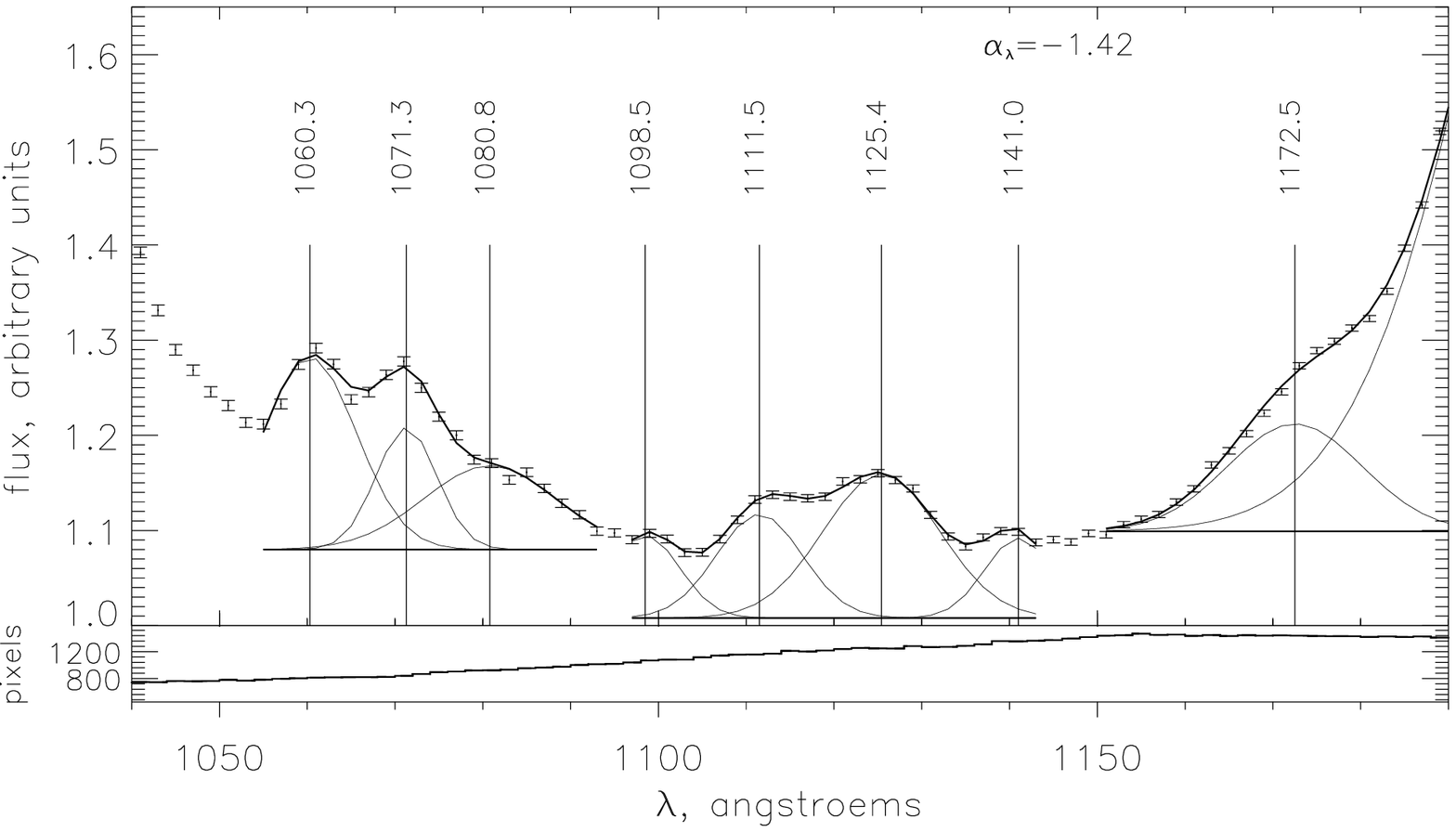,width=.99\linewidth}
\caption{Composite spectra 5$-$8 with $\alpha_{\lambda}$ (from top to bottom): $-1.19$, $-1.35$, $-1.42$, $-1.42$  (blue part).}
\label{fig:spec-5-8-a}
\end{figure}

\begin{figure}
 \centering
 \epsfig{figure=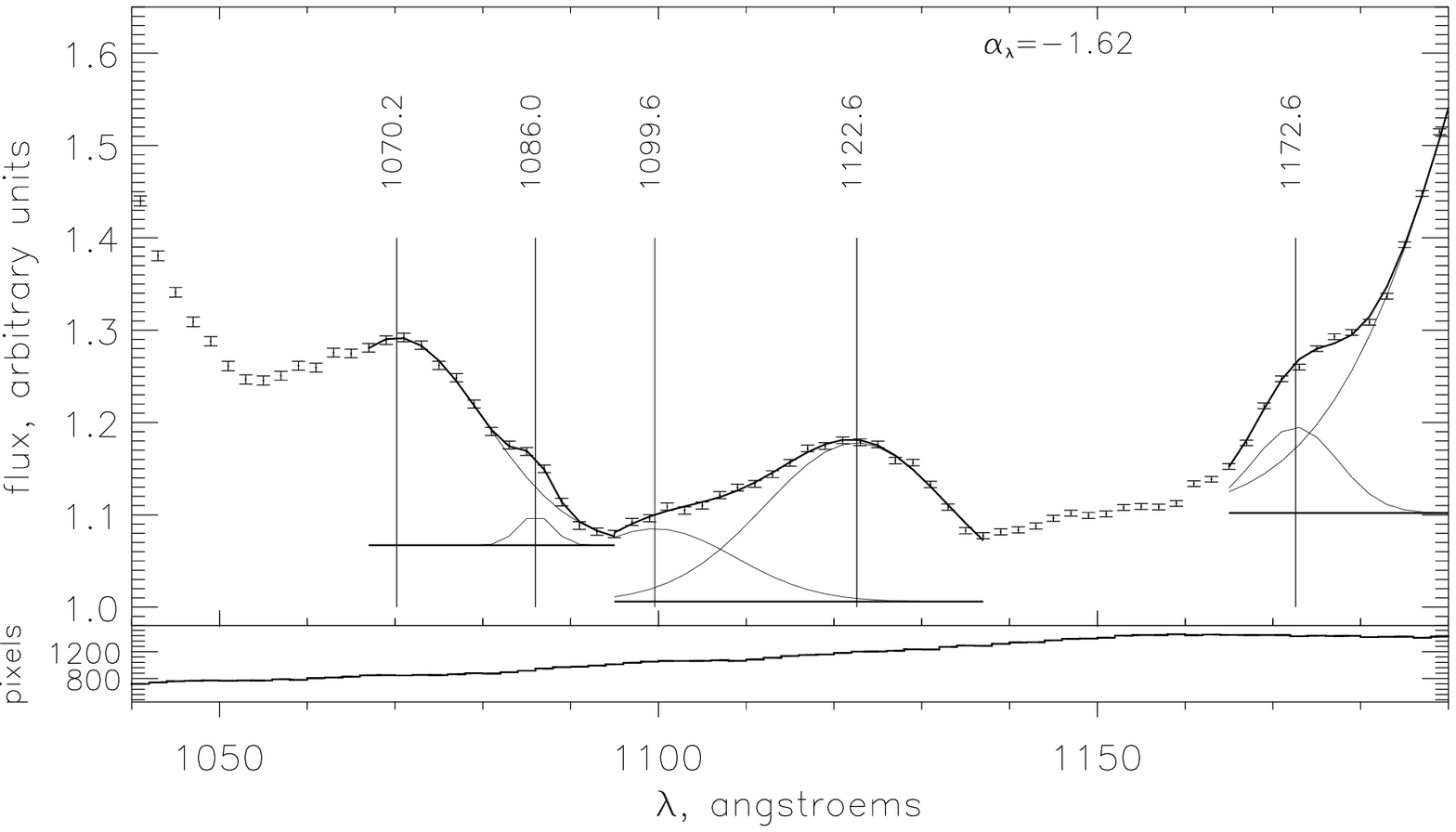,width=.99\linewidth}
 \epsfig{figure=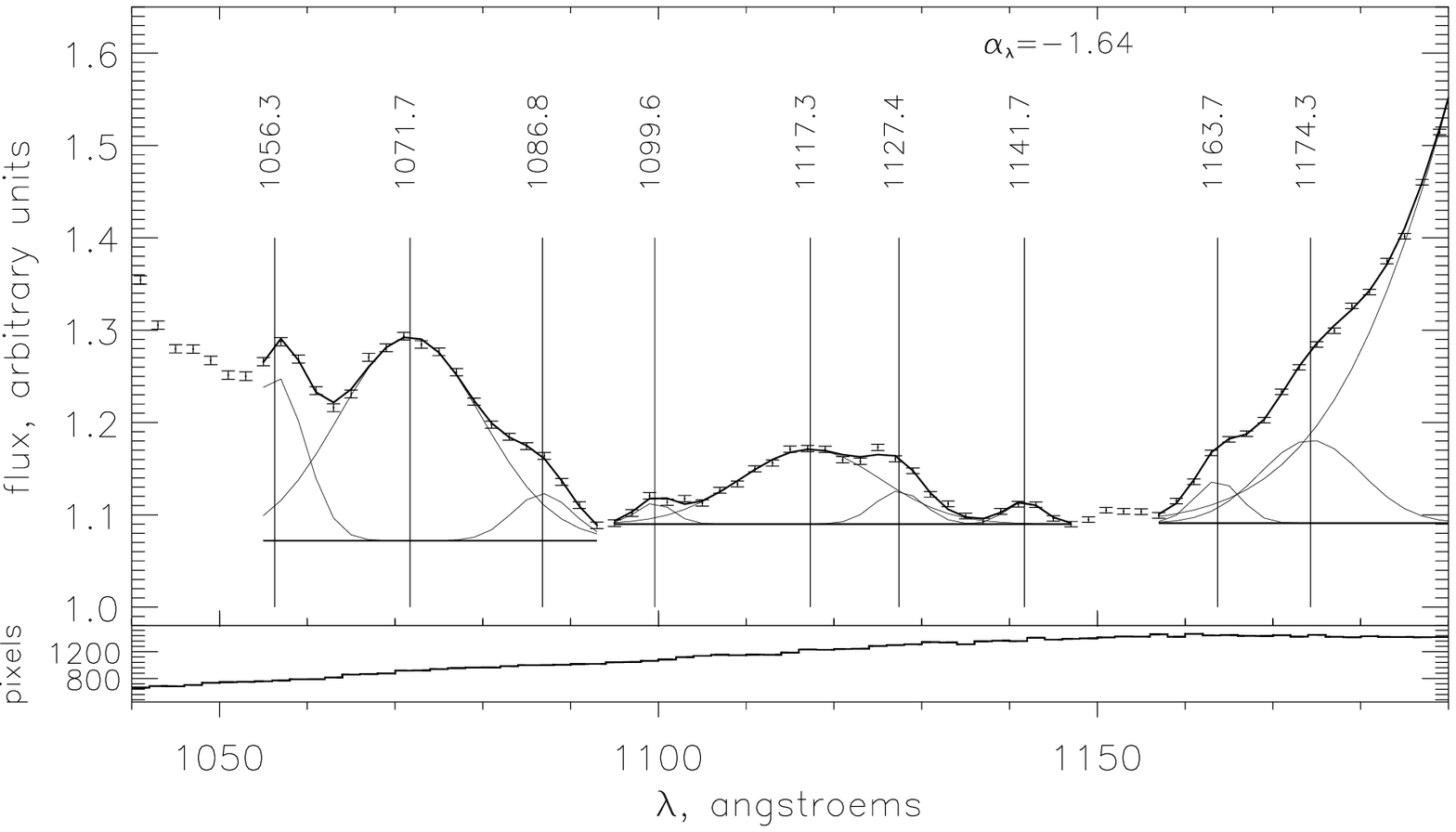,width=.99\linewidth}
 \epsfig{figure=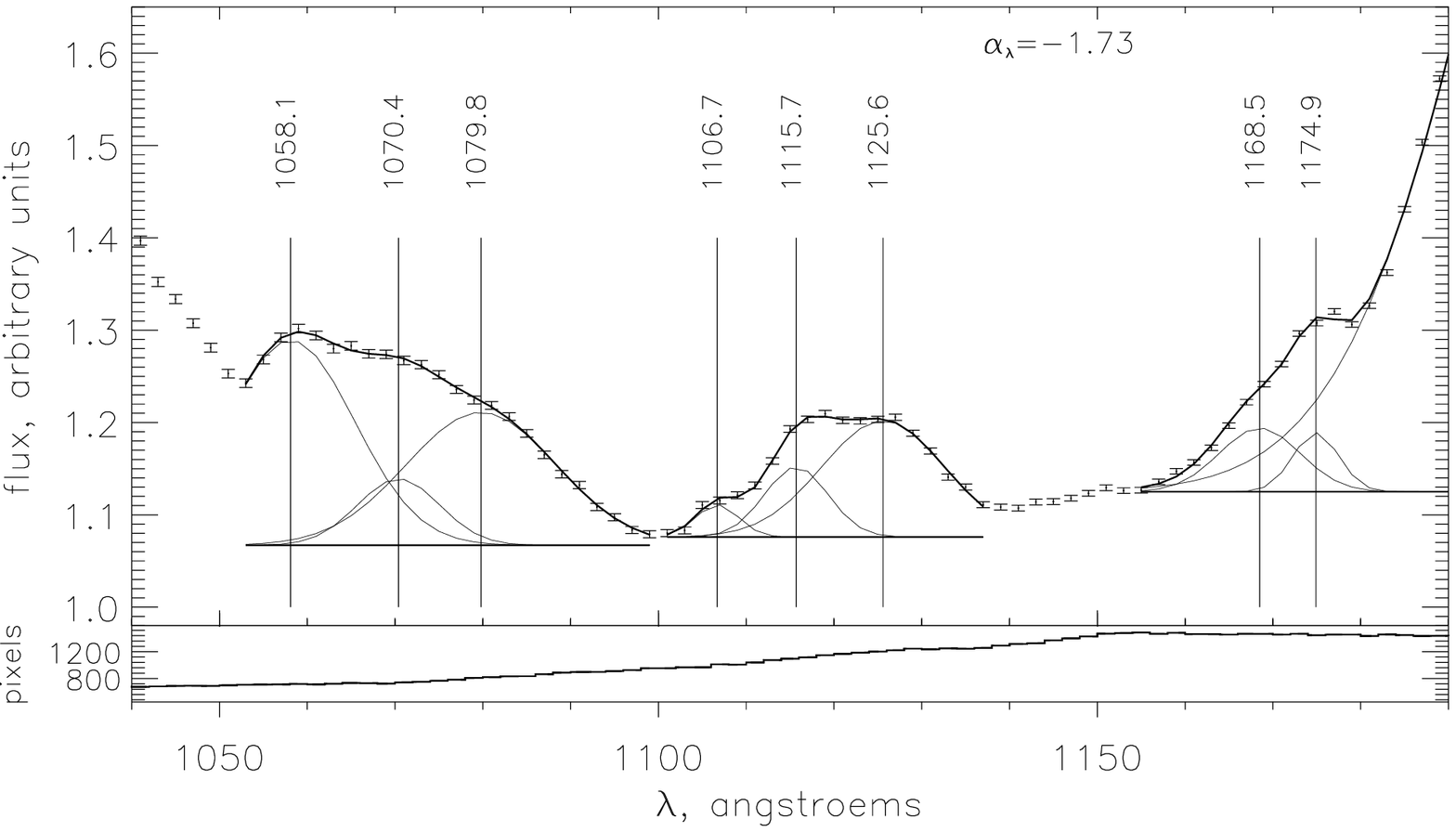,width=.99\linewidth}
 \epsfig{figure=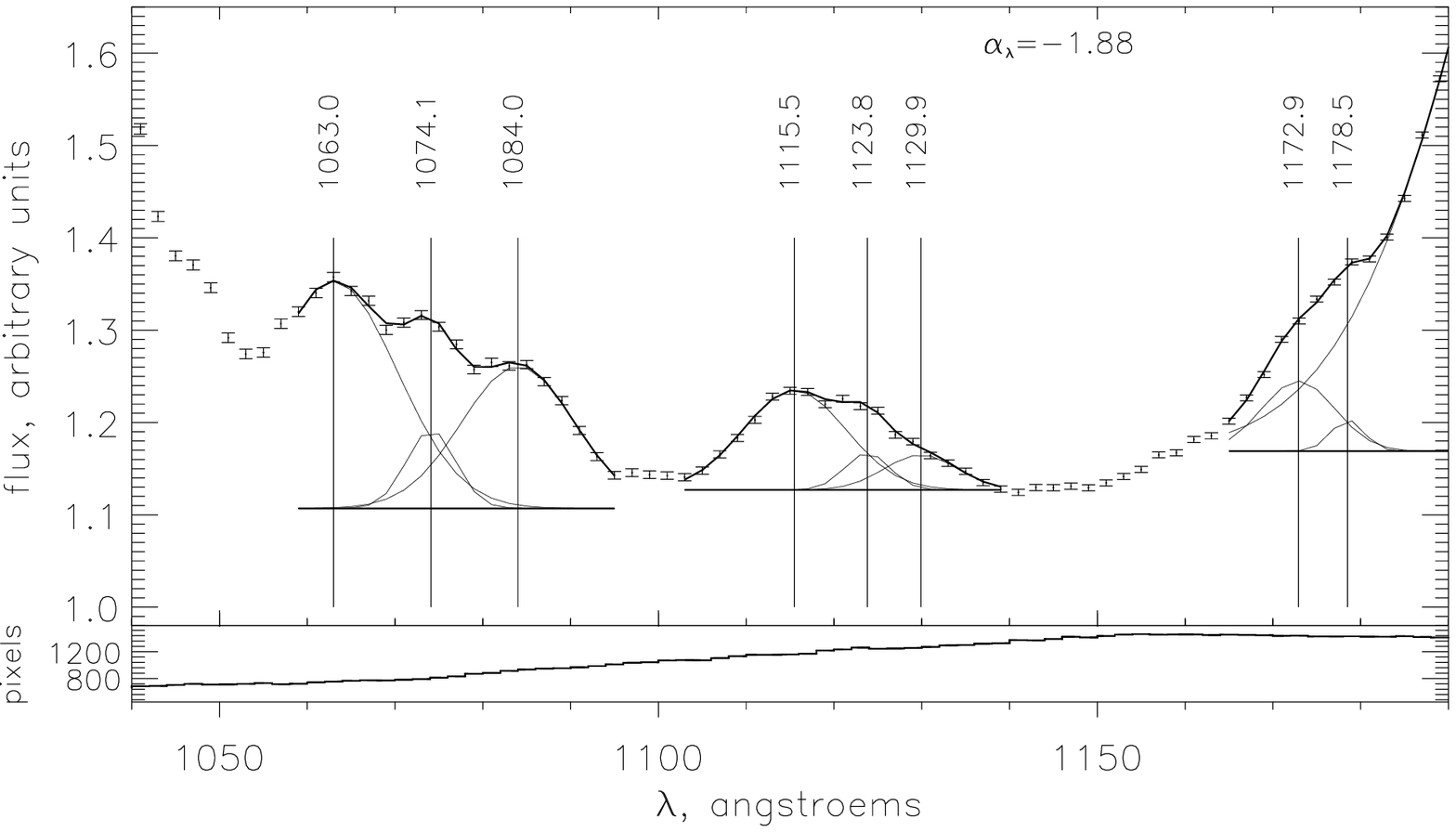,width=.99\linewidth}
 \caption{Composite spectra 9$-$12 with $\alpha_{\lambda}$ (from top to bottom): $-1.62$, $-1.64$, $-1.73$, $-1.88$  (blue part).}
 \label{fig:spec-9-12-a}
 \end{figure}

\begin{figure}
 \centering
 \epsfig{figure=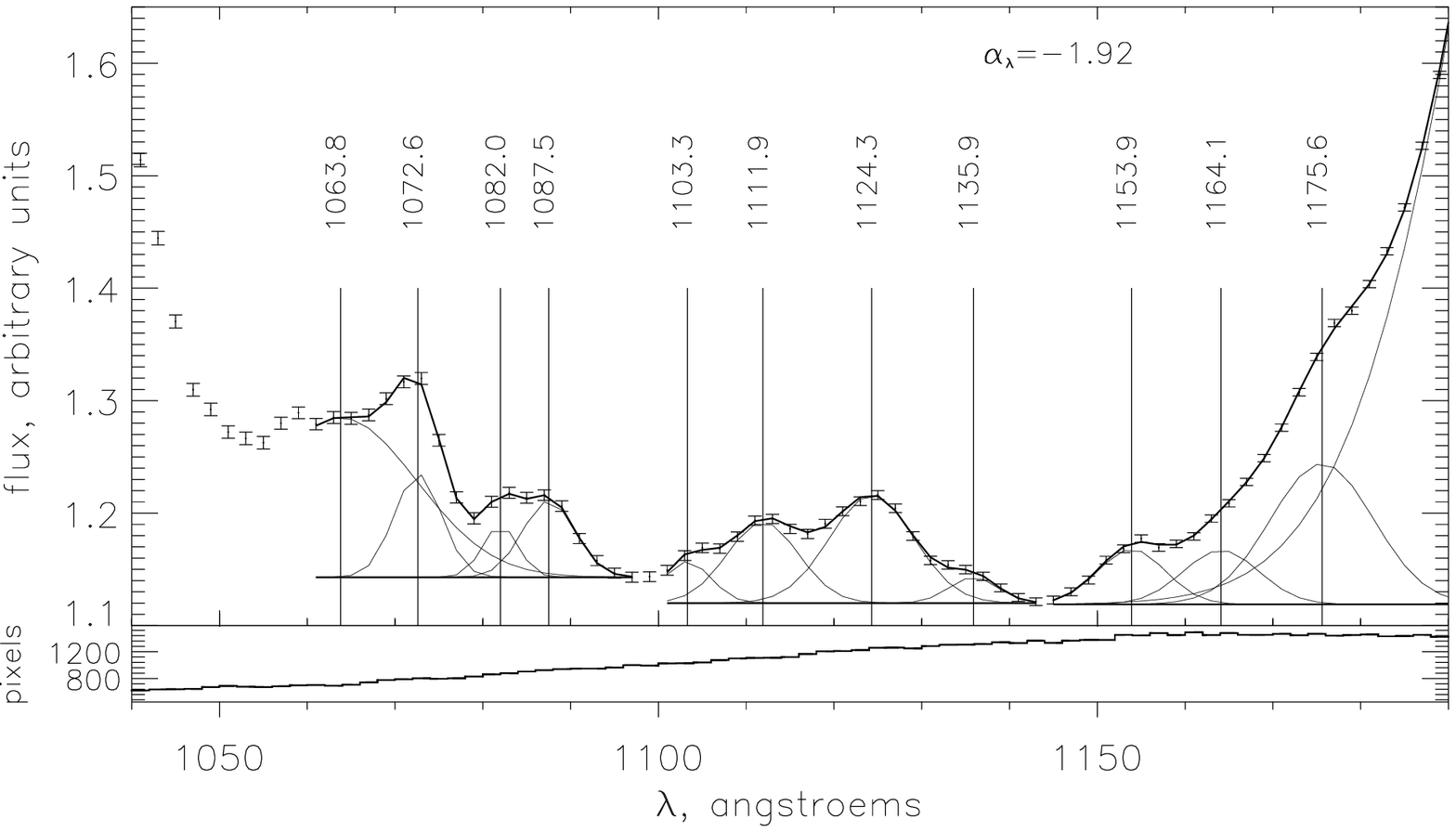,width=.99\linewidth}
 \epsfig{figure=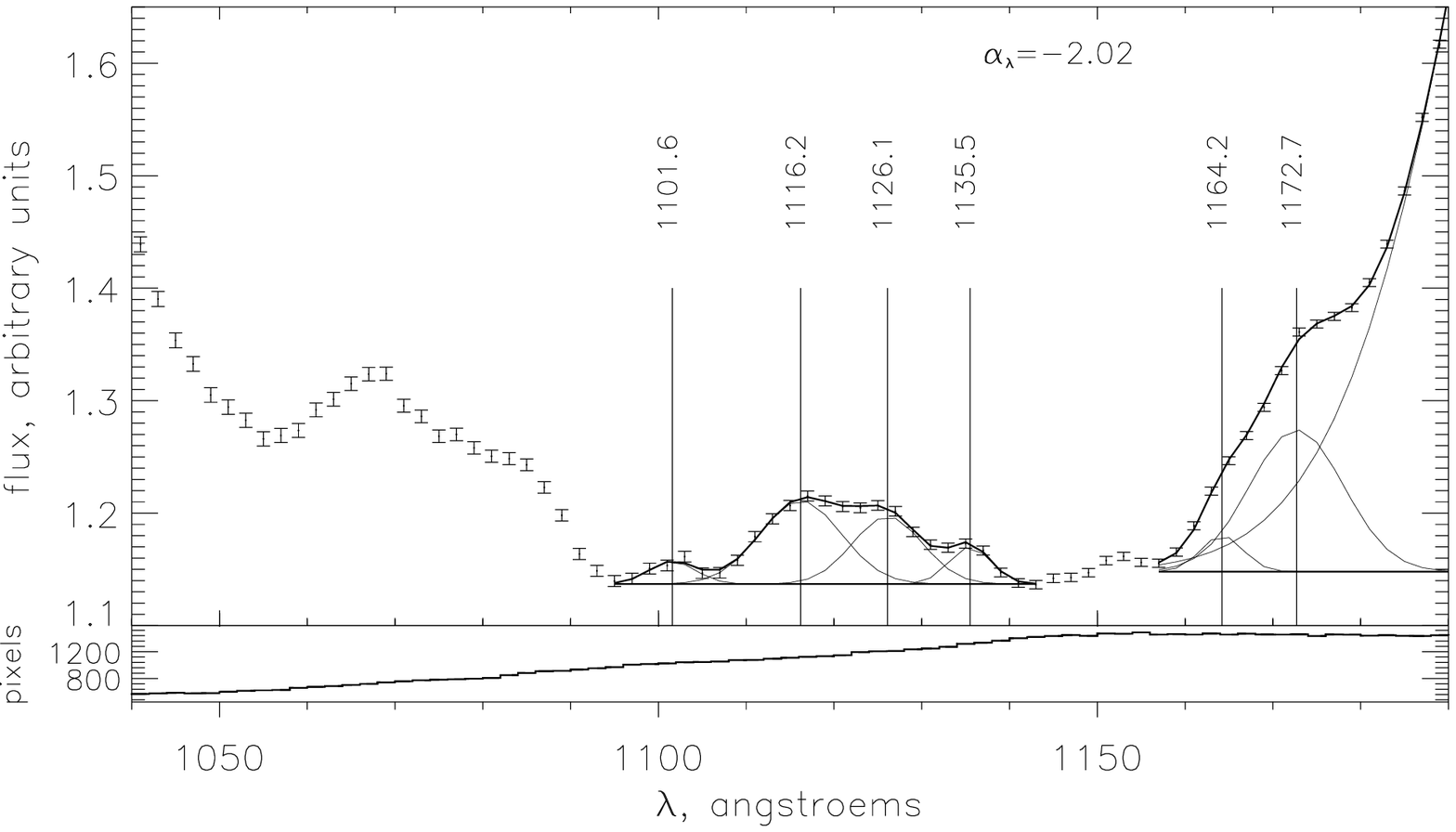,width=.99\linewidth}
 \epsfig{figure=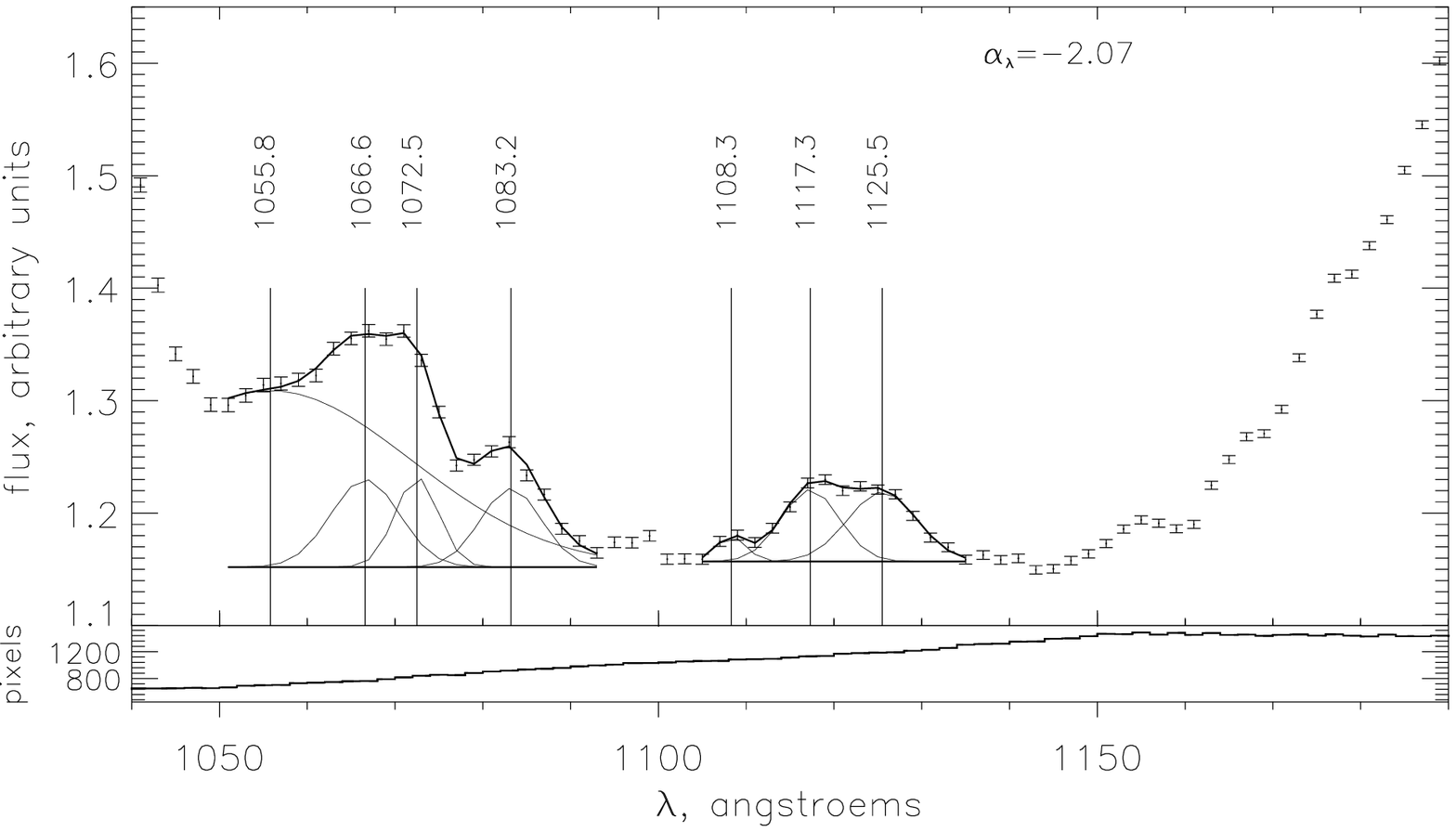,width=.99\linewidth}
 \epsfig{figure=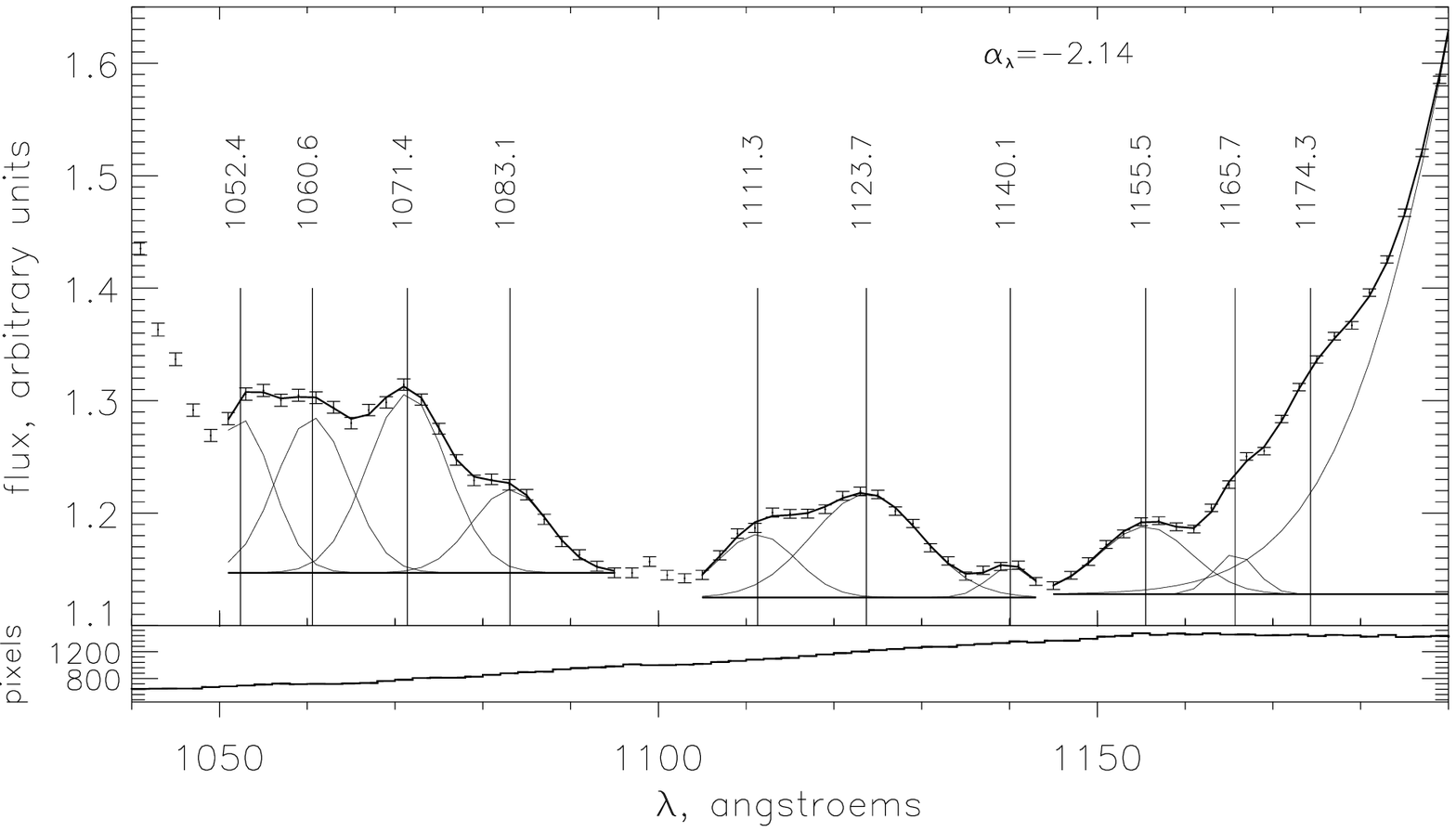,width=.99\linewidth}
 \caption{Composite spectra 13$-$16 with $\alpha_{\lambda}$ (from top to bottom): $-1.92$, $-2.02$, $-2.07$, $-2.14$  (blue part).}
 \label{fig:spec-13-16-a}
 \end{figure}

\indent\indent The obtained values of the line parameters along with their 1$\sigma$ marginalized errors are presented in Tables~\ref{tab-spec-1}-\ref{tab-spec-16}, the values of constant $b$ for ranges 1--4 and $d$ for range 5 are presented in Table~\ref{tab:b-param}. The spectra along with the best fits for each range, separate Gaussians and continuum level are shown in Figures~\ref{fig:spec-1-4-a}-\ref{fig:spec-13-16-b}. In the last column of Tables~\ref{tab-spec-1}-\ref{tab-spec-16} the lower 3$\sigma$ marginalized limits a-3$\sigma$ for the amplitudes of each line are presented. All these limits are $>0$, this allows us to claim that all the `lines' are really detected in the composite spectra at least at 3$\sigma$ confidence level.
  
\subsection{Discussion}\label{sec:5.5}

\indent\indent As one can see from Table~\ref{tab:b-param} in most cases the values of continuum parameter $b$ for ranges 1--3 within one spectrum vary from one range to another within errors. Therefore, continuum level within the range between Ly$\beta$ and Ly$\alpha$ emission lines, utilized for studies of the Ly$\alpha$-forest, can be considered as a constant rather than having the same power-law form as that redward from Ly$\alpha$ emission line. This result agrees well with results from the composite UV spectra of quasars \citep{telfer+02,zheng+97}, which evidence for much steeper continuum in the Ly$\alpha$-forest region, than that from $\lambda>1216$\,\AA\ range. 

\begin{figure}
\centering
\epsfig{figure=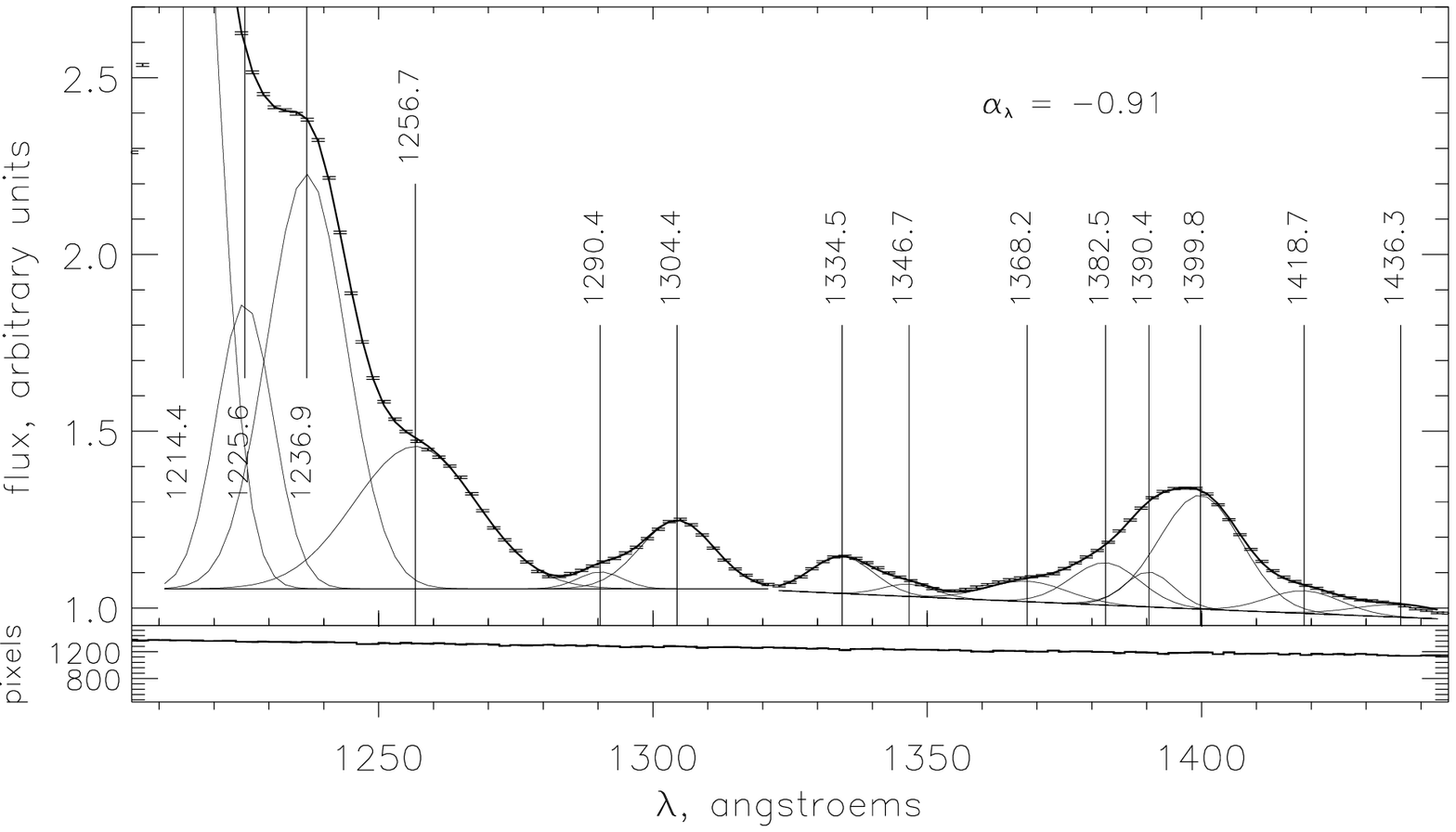,width=.99\linewidth}
\epsfig{figure=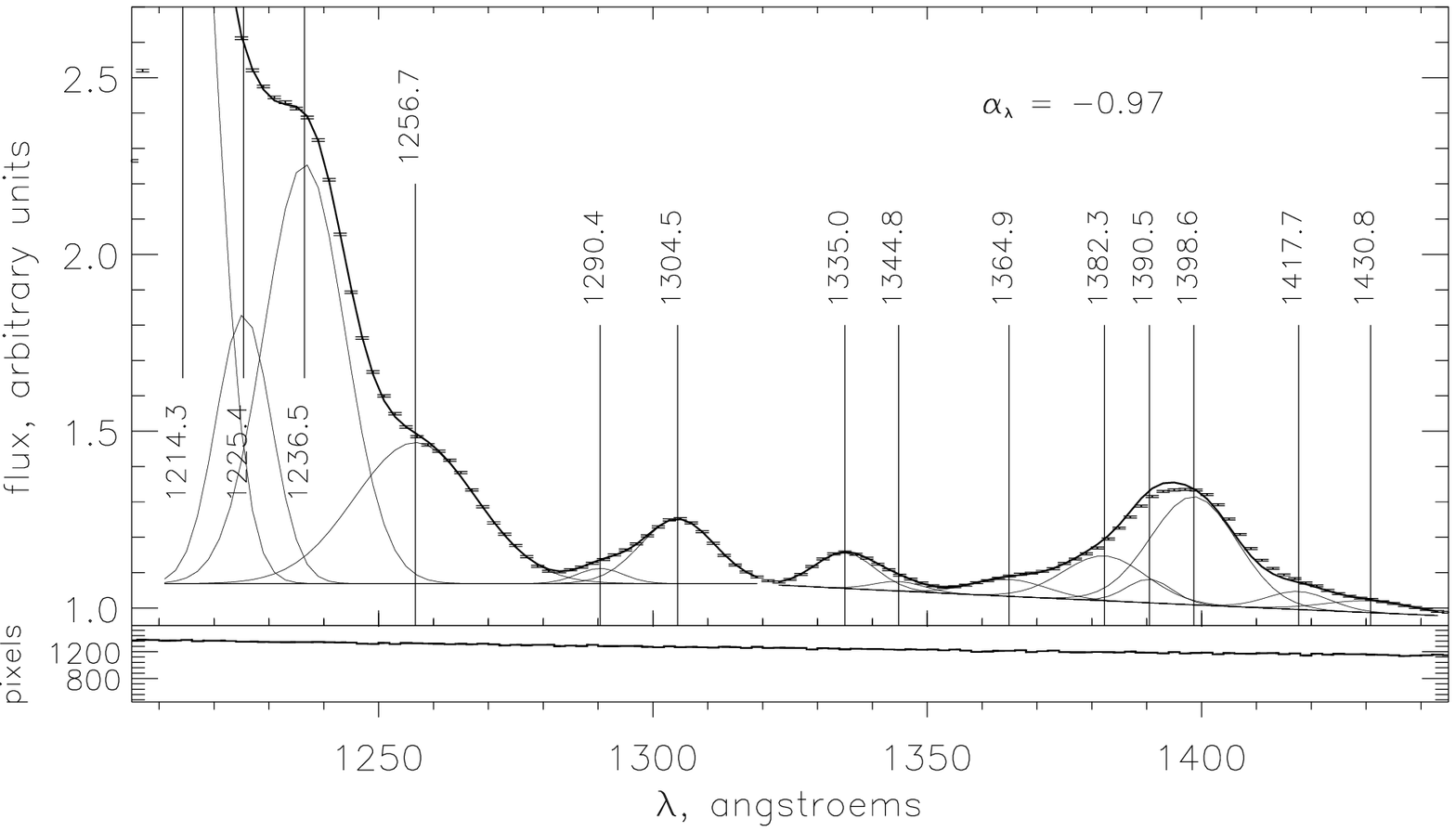,width=.99\linewidth}
\epsfig{figure=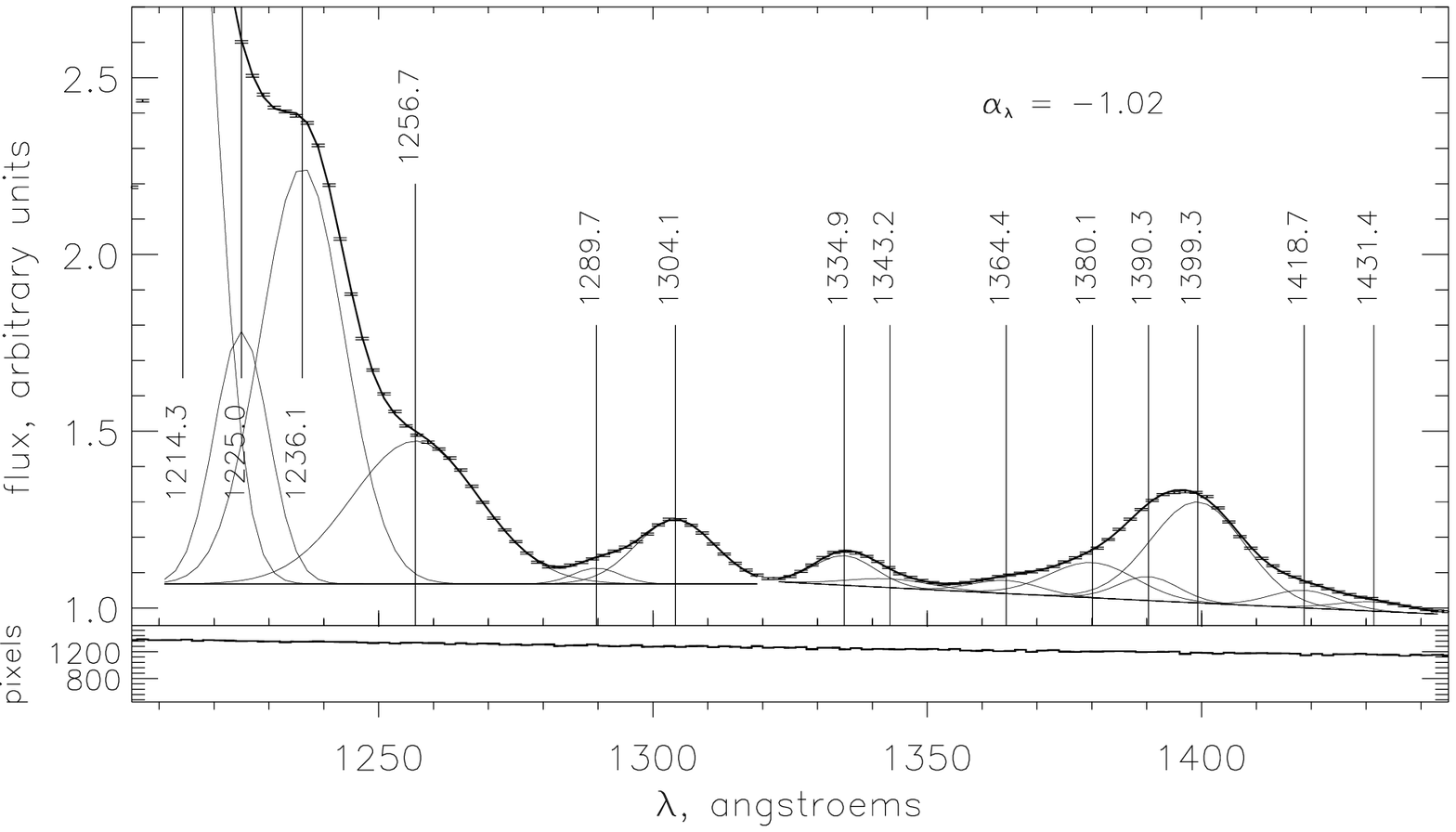,width=.99\linewidth}
\epsfig{figure=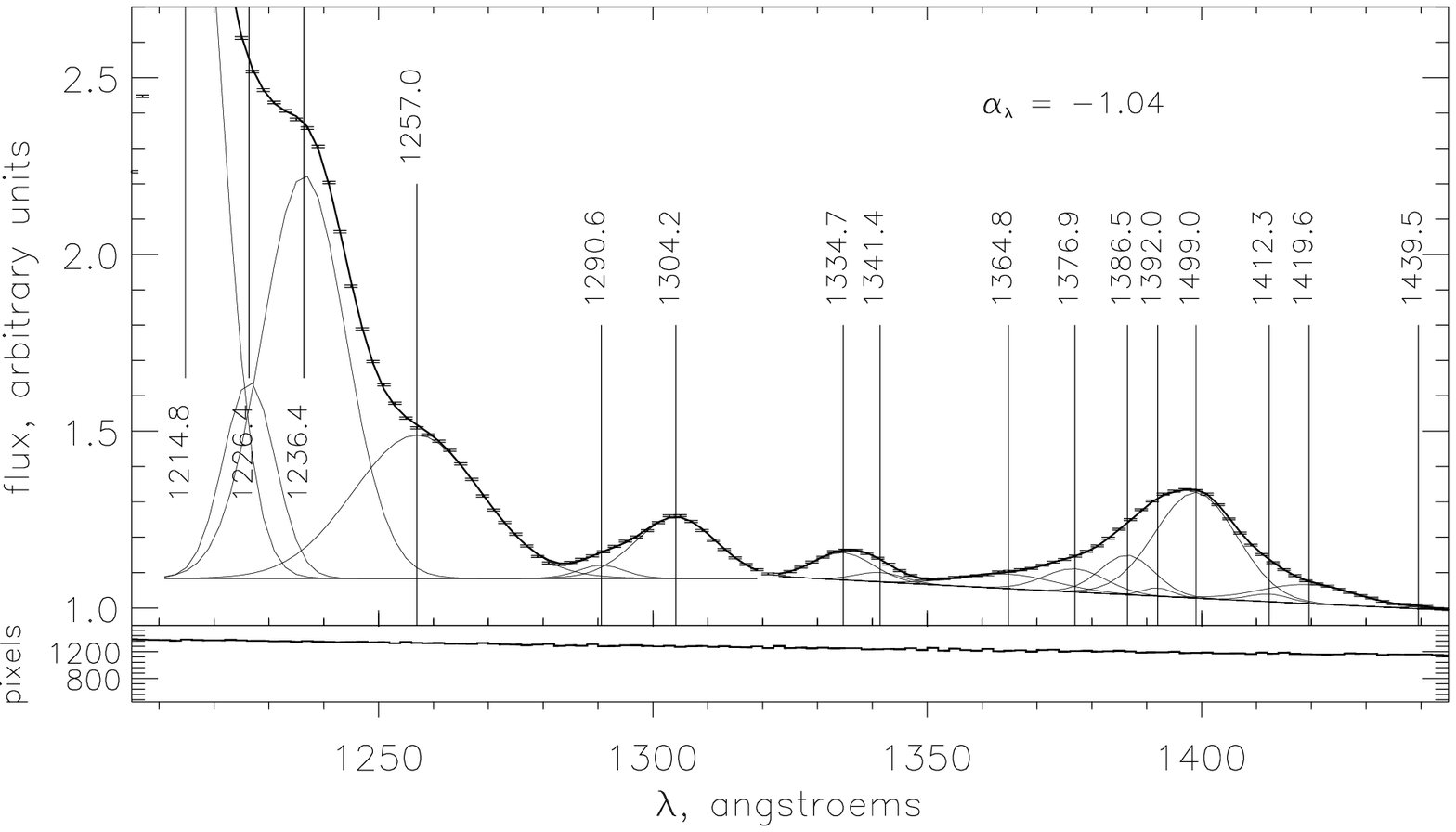,width=.99\linewidth}
\caption{Composite spectra 1$-$4 with $\alpha_{\lambda}$ (from top to bottom): $-0.91$, $-0.97$, $-1.02$, $-1.04$  (red part).}
\label{fig:spec-1-4-b}
\end{figure}

\begin{figure}
\centering
\epsfig{figure=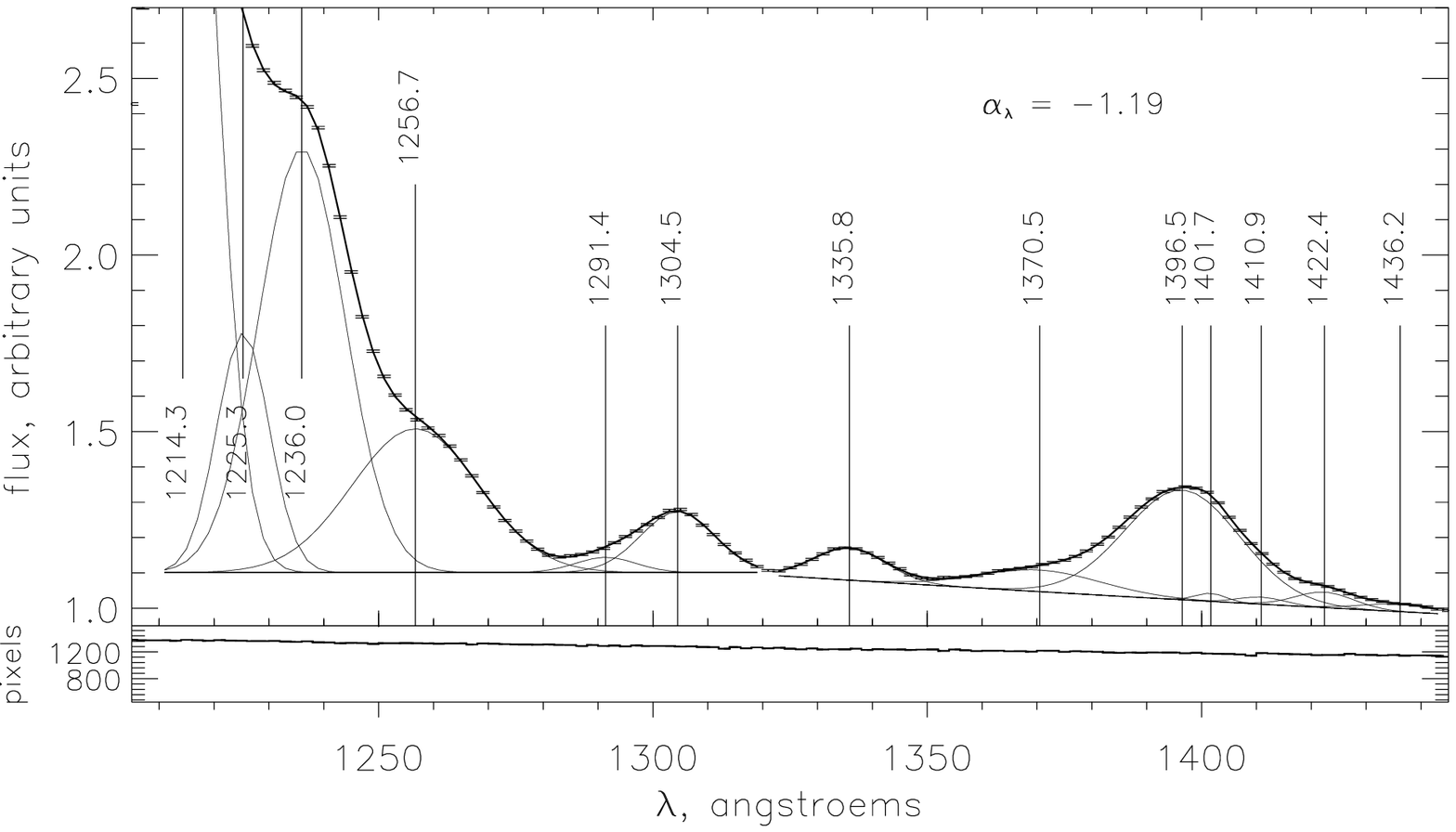,width=.99\linewidth}
 \epsfig{figure=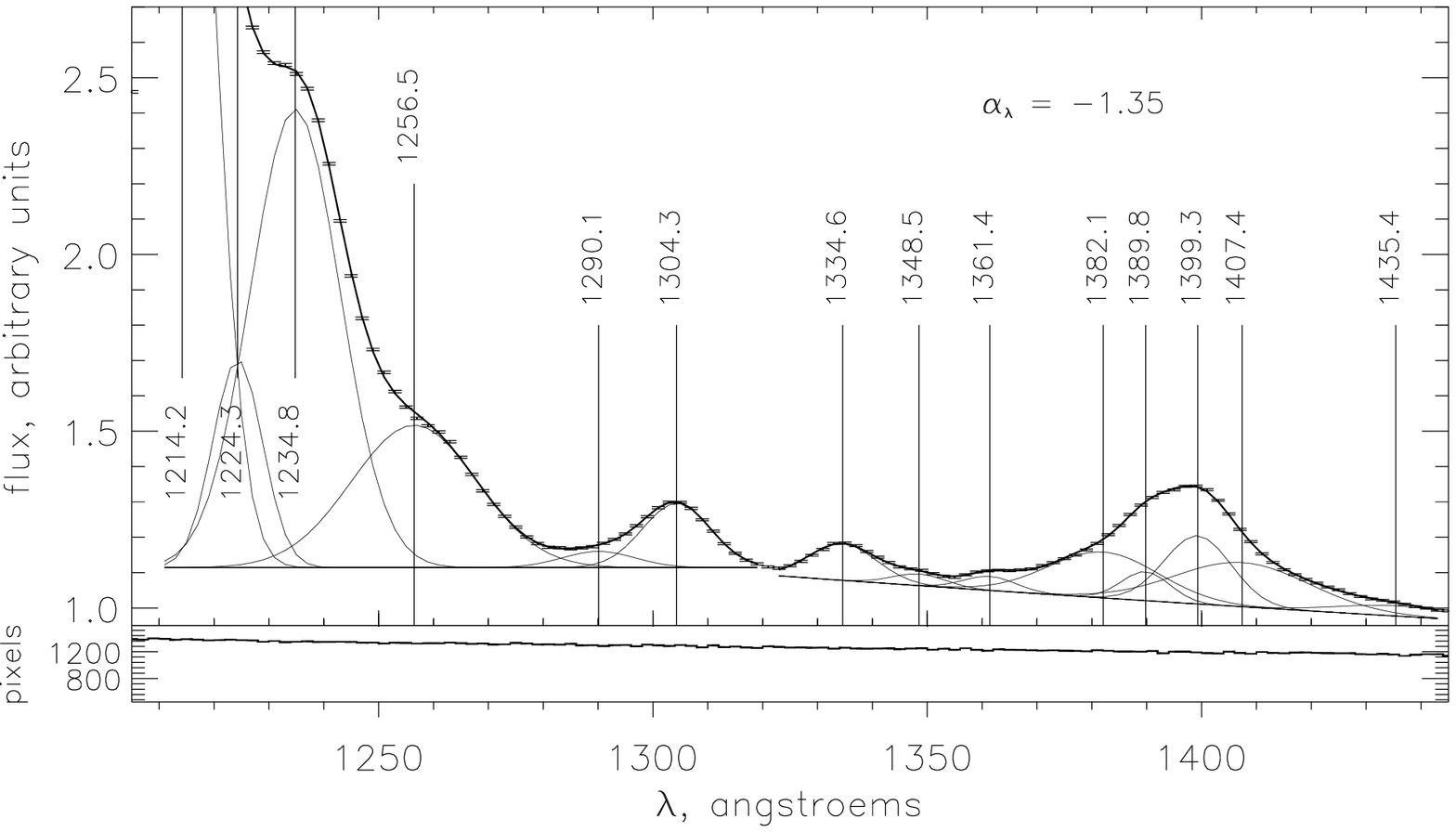,width=.99\linewidth}
 \epsfig{figure=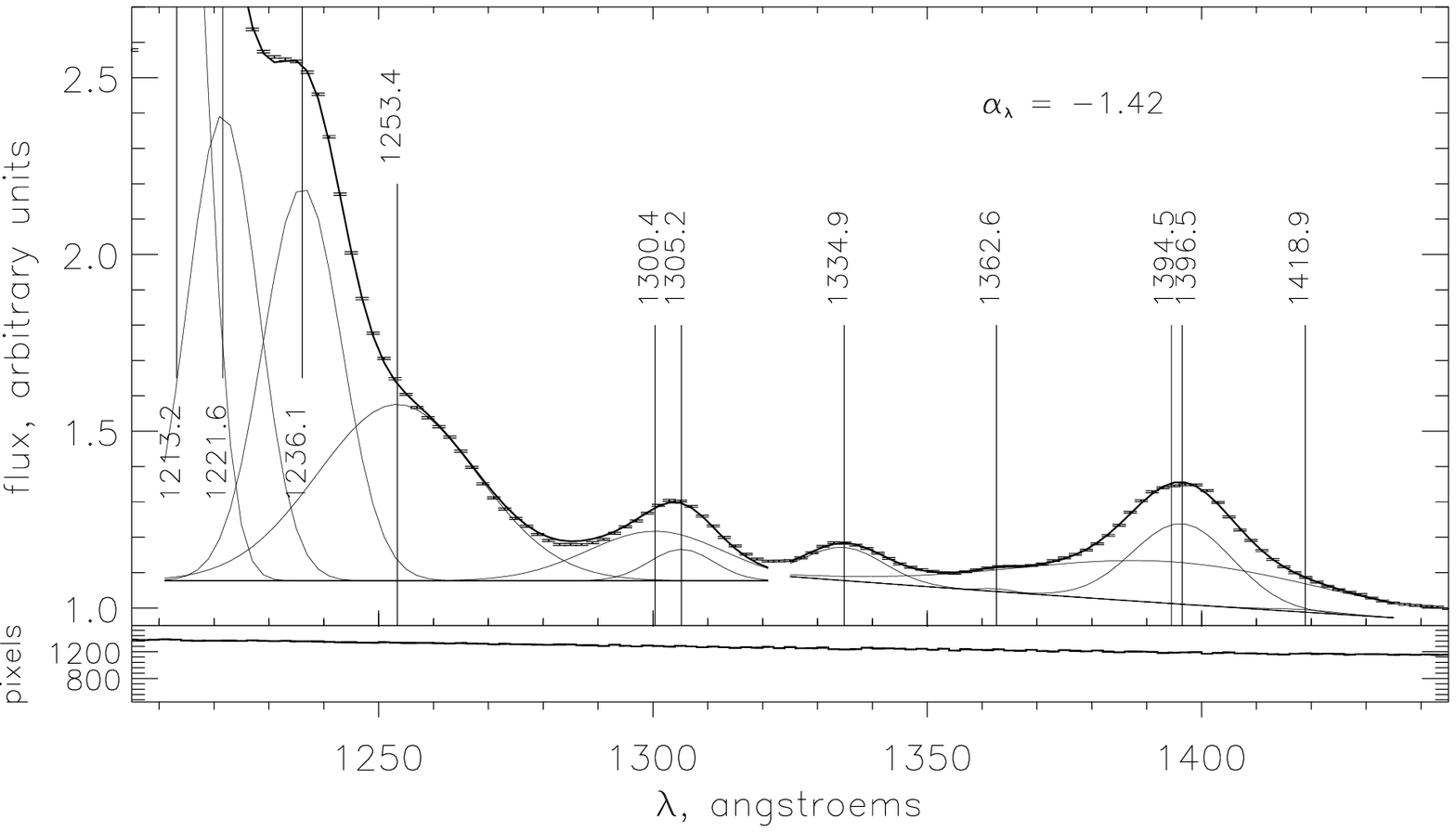,width=.99\linewidth}
 \epsfig{figure=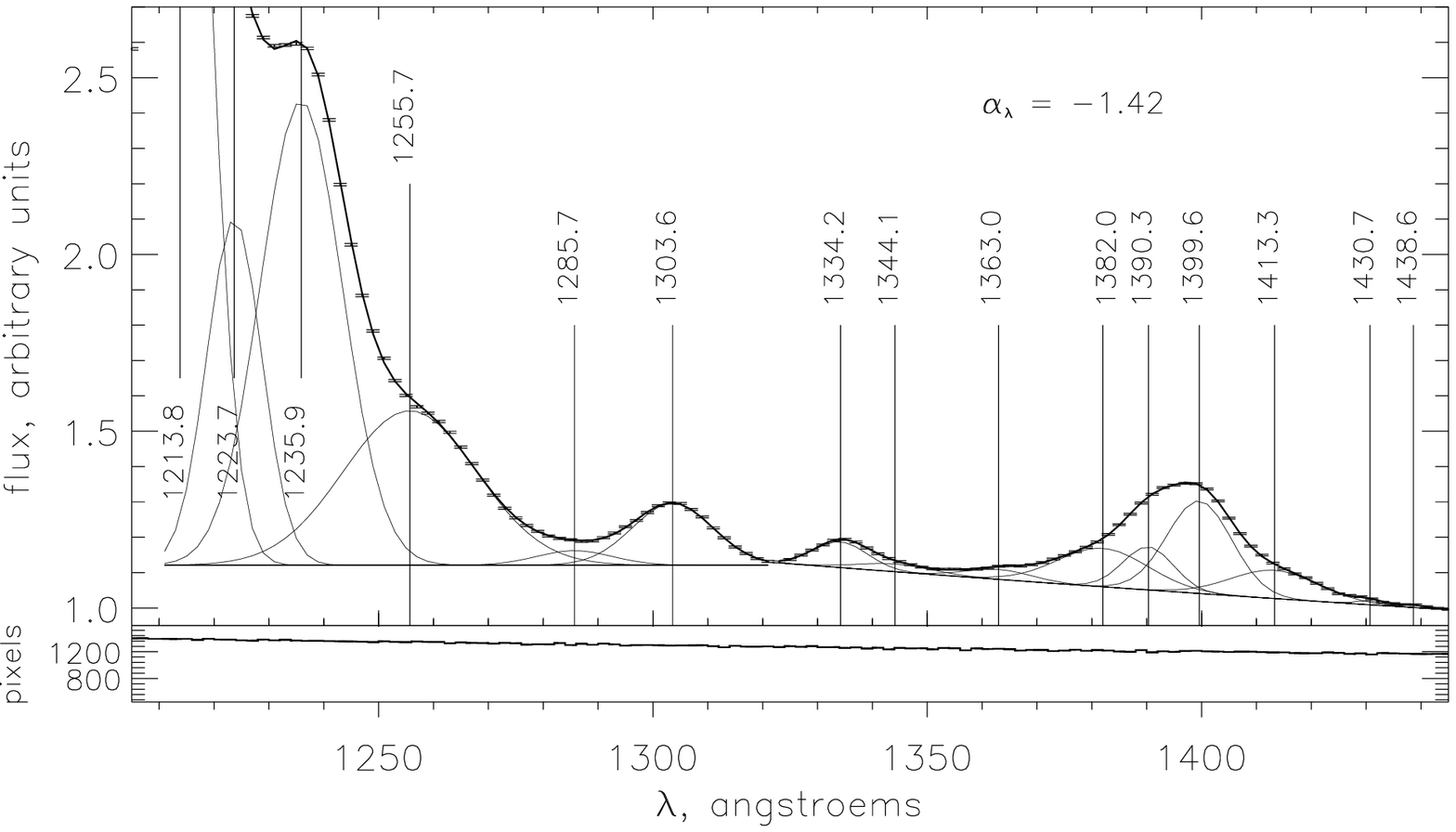,width=.99\linewidth}
\caption{Composite spectra 5$-$8 with $\alpha_{\lambda}$ (from top to bottom): $-1.19$, $-1.35$, $-1.42$, $-1.42$  (red part).}
\label{fig:spec-5-8-b}
\end{figure}

\begin{figure}
 \centering
 \epsfig{figure=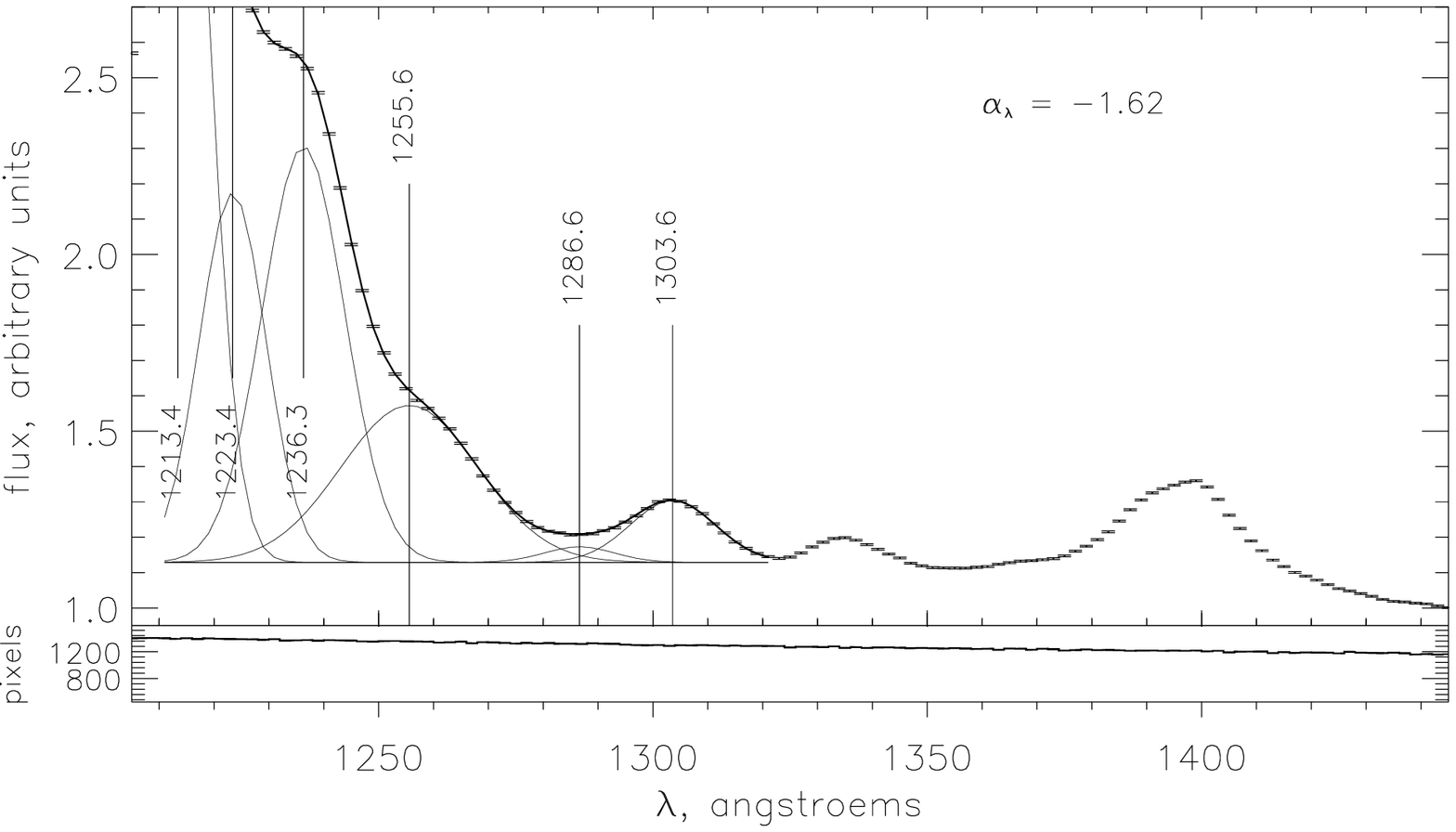,width=.99\linewidth}
 \epsfig{figure=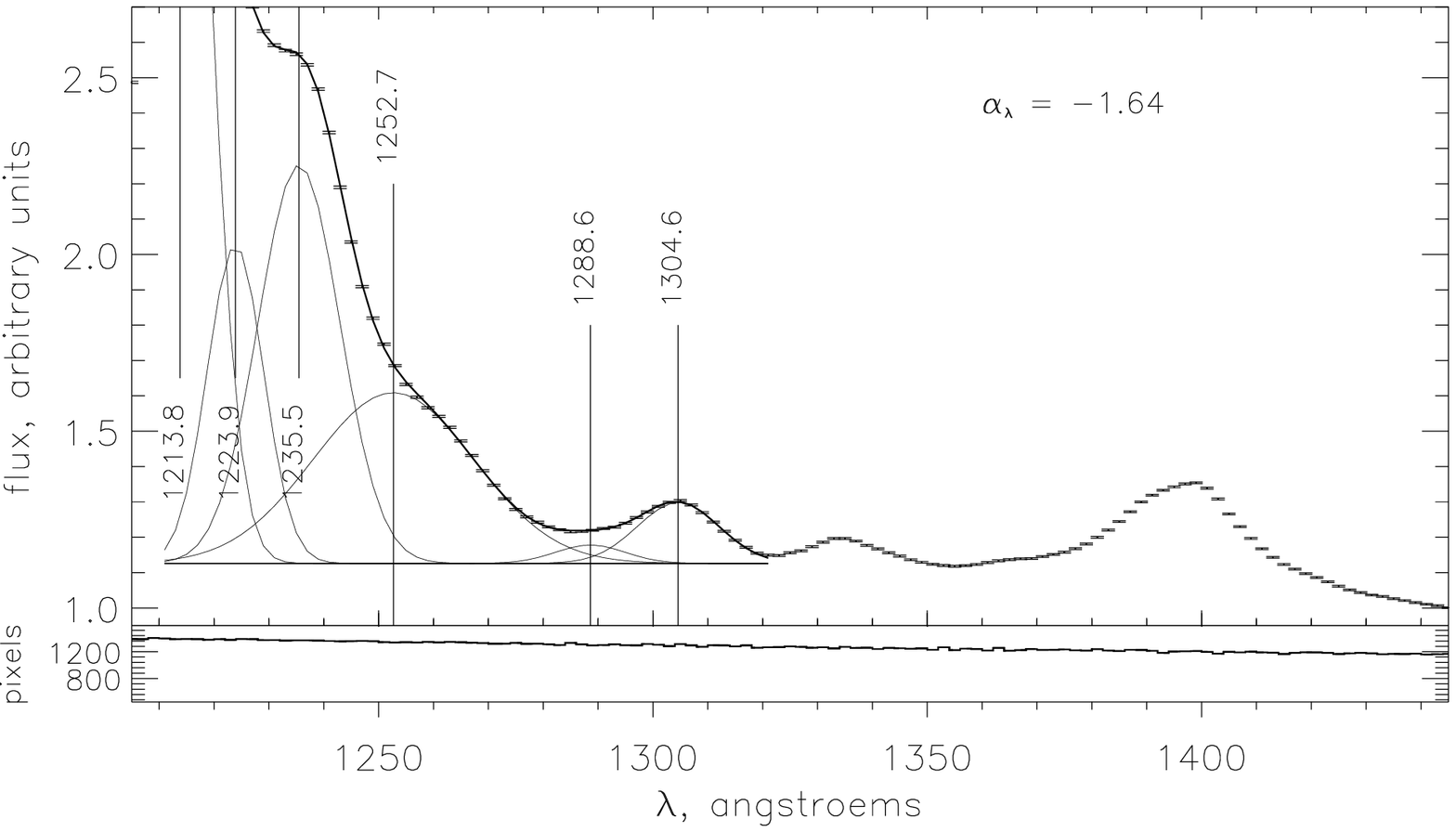,width=.99\linewidth}
 \epsfig{figure=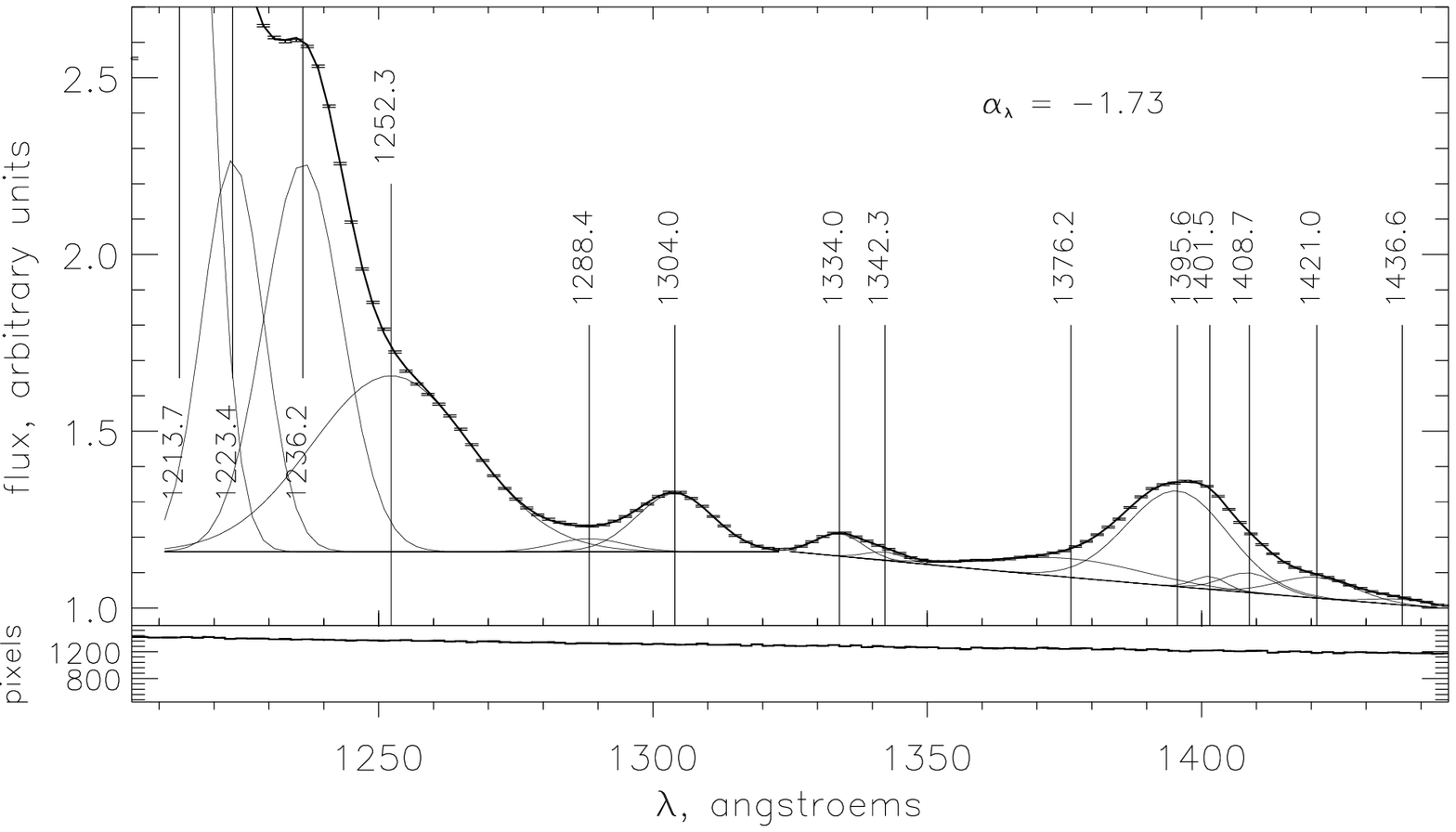,width=.99\linewidth}
 \epsfig{figure=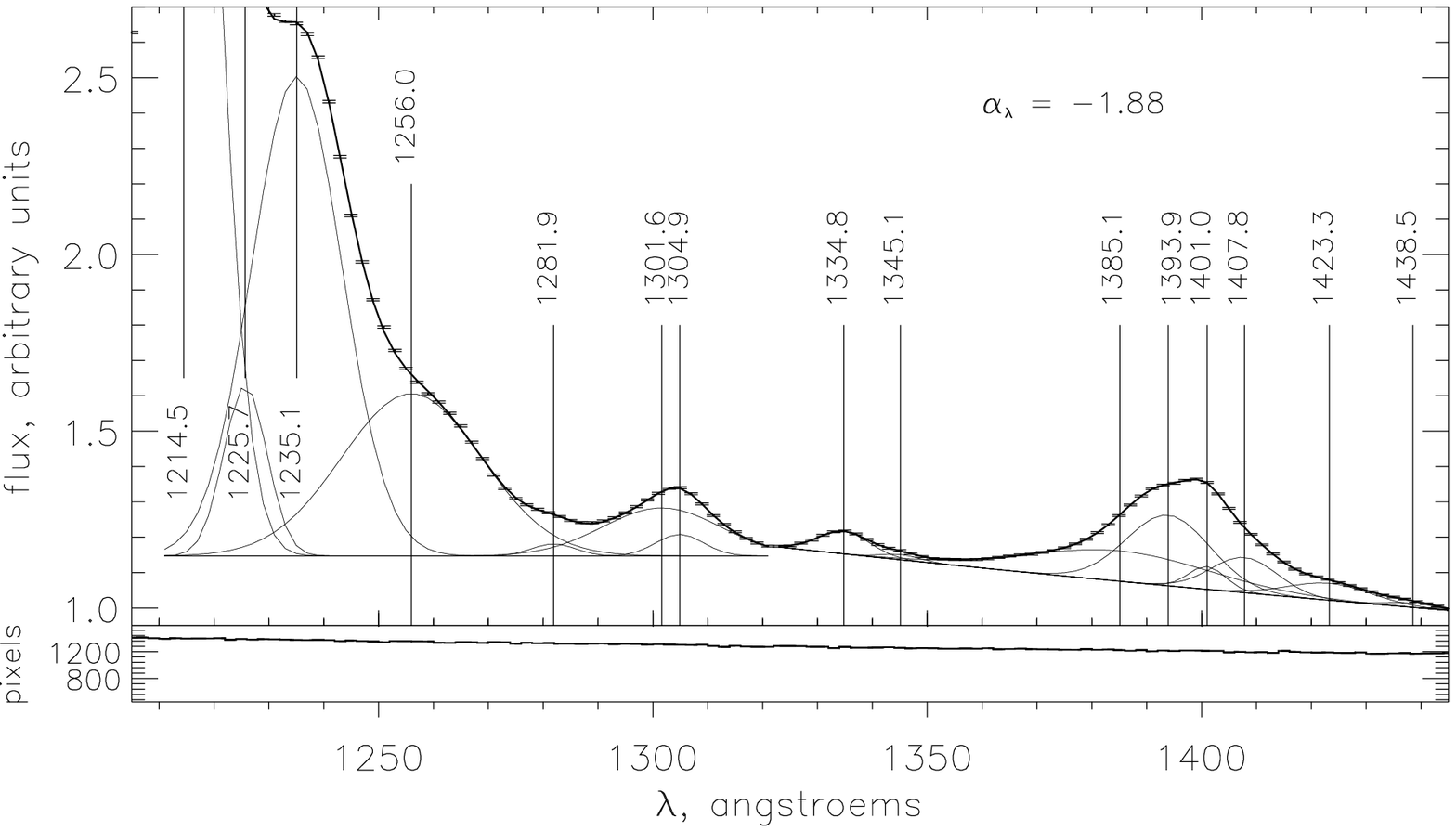,width=.99\linewidth}
 \caption{Composite spectra 9$-$12 with $\alpha_{\lambda}$ (from top to bottom): $-1.62$, $-1.64$, $-1.73$, $-1.88$  (red part).}
 \label{fig:spec-9-12-b}
 \end{figure}

\begin{figure}
 \centering
 \epsfig{figure=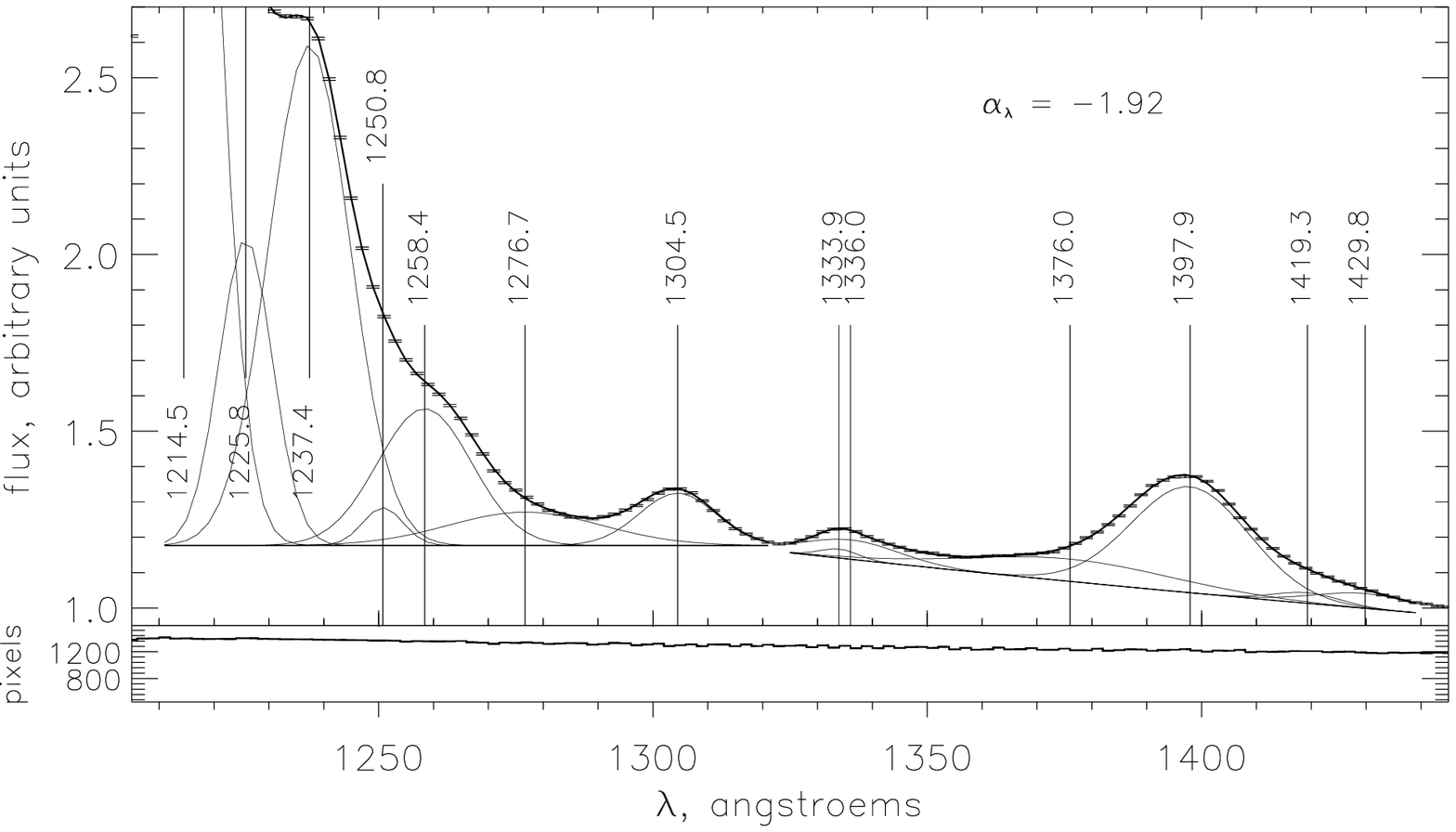,width=.99\linewidth}
 \epsfig{figure=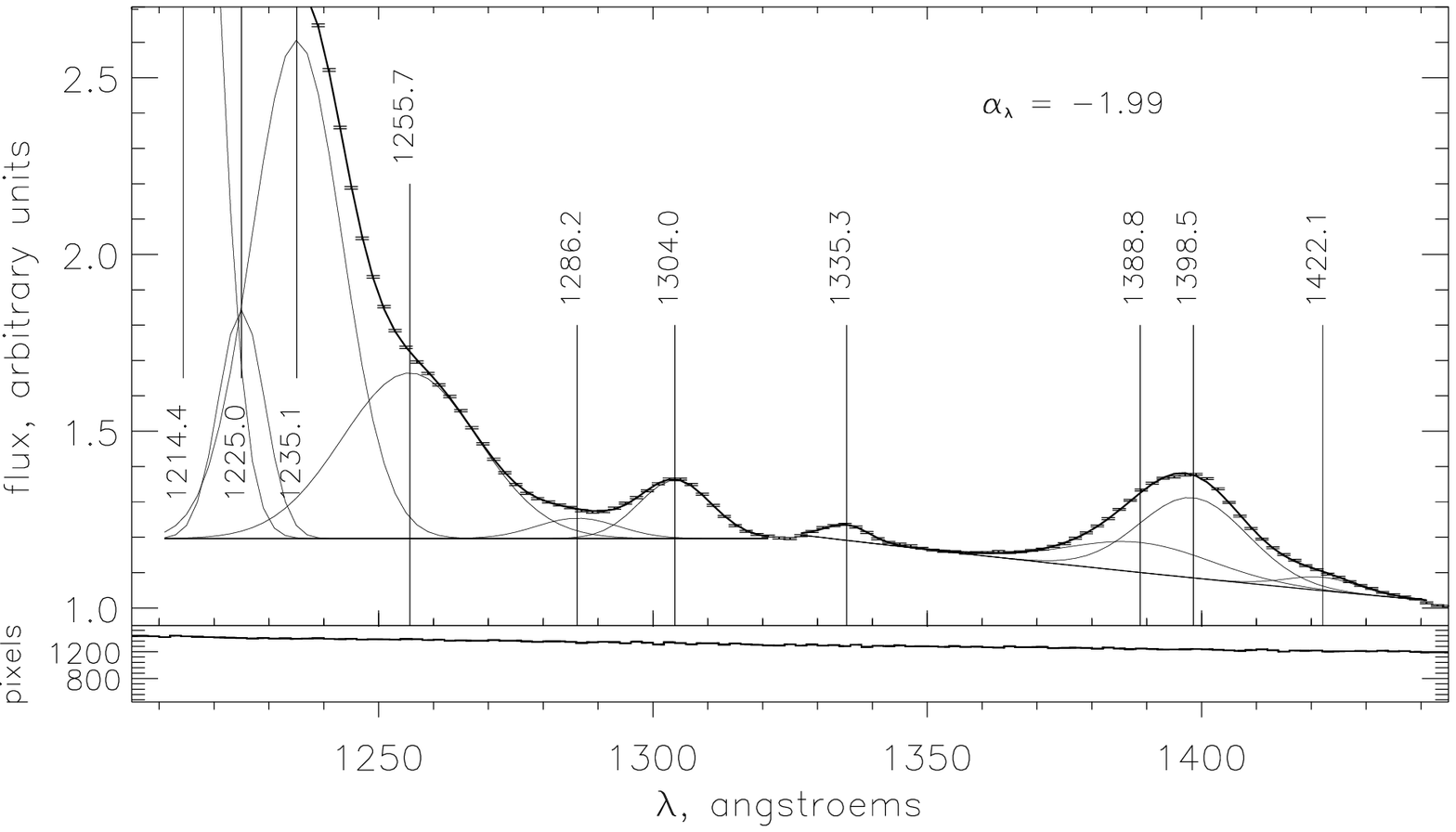,width=.99\linewidth}
 \epsfig{figure=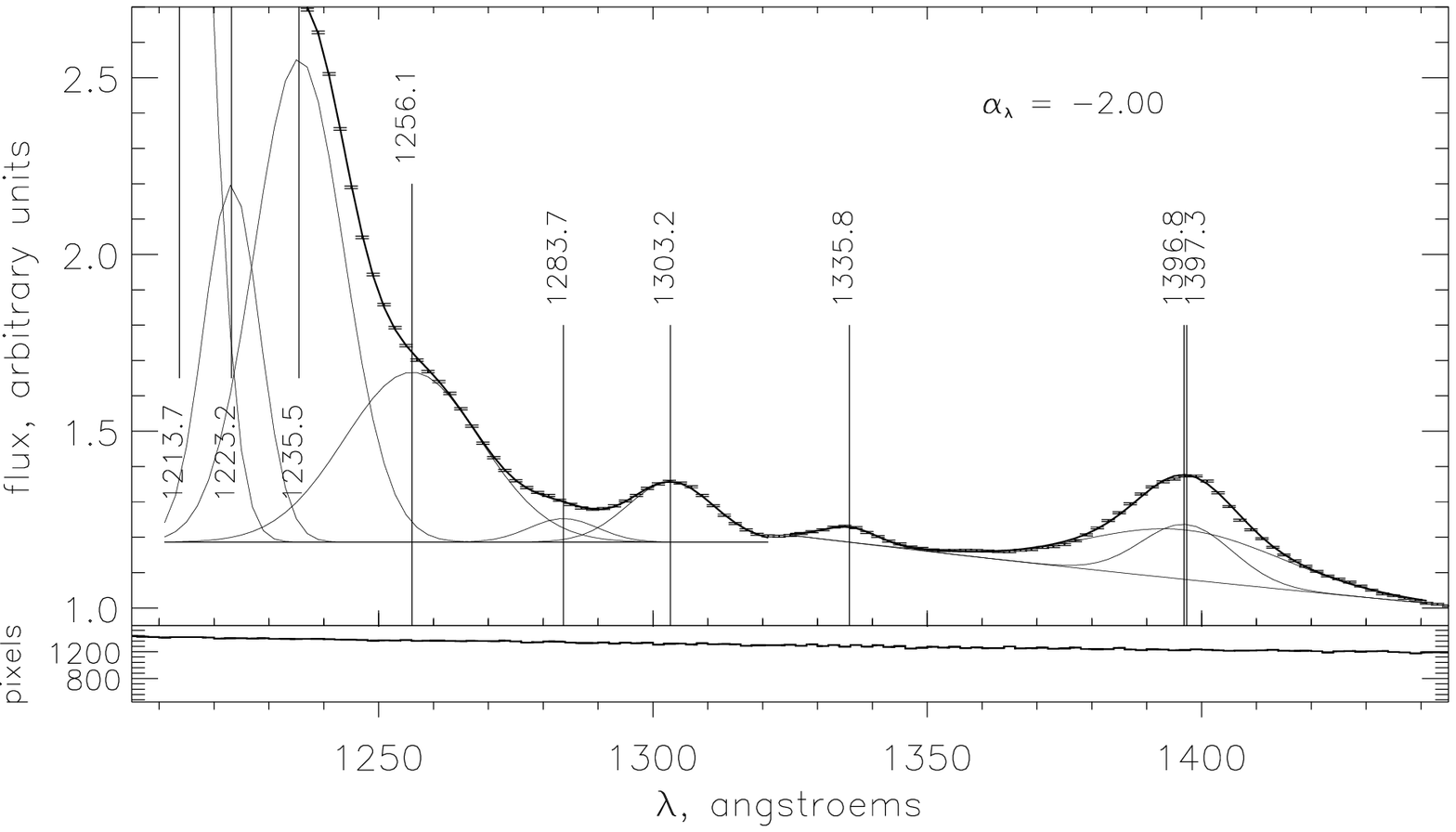,width=.99\linewidth}
 \epsfig{figure=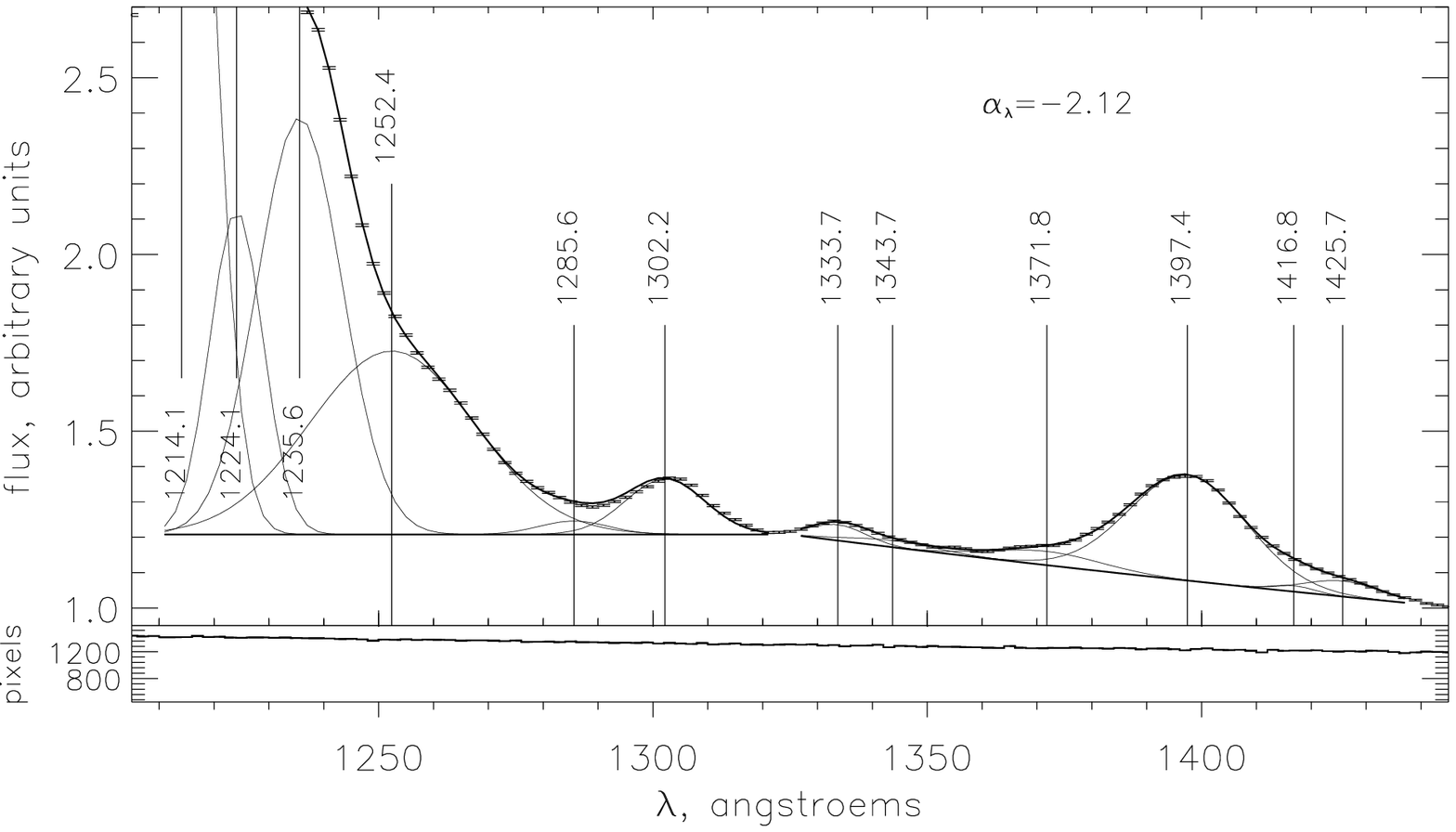,width=.99\linewidth}
 \caption{Composite spectra 13$-$16 with $\alpha_{\lambda}$ (from top to bottom): $-1.92$, $-2.02$, $-2.07$, $-2.14$  (red part).}
 \label{fig:spec-13-16-b}
 \end{figure}

The number of emission lines, `detected' in both blue and red parts of UV bump, is larger than the previously known one. It means that differentiation of spectra according to their spectral index makes sense and indeed helps to reveal new emission features, which are not seen in composite spectra compiled neglecting this difference. It is clearly seen from Fig.~\ref{fig:spec-1-4-b}, \ref{fig:spec-5-8-b}, \ref{fig:spec-9-12-b} and \ref{fig:spec-13-16-b} that in some cases the wings of the emission feature in the 5-th wavelength region are fitted with one or two very broad but low-amplitude Gaussians. Most probably these Gaussians are `artificial' and serve as a fit for superposition of a number of weaker lines. On the other hand, in the continuum level in case of first three wavelength regions the continuum level varies from one range to another, thus the values of $sigma$ parameter, which is proportional to FWHM, cannot be considered as the `true' values and used for further analysis, e.\.g. for studies of FWHM variations with the spectral index. The values of Ly$\alpha$-line parameters from both red and blue parts are presented only for information about fits in general and cannot be compared or considered as the true Ly$\alpha$ line parameters.
 
\section{Conclusions}\label{sec:6}

\indent\indent We compiled 16 composite spectra from subsamples of individual SDSS DR7 quasar spectra with different spectral indices $\alpha_{\lambda}$ within the wavelength range  1270--1480\,\AA\ and studied the possible effects caused by neglect of this discretization when using composite spectra of quasars in different fields. The main results of the present work are the following:\\
\indent(i) the redshifts measured for a test sample of high signal-to-noise ratio quasar spectra using these composites as templates appear to be systematically higher than those calculated with a traditional template, compiled from spectra with different $\alpha_{\lambda}$, with 1.5 times smaller errors in the former case; \\
\indent(ii) the difference in $\alpha_{\lambda}$ in individual spectra used for compilation of composites can yield the mean transmission uncertainty up to 20\%; \\
\indent(iii) a number of emission lines indistinguishable in ordinary composites, but seen in individual high-resolution spectra, can be detected in such composites; \\
\indent(iv) it is confirmed that continuum level within the range between Ly$\beta$ and Ly$\alpha$ emission lines, utilized for studies of the Ly$\alpha$-forest, can be considered as a constant rather than having the same power-law form as that redward from Ly$\alpha$ emission line, that agrees well with steeper continuum index obtained from the composite UV quasar spectra \citep{telfer+02,zheng+97}, and was discussed previously \citep{desjacques_07};\\
\indent(v) it is also shown, that there is no dependence of $\alpha_{\lambda}$ on quasar luminosity in SDSS $u$, $g$, $r$ and $i$ bands, and monochromatic luminosity At $1450$\,\AA. It confirms results of \citet{yip+04} and \citet{vandenberk+04}, who found no relation between luminosity and spectral index, when analysing composite spectra with different luminosity.

Detailed analysis in our previous study of the ranges redward of the Ly$\alpha$ emission lines, which is free from the Ly$\alpha$-forest, shown that there is also no evidence for spectral index dependence of equivalent width of emission lines \citep{torbaniuk+12}. The absence of dependence of the UV-bump shape on luminosity in the bands mentioned above requires further study of this region for understanding the physics behind the difference in the UV-bump shape in spectra of different quasars. It is worth to note, that the absolute magnitudes in given bands used as a characteristic of luminosity include K-correction, which itself is determined within the frame of some model of the spectrum general shape (e.\,g. with the spectral index of $\alpha_{\lambda}=-1.5$).

The proposed approach can be applied for generation of new templates for more precise quasar redshift measurements with the common cross-correlation technique used in redshift surveys, more precise theoretical determination of K-correction and colour-indices, as far as for determination of continuum and mean transmission in Ly$\alpha$ forest studies.

\section*{Acknowledgements}
 
\indent\indent The authors are thankful to Mariangela Bernardi, Ravi K. Sheth, Oleg Ruchayskiy, Alexey Boyarsky and Ievgen Vovk for fruitful discussions. The authors also acknowledge the usage of \texttt{CosmoMC} package. This work has been supported by Swiss National Science Foundation  (SCOPES grant No 128040). 

The authors are also thankful to the Sloan Digital Sky Survey team. Funding for the SDSS and SDSS-II has been provided by the Alfred P.\,Sloan Foundation, the Participating Institutions, the National Science Foundation, the U.S. Department of Energy, the National Aeronautics and Space Administration, the Japanese Monbukagakusho, the Max Planck Society, and the Higher Education Funding Council for England. The SDSS Web Site is http://www.sdss.org/. The SDSS is managed by the Astrophysical Research Consortium for the Participating Institutions. The Participating Institutions are the American Museum of Natural History, Astrophysical Institute Potsdam, University of Basel, University of Cambridge, Case Western Reserve University, University of Chicago, Drexel University, Fermilab, the Institute for Advanced Study, the Japan Participation Group, Johns Hopkins University, the Joint Institute for Nuclear Astrophysics, the Kavli Institute for Particle Astrophysics and Cosmology, the Korean Scientist Group, the Chinese Academy of Sciences (LAMOST), Los Alamos National Laboratory, the Max-Planck-Institute for Astronomy (MPIA), the Max-Planck-Institute for Astrophysics (MPA), New Mexico State University, Ohio State University, University of Pittsburgh, University of Portsmouth, Princeton University, the United States Naval Observatory, and the University of Washington.

\input{ivashchenko.sergijenko.torbaniuk.bbl}

\clearpage
\begin{sidewaystable*}
 \vspace*{130ex} 
\centering
\begin{tabular}{c|c|c|c|c|c|c|c|c|c|c|c|c} 
\multicolumn{13}{l}{}\\
\multicolumn{13}{l}{{\bf Table 3.} Emission lines with their measured rest-frame wavelengths from individual and composite quasar spectra within the range $\sim1050-1340$\,\AA\ from}\\
\multicolumn{13}{l}{previous studies. Asterisk indicates more than one allowed value of the total angular momentum J for that specific term and transition, bracket means}\\
\multicolumn{13}{l}{intercombination transitions.}\\
\multicolumn{13}{l}{}\\
\hline
line                    & S{\sc iv} & Ar{\sc i}+N{\sc ii} & Si{\sc iii}* & Fe{\sc iii} & C{\sc iii}* & Si{\sc ii} & Si{\sc iii} & Ly $\alpha$ & O~{\sc v} & N~{\sc v} & Si~{\sc ii}$^{*}$ & Si~{\sc ii} \\
\hline
$\lambda_{lab}$,\,\AA   &           & 1066.66+   & 1111.59      &             & 1175.67     & 1194.12    &  	          & 1215.67     & 1218.3    & 1238.8+   & 1248.4+           & 1260.4+\\
			&           & +1085.12     & 	           &             & 	       & 	    &  	          & 	        & 	    & +1242     & +1251.1           & +1264.7+ \\
			&           & 	       	    & 	           &             & 	       & 	    &  	          & 	        & 	    &           &                   & +1265.0 \\
\hline
\citet{francis+91}     & 	& 	            &        & 	 & 	   & 	    &      & 1216    &	      & 1240     &        & \\
\citet{brotherton+94}  & 	& 	            &        & 	 & 	   & 	    &      & 1217.2  &	      & 1241.8   &        & \\
\citet{Laor+94}        & 	& 	            &        & 	 & 	   &	    &      & 1215.67 &	      & 1238.82+ & 	  & \\
		       & 	& 	            &        & 	 & 	   &	    &      & 	     &	      & +1242.8  & 	  & \\
\citet{Laor+1995}      & 	& 	            &        & 	 &         & 	    &      & 1215.67 &	      & 1240.15  &        & 1263.31 \\
\citet{laor+97}        & 	& 	            &        & 	 & 	   & 	    &      & 1214.2  &        & 1236.7+  & 1249.5 & 1260.5+ \\
		       & 	& 	            &        & 	 & 	   & 	    &      & 1214.2  &	      & +1239.1  & 1249.5 & +1264.5 \\
\citet{zheng+97}       & 	& tent.             &        &          &         & 	    &      &         &        &          &        & \\
\citet{vandenberk+01}  & 	& 1065.10           &        & 1117.26  & 1175.35 & 	    &      & 1216.25 &        & 1239.85  &        & 1265.22 \\
\citet{brotherton+01}  & 	& 	            &        & 	 & 	   & 	    &      & 1216    &	      & 1240     &	  & \\
\citet{Vestergaard+01} & 	&	            &        &		 &	   & 	    &      & 1210.8+ & 1219.0 & 1238.8   & 1249.7 & 1257.2+ \\
		       & 	&	            &        &	 	 &	   & 	    &      & +1214.4+&        & 1242.8   & 1250.7 & +1260.7+ \\
		       & 	&	            &        &	 	 &	   & 	    &      & +1216.9 &        &          &        & +1265.2 \\
\citet{telfer+02}      & 	& 1065     	     &        & 1123     &  1176   & 1195   &      & 	     &        &          & 	  & \\
\citet{scott2+04}      & 1062+  &           1084 &        & 	 & 	   & 	    &      &	     &        & 	 &        & \\
		       & +1073  &                  &        & 	 & 	   & 	    &      &	     &        & 	 &        & \\
\citet{tytler+04}      & 	& 1070.95        &        & 1123.3   & 1175.88 & 	    &      &	     &	      &	         &	  & \\
\citet{leighly+07}     & 	& 	    1084.2 & 1110.5 &          & 1175.4  & 1193.6 &      &         &	      & 	 &	  & \\
\citet{binette+08}     & 1067   & 	    1084 &        & 1123     & 1176    & 1194   & 1207 &	     &	      & 	 &        & \\
\hline
\end{tabular}
\vspace*{-1ex}
\begin{tabular}{c|c|c|c|c|c|c|c|c|c|c} 
\hline
line                   & Si~{\sc iii}$^{*}$ & O~{\sc i}+Si~{\sc ii} & C~{\sc ii} & Ca~{\sc ii} & O~{\sc iv} & Fe~{\sc v} & Fe~{\sc iii} & Fe~{\sc v} & O~{\sc i} & Si~{\sc iv}+O~{\sc iv}]  \\
\hline
$\lambda_{lab}$,\,\AA  & 1295.5+1298.9      & 1302.2+1304.6+        & 1335.3     & 1342.3      & 1343.5     & 1343.1     & 1343.2       &     1345.6 & 1355.6    & 1395.5+1399.8+ \\
   		       &                    & +1306.0+1309.3        & 	         & 	       & 	    & 	         & 		&            &           & +1401.8+1407.4 \\
\hline
\citet{francis+91}     &               & 1302            & 1335          &        &        &        &        &        &        & 1400 \\
\citet{brotherton+94}  &               & 1305.5+1306.3   & 1338.5+1339.8 &        &        &        &        &        &        & 1399.2+1401.8 \\
\citet{Laor+94}        &               & 1303.49         & 1335.3        &        &        &        &        &        &        & 1399.61 \\
\citet{Laor+1995}      &               & 	         &               &        &        &        &        &        &        & 1396.75+1402.46 \\
\citet{laor+97}        &               & 1302.6+1305.5+  & 1334.3        &        &        &        &        &        &        & 1392.1+1400.6 \\
                       &               & +1309.2         &               &        &        &        &        &        &        & \\
\citet{zheng+97}       &               &                 &               &        &        &        &        &        &        & \\
\citet{vandenberk+01}  &               & 1305.42         & 1336.6        &        &        &        &        &        &        & 1348.33 \\
\citet{brotherton+01}  &               & 1302            & 1335          &        &        &        &        &        &        & 1400 \\
\citet{Vestergaard+01} & 1293.9+1296.5 & 1299.0+1302.1+  & 1330.8+1334.0 & 1343.9 & 1343.9 & 1343.9 & 1343.9 & 1343.9 & 1343.9 & 1387.4+1392.2+ \\
		       &               & +1305.5+1309.3  &               &        &        &        &        &        &        & +1397.5+1401.2 \\
\hline
\end{tabular}
\end{sidewaystable*}

\begin{sidewaystable*}
 \vspace*{-125ex}
\centering
\begin{tabular}{c|c|c|c|c|c|c|c|c}
\multicolumn{8}{l}{{\bf Table 4.} Emission lines detected in spectra within ranges 1-3. Lines in brackets are dublets. Arrows stand for failed fit.}\\
\multicolumn{8}{l}{}\\
\hline
\hline
 &  1 &  2 & 3 &  4 &  5 & 6 & 7 \\
\hline
 &   X$_1$    &  X$_2$+Ar{\sc i}+X$_3$  &   N{\sc ii}+X$_4$   & X$_5$+X$_6$   & FeIII-multiplet  & X$_7$ & X$_8$+C{\sc iii}$^*$+X$_9$ & X$_{10}$+Ly$\alpha$ \\
\hline
\hline
1 &    tent.    & 1058.4+[1071.8] &  1084.9 & 1096.3 & 1111.8+1125.2 &  & 1161.6+1173.9+1182.1 & 1214.4\\
\hline
2 &     tent.   & 1059.6+[1072.4] &  1084.6 & 1096.4 & 1111.5+1123.3 & tent. & 1161.8+1174.8+1182.4 & 1214.3\\
\hline
3 &     tent.   & 1059.5+[1073.9] &  1086.1 & 1096.9 & 1118.3+1125.7+1134.4 & tent. & 1161.4+[1176.5] & 1214.3\\
\hline
4 &     tent.   & 1059.5+[1071.0] & 1089.3 & tent. &  1117.0+1126.2+1136.5 & tent. & tent.+1175.2+1185.3 & 1214.8\\
\hline
5 &     tent.   & tent.+ [1070.7] & 1087.6 &  &  1115.2+1124.4+1136.2 &  1152.3 & 1171.9 & 1214.3\\
\hline
6 &      $\rightarrow$    & $\rightarrow$ & $\rightarrow$  & 1100.9 &  1116.9+1125.0+1132.4 & tent. & tent.+[1174.2] & 1214.2\\
\hline
7 &     $\rightarrow$     &  1061.6+[1071.1] & 1079.3+1091.0  & & 1108.4+1116.6+1127.0+1143.9 & & tent.+[1172.1] & 1213.2\\
\hline
8 &     $\rightarrow$     & 1060.3+[1071.3] &  1080.8 & 1098.5 & 1111.5+1125.4+1141.0 & tent. & 1172.5 & 1213.8\\
\hline
9 &     $\rightarrow$     &  tent.+[1070.2] &  1086.0 &   1099.6 & 1122.6 & tent. & 1172.6 & 1213.4 \\
\hline
10 &   $\rightarrow$   &  1056.3+[1071.7] &  1086.8 &   1099.6 & 1117.3+1127.4+1141.7 & tent. & 1163.7+[1174.2] & 1213.8\\
\hline
11 &   $\rightarrow$   &   1058.1+[1070.4] & 1079.9 &  & 1106.7+1115.7+ 1125.6 & tent. & [1168.5]+1174.9 & 1213.7\\
\hline
12 &  $\rightarrow$   &    [1063.0]+1074.1 &  1084.0 &  & 1115.5+1123.8+1129.0 & & tent.+1172.9+1178.5 & 1214.5\\
\hline
13 &  $\rightarrow$   &    [1063.8]+1072.6 &  1082.0+1087.5 & 1103.3 & 1111.9+1124.3+1135.9 & 1153.9 & 1164.1+[1172.7] & 1214.5\\
\hline
14 &     $\rightarrow$     &  $\rightarrow$ &  $\rightarrow$ &  1101.6 & 1116.2+1126.2+1135.5 & tent. & 1172.0 & 1214.4\\
\hline
15 & 1055.8 &  [1066.6]+1072.5 & 1083.2 & tent & 1108.3+1117.3+1125.5+tent. & $\leftarrow$ & $\leftarrow$ & $\leftarrow$\\
\hline
16 & 1052.4 & 1060.6+[1071.4] & 1083.1 & tent. &  1111.3+1123.7+1140.1 & 1155.5 & 1165.7+[1174.3] & 1214.1\\
\hline
\end{tabular}
  \end{sidewaystable*}

  \begin{sidewaystable*}
 \centering
    \vspace*{125ex}
 \begin{tabular}{c|c|c|c|c|c|c|}
 \multicolumn{7}{l}{{\bf Table 5.}  Emission lines detected in spectra within ranges 4-5. Arrows stand for failed fit.}\\
 \multicolumn{7}{l}{}\\
 \hline
 \hline
  &  1 &  2 & 3 &  4 &  5 & 6  \\
 \hline
   & Ly$\alpha+$O{\sc iv} & N{\sc v} & Si{\sc ii}$^\ast+$Si{\sc ii} & Si{\sc iii}$^\ast+$O{\sc i}$+$Si{\sc ii} & C{\sc ii}$+$O{\sc iv}$+$Ca{\sc ii} & Si{\sc iv}$+$O{\sc iv}$]$ \\
 \hline
 \hline
1 & 1214.4$+$1225.6 & 1236.9 & 1256.7 & 1290.4$+$1304.4 & 1334.5$+$1346.7 & 1368.2$+$1382.5$+$1390.4$+$1399.8$+$1418.7$+$1436.3 \\
\hline
2 & 1214.3$+$1225.4 & 1236.5 & 1256.7 & 1290.4$+$1304.5 & 1335.0$+$1344.8 & 1364.9$+$1382.3$+$1390.5$+$1399.6$+$1417.7$+$1430.8 \\
\hline
3 & 1214.3$+$1225.0 & 1236.1 & 1256.7 & 1289.7$+$1304.1 & 1334.9$+$1343.2 & 1364.4$+$1380.1$+$1390.3$+$1399.3$+$1418.7$+$1431.4 \\
\hline
4 & 1214.8$+$1226.4 & 1236.4 & 1257.0 & 1290.6$+$1304.2 & 1334.7$+$1341.4 & 1364.8$+$1376.9$+$1386.5$+$1392.0$+$1399.0$+$1412.3$+$1419.6$+$1439.5 \\
\hline
5 & 1214.3$+$1225.3 & 1236.0 & 1256.7 & 1291.4$+$1304.5 & 1335.8 & 1370.5$+$1396.5$+$1401.7$+$1410.9$+$1422.4$+$1436.4 \\
\hline
6 & 1214.2$+$1224.3 & 1234.8 & 1256.5 & 1290.1$+$1304.4 & 1334.6$+$1348.5 & 1361.4$+$1382.1$+$1389.8$+$1399.3$+$1407.4$+$1435.4 \\
\hline
7 & 1213.2$+$1221.6 & 1236.1 & 1253.4 & 1300.4$+$1305.2 & 1334.9 & 1362.6$+$1394.5$+$1396.5$+$1418.9 \\
\hline
8 & 1213.8$+$1223.7 & 1235.9 & 1255.7 & 1285.7$+$1303.6 & 1334.2$+$1344.1 & 1363.0$+$1382.0$+$1390.3$+$1399.6$+$1413.3$+$1430.7$+$1438.6 \\
\hline
9 & 1213.4$+$1223.4 & 1236.3 & 1255.6 & 1286.6$+$1303.6 & $\leftarrow$ & $\leftarrow$ \\
\hline
10 & 1213.8$+$1223.9 & 1235.5 & 1252.7 & 1288.6$+$1304.6 & $\leftarrow$ & $\leftarrow$ \\
\hline
11 & 1213.7$+$1223.4 & 1236.2 & 1252.3 & 1288.4$+$1304.0 & 1334.0$+$1342.3 & 1376.2$+$1395.6$+$1401.5$+$1408.7$+$1421.0$+$1436.6 \\
\hline
12 & 1214.5$+$1225.7 & 1235.1 & 1256.0 & 1281.9$+$1301.6$+$1304.9 & 1334.8$+$1345.1 & 1385.1$+$1393.9$+$1401.0$+$1407.8$+$1423.3$+$1438.5 \\
\hline
13 & 1214.5$+$1225.8 & 1237.4 & 1250.8$+$1258.4 & 1276.7$+$1304.5 & 1333.9$+$1336.0 & 1376.0$+$1397.9$+$1419.3$+$1429.8 \\
\hline
14 & 1214.4$+$1225.0 & 1235.1 & 1255.7 & 1286.2$+$1304.0 & 1335.3 & 1388.8$+$1398.5$+$1422.1 \\
\hline
15 & 1213.7$+$1223.2 & 1235.5 & 1256.1 & 1283.7$+$1303.2 & 1335.8 & 1396.8$+$1397.3 \\
\hline
16 & 1214.1$+$1224.1 & 1235.6 & 1252.4 & 1285.6$+$1302.2 & 1333.7$+$1343.7 & 1371.8$+$1397.4$+$1416.8$+$1425.7 \\
 \hline
 \end{tabular}
 \end{sidewaystable*}

\clearpage
\setcounter{table}{5}
\begin{table*}
\begin{minipage}[t]{0.3\linewidth}\centering
\caption{ Parameters of emission lines for spectrum 1 with $\alpha_{\lambda}=-0.91$.}\label{tab-spec-1}
\vspace*{-2ex}
\fontsize{7}{7}\selectfont
 \begin{tabular}{p{0.7cm}p{1.3cm}p{1.2cm}p{0.6cm}}
\hline
$\lambda_{0}$,\,\AA   & a & w,\,\AA & a-3$\sigma$\\
\hline
1058.4 & $0.139^{+0.015}_{-0.027}$ & $5.81^{+0.40}_{-0.74}$ & $0.105$\\
1071.8 & $0.191^{+0.009}_{-0.017}$ & $6.76^{+0.41}_{-0.73}$ & $0.169$\\
1084.9 & $0.094^{+0.014}_{-0.025}$ & $3.92^{+0.50}_{-0.88}$ & $0.063$\\
1096.3 & $0.025^{+0.005}_{-0.010}$ & $1.40^{+0.43}_{-0.72}$ & $0.012$\\
1111.8 & $0.047^{+0.003}_{-0.006}$ & $5.41^{+0.40}_{-0.72}$ & $0.038$\\
1125.2 & $0.091^{+0.003}_{-0.005}$ & $7.82^{+0.37}_{-0.67}$ & $0.084$\\
1161.6 & $0.046^{+0.007}_{-0.013}$ & $3.39^{+0.67}_{-1.13}$ & $0.029$\\
1173.9 & $0.088^{+0.006}_{-0.010}$ & $5.21^{+0.44}_{-0.81}$ & $0.075$\\
1182.1 & $0.031^{+0.008}_{-0.016}$ & $1.78^{+0.56}_{-1.13}$ & $0.012$\\
1214.1 & $1.111^{+0.025}_{-0.047}$ & $17.33^{+0.27}_{-0.53}$ & $1.051$\\
\hline
1214.1   & $2.831^{+0.007}_{-0.014}$ & $5.61^{+0.02}_{-0.04}$ & $2.813$\\
1225.6   & $0.809^{+0.010}_{-0.019}$ & $5.18^{+0.05}_{-0.09}$ & $0.784$\\
1236.9   & $1.172^{+0.005}_{-0.011}$ & $7.12^{+0.04}_{-0.07}$ & $1.158$\\
1256.7   & $0.404^{+0.006}_{-0.011}$ & $11.21^{+0.18}_{-0.35}$ & $0.390$\\
1290.4   & $0.048^{+0.005}_{-0.009}$ & $4.07^{+0.39}_{-0.72}$ & $0.036$\\
1304.4   & $0.194^{+0.005}_{-0.010}$ & $6.78^{+0.23}_{-0.44}$ & $0.181$\\
1334.5  & $0.118^{+0.003}_{-0.005}$ & $6.45^{+0.14}_{-0.27}$ & $0.112$\\
1346.7	& $0.035^{+0.001}_{-0.002}$ & $5.04^{+0.23}_{-0.44}$ & $0.032$\\
1368.2  & $0.077^{+0.003}_{-0.005}$ & $9.136^{+0.34}_{-0.64}$ & $0.071$\\
1382.5  & $0.112^{+0.002}_{-0.005}$ & $5.96^{+0.09}_{-0.18}$ & $0.106$\\
1390.4  & $0.115^{+0.002}_{-0.004}$ & $4.77^{+0.10}_{-0.19}$ & $0.110$\\
1399.8  & $0.326^{+0.003}_{-0.006}$ & $7.49^{+0.06}_{-0.11}$ & $0.318$\\
1418.7  & $0.080^{+0.002}_{-0.004}$ & $7.67^{+0.16}_{-0.30}$ & $0.075$\\
1436.3  & $0.043^{+0.003}_{-0.005}$ & $7.11^{+0.34}_{-0.65}$ & $0.036$\\

\hline
\end{tabular}
\end{minipage}
\hfil
\begin{minipage}[t]{0.3\linewidth}\centering
\caption{ Parameters of emission lines for spectrum 2 with $\alpha_{\lambda}=-0.97$.}\label{tab-spec-2}
\vspace*{-2ex}
\fontsize{7}{7}\selectfont
 \begin{tabular}{p{0.7cm}p{1.3cm}p{1.2cm}p{0.6cm}}
\hline
$\lambda_{0}$,\,\AA   & a & w,\,\AA & a-3$\sigma$\\
\hline
1059.6   & $0.154^{+0.013}_{-0.024}$ & $5.68^{+0.40}_{-0.71}$ & $0.125$\\
1072.4   & $0.197^{+0.007}_{-0.013}$ & $5.53^{+0.20}_{-0.36}$ & $0.180$\\
1084.6   & $0.106^{+0.013}_{-0.022}$ & $4.45^{+0.49}_{-0.85}$ & $0.078$\\
1096.4   & $0.017^{+0.004}_{-0.008}$ & $1.99^{+0.56}_{-1.25}$ & $0.006$\\
1111.5   & $0.030^{+0.003}_{-0.006}$ & $3.10^{+0.30}_{-0.56}$ & $0.022$\\
1123.3   & $0.096^{+0.003}_{-0.006}$ & $9.38^{+0.47}_{-0.84}$ & $0.089$\\
1161.8   & $0.057^{+0.014}_{-0.020}$ & $4.14^{+1.09}_{-1.60}$ & $0.033$\\
1174.8   & $0.104^{+0.008}_{-0.014}$ & $4.90^{+0.36}_{-0.70}$ & $0.088$\\
1182.4   & $0.035^{+0.006}_{-0.014}$ & $2.19^{+0.38}_{-0.73}$ & $0.018$\\
1214.3   & $1.112^{+0.024}_{-0.046}$ & $17.08^{+0.37}_{-0.64}$ & $1.053$\\
\hline 
1214.3& $2.715^{+0.007}_{-0.014}$ & $5.58^{+0.02}_{-0.05}$ & $2.696$\\
1225.4& $0.768^{+0.010}_{-0.019}$ & $5.06^{+0.05}_{-0.09}$ & $0.742$\\
1236.5& $1.186^{+0.005}_{-0.010}$ & $7.33^{+0.04}_{-0.07}$ & $1.172$\\
1256.7& $0.403^{+0.006}_{-0.011}$ & $11.18^{+0.18}_{-0.36}$ & $0.389$\\
1290.4& $0.043^{+0.005}_{-0.009}$ & $4.29^{+0.47}_{-0.87}$ & $0.031$\\
1304.5& $0.185^{+0.005}_{-0.011}$ & $6.79^{+0.24}_{-0.45}$ & $0.171$\\
1335.0& $0.100^{+0.002}_{-0.003}$ & $5.46^{+0.13}_{-0.23}$ & $0.096$\\
1344.8& $0.026^{+0.001}_{-0.002}$ & $4.75^{+0.34}_{-0.58}$ & $0.023$\\
1364.9& $0.048^{+0.002}_{-0.004}$ & $6.76^{+0.26}_{-0.49}$ & $0.043$\\
1382.3& $0.127^{+0.002}_{-0.003}$ & $7.57^{+0.10}_{-0.19}$ & $0.123$\\
1390.5& $0.067^{+0.002}_{-0.004}$ & $4.13^{+0.10}_{-0.19}$ & $0.062$\\
1399.6& $0.304^{+0.002}_{-0.004}$ & $7.89^{+0.06}_{-0.11}$ & $0.300$\\
1417.7& $0.0516^{+0.001}_{-0.002}$ & $5.64^{+0.17}_{-0.32}$ & $0.049$\\
1430.8& $0.0326^{+0.002}_{-0.004}$ & $8.18^{+0.43}_{-0.81}$ & $0.028$\\
\hline
\end{tabular}
\end{minipage}
\hfil
\begin{minipage}[t]{0.3\linewidth}\centering
 \caption{ Parameters of emission lines for spectrum 3 with $\alpha_{\lambda}=-1.02$.}\label{tab-spec-3}
 \vspace*{-2ex}
 \fontsize{7}{7}\selectfont
 \begin{tabular}{p{0.7cm}p{1.3cm}p{1.2cm}p{0.6cm}}
\hline
$\lambda_{0}$,\,\AA   & a & w,\,\AA & a-3$\sigma$\\
\hline
1059.5& $0.166^{+0.007}_{-0.013}$ & $6.94^{+0.31}_{-0.56}$ & $0.149$\\
1073.9& $0.191^{+0.008}_{-0.015}$ & $5.79^{+0.15}_{-0.29}$ & $0.171$\\
1086.1& $0.084^{+0.007}_{-0.013}$ & $3.37^{+0.21}_{-0.41}$ & $0.067$\\
1096.9& $0.028^{+0.007}_{-0.013}$ & $3.05^{+0.88}_{-1.43}$ & $0.012$\\
1118.3& $0.068^{+0.003}_{-0.006}$ & $8.23^{+0.32}_{-0.61}$ & $0.060$\\
1125.7& $0.029^{+0.003}_{-0.008}$ & $2.59^{+0.27}_{-0.56}$ & $0.017$\\
1134.4& $0.031^{+0.002}_{-0.004}$ & $4.12^{+0.48}_{-0.89}$ & $0.025$\\
1161.4& $0.030^{+0.008}_{-0.017}$ & $2.45^{+0.67}_{-1.26}$ & $0.007$\\
1176.5& $0.101^{+0.011}_{-0.020}$ & $7.60^{+0.62}_{-1.19}$ & $0.074$\\
1214.3& $1.142^{+0.040}_{-0.078}$ & $16.08^{+0.29}_{-0.55}$ & $1.040$\\
\hline
1214.3& $2.544^{+0.008}_{-0.015}$ & $5.62^{+0.03}_{-0.06}$ & $2.524$\\
1225.0& $0.713^{+0.011}_{-0.021}$ & $4.96^{+0.05}_{-0.10}$ & $0.685$\\
1236.1& $1.177^{+0.005}_{-0.010}$ & $7.55^{+0.04}_{-0.07}$ & $1.164$\\
1256.7& $0.404^{+0.006}_{-0.012}$ & $11.63^{+0.20}_{-0.39}$ & $0.388$\\
1289.7& $0.044^{+0.005}_{-0.009}$ & $4.13^{+0.43}_{-0.78}$ & $0.032$\\
1304.1& $0.181^{+0.006}_{-0.012}$ & $7.24^{+0.29}_{-0.54}$ & $0.166$\\
1334.9& $0.083^{+0.002}_{-0.004}$ & $5.75^{+0.24}_{-0.47}$ & $0.078$\\
1343.2& $0.028^{+0.004}_{-0.005}$ & $8.83^{+1.72}_{-3.02}$ & $0.022$\\
1364.4& $0.038^{+0.003}_{-0.007}$ & $6.18^{+0.51}_{-0.88}$ & $0.029$\\
1380.1& $0.104^{+0.006}_{-0.011}$ & $7.91^{+0.37}_{-0.64}$ & $0.088$\\
1390.3& $0.066^{+0.005}_{-0.010}$ & $5.37^{+0.32}_{-0.55}$ & $0.052$\\
1399.3& $0.290^{+0.006}_{-0.011}$ & $8.57^{+0.12}_{-0.24}$ & $0.275$\\
1418.7& $0.050^{+0.002}_{-0.004}$ & $6.06^{+0.31}_{-0.55}$ & $0.045$\\
1431.4& $0.030^{+0.005}_{-0.008}$ & $7.82^{+1.35}_{-2.15}$ & $0.020$\\
\hline
\end{tabular}
\end{minipage}
\end{table*}
\begin{table*}
\begin{minipage}[t]{0.3\linewidth}\centering
\caption{ Parameters of emission lines for spectrum 4 with $\alpha_{\lambda}=-1.04$.}\label{tab-spec-4}
\vspace*{-2ex}
\fontsize{7}{7}\selectfont
 \begin{tabular}{p{0.7cm}p{1.3cm}p{1.2cm}p{0.6cm}}
\hline
$\lambda_{0}$,\,\AA   & a & w,\,\AA & a-3$\sigma$\\
\hline
1059.5	& $0.099^{+0.008}_{-0.014}$ & $2.43^{+0.16}_{-0.31}$ & $0.081$\\
1071.0	& $0.239^{+0.014}_{-0.027}$ & $9.54^{+0.33}_{-0.64}$ & $0.204$\\
1089.3	& $0.054^{+0.011}_{-0.021}$ & $4.06^{+0.60}_{-1.10}$ & $0.027$\\
1117.0  & $0.124^{+0.005}_{-0.010}$ & $8.32^{+0.26}_{-0.52}$ & $0.111$\\
1126.2	& $0.055^{+0.003}_{-0.005}$ & $3.54^{+0.16}_{-0.31}$ & $0.048$\\
1136.5	& $0.085^{+0.004}_{-0.008}$ & $4.69^{+0.20}_{-0.39}$ & $0.074$\\
1175.2	& $0.093^{+0.007}_{-0.013}$ & $6.04^{+0.58}_{-0.99}$ & $0.076$\\
1185.3	& $0.026^{+0.007}_{-0.015}$ & $1.58^{+0.78}_{-1.36}$ & $0.007$\\
1214.8	& $1.236^{+0.042}_{-0.076}$ & $16.07^{+0.29}_{-0.72}$ & $1.140$\\
\hline
1214.8	& $2.465^{+0.008}_{-0.014}$ & $6.14^{+0.03}_{-0.06}$ & $2.446$\\
1226.4	& $0.554^{+0.010}_{-0.020}$ & $4.74^{+0.06}_{-0.11}$ & $0.528$\\
1236.4	& $1.142^{+0.005}_{-0.010}$ & $7.62^{+0.04}_{-0.07}$ & $1.129$\\
1257.0	& $0.405^{+0.006}_{-0.012}$ & $11.54^{+0.21}_{-0.39}$ & $0.389$\\
1290.6	& $0.038^{+0.005}_{-0.009}$ & $3.94^{+0.44}_{-0.83}$ & $0.026$\\
1304.2	& $0.173^{+0.006}_{-0.011}$ & $7.28^{+0.32}_{-0.58}$ & $0.159$\\
1334.7	& $0.077^{+0.001}_{-0.003}$ & $5.50^{+0.11}_{-0.21}$ & $0.074$\\
1341.4	& $0.026^{+0.001}_{-0.002}$ & $4.28^{+0.23}_{-0.43}$ & $0.024$\\
1364.8	& $0.041^{+0.002}_{-0.003}$ & $8.91^{+0.35}_{-0.63}$ & $0.037$\\
1376.9	& $0.065^{+0.002}_{-0.003}$ & $5.77^{+0.17}_{-0.31}$ & $0.061$\\
1386.5	& $0.114^{+0.003}_{-0.006}$ & $4.95^{+0.09}_{-0.16}$ & $0.106$\\
1392.0	& $0.028^{+0.002}_{-0.004}$ & $2.82^{+0.16}_{-0.24}$ & $0.024$\\
1399.0	& $0.297^{+0.001}_{-0.002}$ & $7.31^{+0.08}_{-0.17}$ & $0.294$\\
1412.3	& $0.024^{+0.002}_{-0.004}$ & $3.71^{+0.26}_{-0.51}$ & $0.019$\\
1419.6	& $0.054^{+0.001}_{-0.003}$ & $8.73^{+0.20}_{-0.37}$ & $0.050$\\
1439.5	& $0.006^{+0.001}_{-0.002}$ & $2.37^{+0.53}_{-0.91}$ & $0.004$\\
\hline
\end{tabular}
\end{minipage}
\hfil
\begin{minipage}[t]{0.3\linewidth}\centering
\caption{ Parameters of emission lines for spectrum 5 with $\alpha_{\lambda}=-1.19$.}\label{tab-spec-5}
\vspace*{-2ex}
\fontsize{7}{7}\selectfont
 \begin{tabular}{p{0.7cm}p{1.3cm}p{1.2cm}p{0.6cm}}
\hline
$\lambda_{0}$,\,\AA   & a & w,\,\AA & a-3$\sigma$\\
\hline
1070.7	& $0.221^{+0.004}_{-0.007}$ & $8.60^{+0.22}_{-0.41}$ & $0.213$\\
1087.6	& $0.055^{+0.004}_{-0.009}$ & $3.91^{+0.32}_{-0.62}$ & $0.043$\\
1115.2	& $0.043^{+0.004}_{-0.007}$ & $2.10^{+0.18}_{-0.35}$ & $0.033$\\
1124.4	& $0.101^{+0.004}_{-0.006}$ & $8.28^{+0.60}_{-1.08}$ & $0.094$\\
1136.2	& $0.028^{+0.004}_{-0.007}$ & $2.17^{+0.25}_{-0.48}$ & $0.018$\\
1152.3	& $0.039^{+0.006}_{-0.012}$ & $3.25^{+0.54}_{-1.01}$ & $0.024$\\
1171.9	& $0.108^{+0.007}_{-0.013}$ & $7.94^{+0.39}_{-0.76}$ & $0.091$\\
1214.3	& $1.123^{+0.023}_{-0.045}$ & $17.52^{+0.19}_{-0.37}$ & $1.063$\\
\hline
1214.3	& $2.819^{+0.007}_{-0.013}$ & $5.67^{+0.02}_{-0.04}$ & $2.802$\\
1225.3	& $0.676^{+0.009}_{-0.018}$ & $4.78^{+0.04}_{-0.08}$ & $0.652$\\
1236.0	& $1.203^{+0.005}_{-0.009}$ & $7.80^{+0.03}_{-0.07}$ & $1.191$\\
1256.7	& $0.405^{+0.006}_{-0.011}$ & $11.49^{+0.17}_{-0.33}$ & $0.391$\\
1291.4	& $0.041^{+0.004}_{-0.008}$ & $5.51^{+0.71}_{-1.20}$ & $0.031$\\
1304.5	& $0.171^{+0.005}_{-0.009}$ & $6.77^{+0.28}_{-0.51}$ & $0.159$\\
1335.8	& $0.091^{+0.001}_{-0.003}$ & $6.54^{+0.10}_{-0.20}$ & $0.088$\\
1370.5	& $0.061^{+0.001}_{-0.003}$ & $11.17^{+0.19}_{-0.37}$ & $0.057$\\
1396.5	& $0.310^{+0.001}_{-0.003}$ & $10.45^{+0.04}_{-0.07}$ & $0.307$\\
1401.7	& $0.023^{+0.001}_{-0.002}$ & $2.76^{+0.15}_{-0.30}$ & $0.020$\\
1410.9	& $0.019^{+0.002}_{-0.003}$ & $4.11^{+0.30}_{-0.59}$ & $0.015$\\
1422.4	& $0.043^{+0.001}_{-0.003}$ & $5.33^{+0.13}_{-0.24}$ & $0.040$\\
1436.2	& $0.021^{+0.002}_{-0.003}$ & $7.23^{+0.57}_{-0.99}$ & $0.017$\\
\hline
\end{tabular}
\end{minipage}
\hfil
\begin{minipage}[t]{0.3\linewidth}\centering
\caption{ Parameters of emission lines for spectrum 6 with $\alpha_{\lambda}=-1.35$.}\label{tab-spec-6}
\vspace*{-2ex}
\fontsize{7}{7}\selectfont
 \begin{tabular}{p{0.7cm}p{1.3cm}p{1.2cm}p{0.6cm}}
\hline
$\lambda_{0}$,\,\AA   & a & w,\,\AA & a-3$\sigma$\\
\hline
1100.9	& $0.023^{+0.003}_{-0.005}$ & $1.66^{+0.22}_{-0.38}$ & $0.017$\\
1116.9	& $0.118^{+0.002}_{-0.004}$ & $5.08^{+0.13}_{-0.24}$ & $0.113$\\
1125.0	& $0.040^{+0.003}_{-0.007}$ & $2.57^{+0.13}_{-0.25}$ & $0.030$\\
1132.4	& $0.074^{+0.002}_{-0.005}$ & $4.69^{+0.23}_{-0.43}$ & $0.068$\\
1174.2	& $0.076^{+0.008}_{-0.014}$ & $4.71^{+0.56}_{-0.99}$ & $0.057$\\
1214.2	& $1.122^{+0.032}_{-0.059}$ & $17.05^{+0.25}_{-0.49}$ & $1.047$\\
\hline
1214.2& $2.920^{+0.007}_{-0.014}$ & $5.54^{+0.03}_{-0.05}$ & $2.902$\\
1224.3& $0.586^{+0.011}_{-0.022}$ & $4.55^{+0.05}_{-0.10}$ & $0.557$\\
1234.8& $1.298^{+0.005}_{-0.010}$ & $8.16^{+0.03}_{-0.066}$ & $1.284$\\
1256.5& $0.404^{+0.005}_{-0.011}$ & $11.42^{+0.18}_{-0.37}$ & $0.390$\\
1290.1& $0.046^{+0.004}_{-0.008}$ & $7.03^{+0.86}_{-1.52}$ & $0.036$\\
1304.4& $0.181^{+0.006}_{-0.012}$ & $6.25^{+0.21}_{-0.40}$ & $0.164$\\
1334.6& $0.106^{+0.004}_{-0.007}$ & $6.20^{+0.24}_{-0.42}$ & $0.097$\\
1348.5& $0.033^{+0.002}_{-0.004}$ & $4.86^{+0.35}_{-0.65}$ & $0.028$\\
1361.4& $0.041^{+0.004}_{-0.008}$ & $5.01^{+0.47}_{-0.87}$ & $0.032$\\
1382.1& $0.131^{+0.005}_{-0.008}$ & $9.95^{+0.26}_{-0.50}$ & $0.120$\\
1389.8& $0.081^{+0.004}_{-0.009}$ & $4.37^{+0.13}_{-0.24}$ & $0.069$\\
1399.3& $0.196^{+0.004}_{-0.008}$ & $6.13^{+0.14}_{-0.27}$ & $0.186$\\
1407.4& $0.125^{+0.005}_{-0.010}$ & $12.27^{+0.33}_{-0.60}$ & $0.111$\\
1435.4& $0.030^{+0.004}_{-0.007}$ & $10.77^{+1.19}_{-2.08}$ & $0.021$\\
\hline
\end{tabular}
\end{minipage}
\end{table*}

\begin{table*}
\begin{minipage}[t]{0.3\linewidth}\centering
 \caption{ Parameters of emission lines for spectrum 7 with $\alpha_{\lambda}=-1.42$.}\label{tab-spec-7}
\fontsize{7}{7}\selectfont
\vspace*{-2ex}
 \begin{tabular}{p{0.7cm}p{1.3cm}p{1.2cm}p{0.6cm}}
 \hline
 $\lambda_{0}$,\,\AA   & a & w,\,\AA & a-3$\sigma$\\
 \hline
1061.6	& $0.211^{+0.008}_{-0.014}$ & $7.71^{+0.41}_{-0.75}$ & $0.194$\\
1071.1	& $0.056^{+0.010}_{-0.020}$ & $2.36^{+0.28}_{-0.59}$ & $0.029$\\
1079.3	& $0.137^{+0.007}_{-0.013}$ & $6.00^{+0.45}_{-0.80}$ & $0.120$\\
1091.0	& $0.052^{+0.009}_{-0.017}$ & $3.33^{+0.60}_{-1.07}$ & $0.030$\\
1108.4	& $0.055^{+0.014}_{-0.025}$ & $3.63^{+0.98}_{-1.49}$ & $0.024$\\
1116.6	& $0.093^{+0.007}_{-0.014}$ & $3.81^{+0.35}_{-0.61}$ & $0.075$\\
1127.0	& $0.126^{+0.013}_{-0.022}$ & $4.06^{+0.43}_{-0.72}$ & $0.098$\\
1143.9	& $0.051^{+0.013}_{-0.021}$ & $4.07^{+1.08}_{-1.83}$ & $0.025$\\
1172.1	& $0.086^{+0.009}_{-0.017}$ & $5.52^{+0.54}_{-0.99}$ & $0.064$\\
1213.2	& $1.166^{+0.026}_{-0.049}$ & $16.18^{+0.18}_{-0.36}$ & $1.106$\\
 \hline
1213.2& $2.723^{+0.015}_{-0.028}$ & $4.95^{+0.05}_{-0.09}$ & $2.687$\\
1221.6& $1.319^{+0.020}_{-0.038}$ & $6.40^{+0.08}_{-0.14}$ & $1.270$\\
1236.1& $1.115^{+0.007}_{-0.014}$ & $7.11^{+0.05}_{-0.10}$ & $1.096$\\
1253.4& $0.498^{+0.025}_{-0.042}$ & $14.37^{+0.34}_{-0.71}$ & $0.447$\\
1300.4& $0.136^{+0.044}_{-0.028}$ & $12.37^{+1.76}_{-3.03}$ & $0.098$\\
1305.2& $0.092^{+0.012}_{-0.022}$ & $5.80^{+0.64}_{-1.19}$ & $0.062$\\
1334.9& $0.094^{+0.004}_{-0.007}$ & $7.24^{+0.27}_{-0.51}$ & $0.085$\\
1362.6& $0.008^{+0.002}_{-0.004}$ & $3.26^{+1.02}_{-1.71}$ & $0.003$\\
1394.5& $0.118^{+0.006}_{-0.013}$ & $27.91^{+1.06}_{-2.08}$ & $0.102$\\
1396.5& $0.227^{+0.004}_{-0.009}$ & $9.12^{+0.12}_{-0.24}$ & $0.216$\\
1418.9& $0.006^{+0.002}_{-0.003}$ & $12.28^{+8.38}_{-8.97}$ & $0.002$\\
\hline
 \end{tabular}
\end{minipage}
\hfil
\begin{minipage}[t]{0.3\linewidth}\centering
\caption{ Parameters of emission lines for spectrum 8 with $\alpha_{\lambda}=-1.42$.}\label{tab-spec-8}
\fontsize{7}{7}\selectfont
\vspace*{-2ex}
 \begin{tabular}{p{0.7cm}p{1.3cm}p{1.2cm}p{0.6cm}}
\hline
$\lambda_{0}$,\,\AA   & a & w,\,\AA & a-3$\sigma$\\
\hline
1060.3	& $0.201^{+0.017}_{-0.032}$ & $5.44^{+0.48}_{-0.87}$ & $0.161$\\
1071.3	& $0.126^{+0.013}_{-0.030}$ & $3.38^{+0.29}_{-0.62}$ & $0.077$\\
1080.8	& $0.091^{+0.022}_{-0.038}$ & $7.69^{+1.31}_{-2.66}$ & $0.045$\\
1098.5	& $0.087^{+0.011}_{-0.019}$ & $3.90^{+0.39}_{-0.71}$ & $0.063$\\
1111.5	& $0.109^{+0.010}_{-0.017}$ & $4.98^{+0.19}_{-0.35}$ & $0.088$\\
1125.4	& $0.152^{+0.012}_{-0.021}$ & $6.61^{+0.24}_{-0.44}$ & $0.126$\\
1141.0	& $0.085^{+0.011}_{-0.018}$ & $3.90^{+0.33}_{-0.60}$ & $0.062$\\
1172.5	& $0.112^{+0.006}_{-0.012}$ & $7.71^{+0.51}_{-0.93}$ & $0.097$\\
1213.8	& $1.249^{+0.028}_{-0.055}$ & $16.42^{+0.18}_{-0.36}$ & $1.178$\\
\hline
1213.8	& $3.121^{+0.008}_{-0.016}$ & $5.07^{+0.02}_{-0.04}$ & $3.099$\\
1223.8	& $0.980^{+0.012}_{-0.023}$ & $5.00^{+0.04}_{-0.08}$ & $0.949$\\
1235.9	& $1.317^{+0.006}_{-0.012}$ & $7.42^{+0.04}_{-0.07}$ & $1.301$\\
1255.7	& $0.438^{+0.007}_{-0.013}$ & $12.02^{+0.25}_{-0.52}$ & $0.421$\\
1285.7	& $0.039^{+0.004}_{-0.009}$ & $7.23^{+1.16}_{-1.87}$ & $0.028$\\
1303.6	& $0.173^{+0.007}_{-0.015}$ & $7.30^{+0.31}_{-0.59}$ & $0.153$\\
1334.2	& $0.073^{+0.002}_{-0.004}$ & $4.94^{+0.11}_{-0.20}$ & $0.068$\\
1344.1	& $0.023^{+0.001}_{-0.002}$ & $7.01^{+0.90}_{-1.51}$ & $0.021$\\
1363.0	& $0.028^{+0.002}_{-0.003}$ & $6.20^{+0.56}_{-1.00}$ & $0.024$\\
1382.0	& $0.110^{+0.002}_{-0.003}$ & $8.35^{+0.21}_{-0.37}$ & $0.105$\\
1390.3	& $0.125^{+0.002}_{-0.004}$ & $4.67^{+0.07}_{-0.13}$ & $0.116$\\
1399.6	& $0.261^{+0.002}_{-0.004}$ & $5.89^{+0.04}_{-0.07}$ & $0.256$\\
1413.3	& $0.082^{+0.001}_{-0.003}$ & $8.03^{+0.16}_{-0.30}$ & $0.078$\\
1430.7	& $0.010^{+0.001}_{-0.003}$ & $2.43^{+0.27}_{-0.55}$ & $0.008$\\
1438.6	& $0.008^{+0.001}_{-0.002}$ & $3.23^{+0.72}_{-1.26}$ & $0.005$\\
\hline
\end{tabular}
\end{minipage}
\hfil
\begin{minipage}[t]{0.3\linewidth}\centering
\caption{ Parameters of emission lines for spectrum 9 with $\alpha_{\lambda}=-1.62$.}\label{tab-spec-9}
\fontsize{7}{7}\selectfont
\vspace*{-2ex}
 \begin{tabular}{p{0.7cm}p{1.3cm}p{1.2cm}p{0.6cm}}
\hline
$\lambda_{0}$,\,\AA   & a & w,\,\AA & a-3$\sigma$\\
\hline
1070.2	& $0.226^{+0.003}_{-0.005}$ & $9.92^{+0.24}_{-0.44}$ & $0.219$\\
1086.0	& $0.034^{+0.003}_{-0.007}$ & $1.92^{+0.25}_{-0.46}$ & $0.025$\\
1099.6	& $0.077^{+0.007}_{-0.015}$ & $9.44^{+0.68}_{-1.19}$ & $0.058$\\
1122.6	& $0.169^{+0.011}_{-0.022}$ & $10.38^{+0.39}_{-0.76}$ & $0.139$\\
1172.6	& $0.095^{+0.009}_{-0.017}$ & $4.95^{+0.43}_{-0.80}$ & $0.073$\\
1213.4	& $1.084^{+0.024}_{-0.045}$ & $17.44^{+0.23}_{-0.44}$ & $1.026$\\
\hline
1213.4	& $2.702^{+0.009}_{-0.017}$ & $5.42^{+0.03}_{-0.06}$ & $2.680$\\
1223.4	& $1.045^{+0.012}_{-0.024}$ & $6.05^{+0.06}_{-0.11}$ & $1.013$\\
1236.3	& $1.177^{+0.006}_{-0.013}$ & $7.49^{+0.04}_{-0.07}$ & $1.161$\\
1255.6	& $0.444^{+0.007}_{-0.014}$ & $12.46^{+0.23}_{-0.45}$ & $0.426$\\
1286.6	& $0.044^{+0.004}_{-0.007}$ & $6.36^{+0.64}_{-1.15}$ & $0.034$\\
1303.6	& $0.175^{+0.007}_{-0.014}$ & $7.75^{+0.32}_{-0.60}$ & $0.158$\\
\hline
\end{tabular}
\end{minipage}
 \end{table*}

\begin{table*}
\begin{minipage}[t]{0.3\linewidth}\centering
\caption{ Parameters of emission lines for spectrum 10 with $\alpha_{\lambda}=-1.64$.}\label{tab-spec-10}
\fontsize{7}{7}\selectfont
\vspace*{-2ex}
 \begin{tabular}{p{0.7cm}p{1.3cm}p{1.2cm}p{0.6cm}}
\hline
$\lambda_{0}$,\,\AA   & a & w,\,\AA & a-3$\sigma$\\
\hline
1056.3	& $0.180^{+0.009}_{-0.016}$ & $3.37^{+0.17}_{-0.30}$ & $0.161$\\
1071.7	& $0.222^{+0.009}_{-0.016}$ & $8.10^{+0.29}_{-0.56}$ & $0.203$\\
1086.8	& $0.054^{+0.008}_{-0.015}$ & $3.41^{+0.53}_{-0.94}$ & $0.035$\\
1099.6	& $0.024^{+0.005}_{-0.009}$ & $2.30^{+0.59}_{-1.19}$ & $0.012$\\
1117.3	& $0.083^{+0.004}_{-0.008}$ & $8.20^{+0.50}_{-0.94}$ & $0.073$\\
1127.4	& $0.036^{+0.004}_{-0.008}$ & $2.91^{+0.34}_{-0.61}$ & $0.026$\\
1141.7	& $0.025^{+0.005}_{-0.010}$ & $2.49^{+0.65}_{-1.11}$ & $0.012$\\
1163.7	& $0.044^{+0.006}_{-0.013}$ & $2.70^{+0.44}_{-0.80}$ & $0.028$\\
1174.3	& $0.090^{+0.005}_{-0.009}$ & $5.84^{+0.46}_{-0.81}$ & $0.078$\\
1213.8	& $1.114^{+0.020}_{-0.037}$ & $17.88^{+0.28}_{-0.56}$ & $1.068$\\
\hline
1213.8	& $2.899^{+0.008}_{-0.016}$ & $5.32^{+0.02}_{-0.05}$ & $2.878$\\
1223.9	& $0.903^{+0.011}_{-0.022}$ & $5.13^{+0.04}_{-0.08}$ & $0.874$\\
1235.5	& $1.128^{+0.005}_{-0.010}$ & $7.44^{+0.05}_{-0.09}$ & $1.115$\\
1252.7	& $0.483^{+0.007}_{-0.014}$ & $14.66^{+0.27}_{-0.51}$ & $0.465$\\
1288.6	& $0.050^{+0.004}_{-0.008}$ & $6.39^{+0.58}_{-1.05}$ & $0.040$\\
1304.6	& $0.171^{+0.007}_{-0.014}$ & $7.50^{+0.33}_{-0.62}$ & $0.153$\\
\hline
\end{tabular}
\end{minipage}
\hfil
\begin{minipage}[t]{0.3\linewidth}\centering
 \caption{ Parameters of emission lines for spectrum 11 with $\alpha_{\lambda}=-1.73$.}\label{tab-spec-11}
 \fontsize{7}{7}\selectfont
\vspace*{-2ex}
 \begin{tabular}{p{0.7cm}p{1.3cm}p{1.2cm}p{0.6cm}}
\hline
$\lambda_{0}$,\,\AA   & a & w,\,\AA & a-3$\sigma$\\
\hline
1058.1	& $0.222^{+0.004}_{-0.007}$ & $7.34^{+0.28}_{-0.50}$ & $0.213$\\
1070.4	& $0.072^{+0.007}_{-0.014}$ & $4.70^{+0.22}_{-0.42}$ & $0.051$\\
1079.9	& $0.145^{+0.004}_{-0.009}$ & $8.59^{+0.25}_{-0.49}$ & $0.133$\\
1106.7	& $0.035^{+0.006}_{-0.011}$ & $2.60^{+0.50}_{-0.84}$ & $0.021$\\
1115.7	& $0.076^{+0.004}_{-0.008}$ & $3.62^{+0.22}_{-0.42}$ & $0.065$\\
1125.6	& $0.128^{+0.007}_{-0.011}$ & $7.00^{+0.34}_{-0.62}$ & $0.114$\\
1168.5	& $0.069^{+0.007}_{-0.014}$ & $4.83^{+0.76}_{-1.38}$ & $0.052$\\
1174.9	& $0.064^{+0.009}_{-0.017}$ & $2.60^{+0.31}_{-0.57}$ & $0.040$\\
1213.7	& $1.212^{+0.024}_{-0.046}$ & $17.38^{+0.23}_{-0.45}$ & $1.151$\\
\hline
1213.7	& $2.984^{+0.007}_{-0.015}$ & $5.08^{+0.03}_{-0.05}$ & $2.964$\\
1223.4	& $1.110^{+0.012}_{-0.024}$ & $5.50^{+0.04}_{-0.09}$ & $1.078$\\
1236.2	& $1.103^{+0.005}_{-0.010}$ & $7.06^{+0.05}_{-0.09}$ & $1.089$\\
1252.3	& $0.495^{+0.005}_{-0.010}$ & $14.55^{+0.19}_{-0.39}$ & $0.482$\\
1288.4	& $0.037^{+0.004}_{-0.008}$ & $6.26^{+0.80}_{-1.42}$ & $0.027$\\
1304.0	& $0.164^{+0.005}_{-0.010}$ & $6.90^{+0.25}_{-0.48}$ & $0.151$\\
1334.0	& $0.065^{+0.001}_{-0.002}$ & $4.34^{+0.11}_{-0.21}$ & $0.062$\\
1342.3	& $0.024^{+0.001}_{-0.002}$ & $3.56^{+0.17}_{-0.31}$ & $0.021$\\
1376.2	& $0.053^{+0.001}_{-0.002}$ & $13.64^{+0.37}_{-0.71}$ & $0.050$\\
1395.6	& $0.271^{+0.001}_{-0.002}$ & $9.18^{+0.05}_{-0.09}$ & $0.269$\\
1401.5	& $0.037^{+0.001}_{-0.002}$ & $3.01^{+0.09}_{-0.18}$ & $0.034$\\
1408.7	& $0.055^{+0.002}_{-0.003}$ & $4.89^{+0.12}_{-0.22}$ & $0.051$\\
1421.0	& $0.059^{+0.001}_{-0.002}$ & $7.08^{+0.16}_{-0.29}$ & $0.056$\\
1436.6	& $0.017^{+0.001}_{-0.003}$ & $5.45^{+0.41}_{-0.78}$ & $0.013$\\
\hline
\end{tabular}
\end{minipage}
\hfil
\begin{minipage}[t]{0.3\linewidth}\centering
\caption{ Parameters of emission lines for spectrum 12 with $\alpha_{\lambda}=-1.88$.}\label{tab-spec-12}
\fontsize{7}{7}\selectfont
\vspace*{-2ex}
 \begin{tabular}{p{0.7cm}p{1.3cm}p{1.2cm}p{0.6cm}}
\hline
$\lambda_{0}$,\,\AA   & a & w,\,\AA & a-3$\sigma$\\
\hline
1063.0	& $0.249^{+0.014}_{-0.025}$ & $7.30^{+0.52}_{-0.98}$ & $0.218$\\
1074.1	& $0.082^{+0.014}_{-0.025}$ & $2.76^{+0.29}_{-0.57}$ & $0.051$\\
1084.0	& $0.157^{+0.014}_{-0.023}$ & $6.56^{+0.57}_{-1.04}$ & $0.128$\\
1115.5	& $0.108^{+0.005}_{-0.009}$ & $6.05^{+0.42}_{-0.73}$ & $0.097$\\
1123.8	& $0.036^{+0.008}_{-0.019}$ & $2.55^{+0.33}_{-0.74}$ & $0.012$\\
1129.0	& $0.038^{+0.005}_{-0.009}$ & $4.53^{+0.83}_{-1.43}$ & $0.027$\\
1172.9	& $0.079^{+0.009}_{-0.014}$ & $4.36^{+0.67}_{-1.13}$ & $0.061$\\
1178.5	& $0.032^{+0.008}_{-0.015}$ & $1.89^{+0.38}_{-0.81}$ & $0.012$\\
1214.5	& $1.185^{+0.019}_{-0.036}$ & $17.34^{+0.26}_{-0.49}$ & $1.139$\\
\hline
1214.5	& $3.251^{+0.015}_{-0.025}$ & $5.85^{+0.03}_{-0.05}$ & $3.219$\\
1225.7	& $0.478^{+0.012}_{-0.024}$ & $3.90^{+0.06}_{-0.12}$ & $0.445$\\
1235.1	& $1.363^{+0.008}_{-0.015}$ & $8.21^{+0.04}_{-0.08}$ & $1.344$\\
1256.0	& $0.460^{+0.014}_{-0.025}$ & $12.54^{+0.30}_{-0.67}$ & $0.429$\\
1281.9	& $0.036^{+0.006}_{-0.012}$ & $4.55^{+1.00}_{-1.77}$ & $0.019$\\
1301.6	& $0.133^{+0.014}_{-0.032}$ & $10.64^{+1.08}_{-1.88}$ & $0.088$\\
1304.9	& $0.063^{+0.009}_{-0.017}$ & $4.43^{+0.61}_{-1.11}$ & $0.041$\\
1334.8	& $0.064^{+0.001}_{-0.002}$ & $4.74^{+0.08}_{-0.15}$ & $0.062$\\
1345.1	& $0.015^{+0.001}_{-0.001}$ & $3.43^{+0.20}_{-0.37}$ & $0.014$\\
1385.1	& $0.087^{+0.001}_{-0.002}$ & $16.27^{+0.26}_{-0.49}$ & $0.083$\\
1393.9	& $0.200^{+0.001}_{-0.002}$ & $7.60^{+0.06}_{-0.12}$ & $0.197$\\
1401.0	& $0.065^{+0.001}_{-0.002}$ & $3.33^{+0.06}_{-0.11}$ & $0.062$\\
1407.8	& $0.101^{+0.002}_{-0.003}$ & $6.04^{+0.06}_{-0.12}$ & $0.096$\\
1423.3	& $0.048^{+0.001}_{-0.001}$ & $7.55^{+0.18}_{-0.34}$ & $0.047$\\
1438.5	& $0.012^{+0.001}_{-0.002}$ & $4.81^{+0.36}_{-0.71}$ & $0.010$\\
\hline
\end{tabular}
\end{minipage}
\end{table*}

\begin{table*}
\begin{minipage}[t]{0.3\linewidth}\centering
\caption{ Parameters of emission lines for spectrum 13 with $\alpha_{\lambda}=-1.92$.}\label{tab-spec-13}
\fontsize{7}{7}\selectfont
\vspace*{-2ex}
 \begin{tabular}{p{0.7cm}p{1.3cm}p{1.2cm}p{0.6cm}}
\hline
$\lambda_{0}$,\,\AA   & a & w,\,\AA & a-3$\sigma$\\
\hline
1063.8	& $0.144^{+0.003}_{-0.006}$ & $8.71^{+0.52}_{-1.07}$ & $0.136$\\
1072.6	& $0.093^{+0.005}_{-0.010}$ & $2.65^{+0.13}_{-0.26}$ & $0.080$\\
1082.0	& $0.044^{+0.005}_{-0.010}$ & $2.07^{+0.24}_{-0.48}$ & $0.030$\\
1087.5	& $0.068^{+0.003}_{-0.006}$ & $3.11^{+0.22}_{-0.40}$ & $0.059$\\
1103.3	& $0.037^{+0.003}_{-0.006}$ & $2.78^{+0.25}_{-0.46}$ & $0.029$\\
1111.9	& $0.070^{+0.003}_{-0.005}$ & $4.08^{+0.17}_{-0.32}$ & $0.063$\\
1124.3	& $0.095^{+0.003}_{-0.005}$ & $4.80^{+0.13}_{-0.25}$ & $0.089$\\
1135.9	& $0.023^{+0.003}_{-0.005}$ & $2.71^{+0.35}_{-0.63}$ & $0.016$\\
1153.9	& $0.049^{+0.004}_{-0.008}$ & $3.87^{+0.37}_{-0.67}$ & $0.039$\\
1164.1	& $0.047^{+0.003}_{-0.007}$ & $4.36^{+0.28}_{-0.53}$ & $0.038$\\
1172.7	& $0.125^{+0.005}_{-0.009}$ & $5.89^{+0.24}_{-0.45}$ & $0.113$\\
1214.5	& $1.471^{+0.020}_{-0.038}$ & $16.84^{+0.13}_{-0.27}$ & $1.422$\\
\hline
1214.5	& $3.211^{+0.004}_{-0.008}$ & $5.63^{+0.01}_{-0.02}$ & $3.200$\\
1225.8	& $0.865^{+0.009}_{-0.018}$ & $4.85^{+0.02}_{-0.04}$ & $0.842$\\
1237.4	& $1.413^{+0.005}_{-0.010}$ & $7.39^{+0.05}_{-0.10}$ & $1.399$\\
1250.8	& $0.104^{+0.006}_{-0.011}$ & $3.67^{+0.16}_{-0.31}$ & $0.090$\\
1258.4	& $0.386^{+0.005}_{-0.009}$ & $8.54^{+0.10}_{-0.19}$ & $0.374$\\
1276.7	& $0.093^{+0.005}_{-0.009}$ & $14.14^{+0.78}_{-1.37}$ & $0.082$\\
1304.5	& $0.145^{+0.005}_{-0.010}$ & $6.90^{+0.21}_{-0.42}$ & $0.132$\\
1333.9	& $0.029^{+0.004}_{-0.007}$ & $3.92^{+0.47}_{-0.84}$ & $0.019$\\
1336.0	& $0.055^{+0.004}_{-0.008}$ & $9.45^{+0.80}_{-1.36}$ & $0.044$\\
1376.0	& $0.069^{+0.007}_{-0.011}$ & $18.68^{+0.71}_{-1.34}$ & $0.055$\\
1397.9	& $0.302^{+0.002}_{-0.004}$ & $10.62^{+0.09}_{-0.19}$ & $0.296$\\
1419.3	& $0.030^{+0.002}_{-0.004}$ & $6.16^{+0.47}_{-0.84}$ & $0.025$\\
1429.8	& $0.046^{+0.005}_{-0.008}$ & $10.20^{+1.14}_{-2.0}$ & $0.036$\\
\hline
\end{tabular}
\end{minipage}
\hfil
\begin{minipage}[t]{0.3\linewidth}\centering
 \caption{ Parameters of emission lines for spectrum 14 with $\alpha_{\lambda}=-2.02$.}\label{tab-spec-14}
 \fontsize{7}{7}\selectfont
\vspace*{-2ex}
 \begin{tabular}{p{0.7cm}p{1.3cm}p{1.2cm}p{0.6cm}}
\hline
$\lambda_{0}$,\,\AA   & a & w,\,\AA & a-3$\sigma$\\
\hline
1101.6	& $0.021^{+0.004}_{-0.008}$ & $2.84^{+0.67}_{-1.11}$ & $0.011$\\
1116.2	& $0.074^{+0.004}_{-0.009}$ & $4.61^{+0.24}_{-0.46}$ & $0.063$\\
1126.2	& $0.060^{+0.004}_{-0.008}$ & $4.05^{+0.32}_{-0.57}$ & $0.050$\\
1135.5	& $0.033^{+0.004}_{-0.009}$ & $2.44^{+0.39}_{-0.71}$ & $0.021$\\
1164.2	& $0.031^{+0.005}_{-0.009}$ & $2.40^{+0.30}_{-0.58}$ & $0.019$\\
1172.0	& $0.125^{+0.003}_{-0.005}$ & $5.42^{+0.25}_{-0.45}$ & $0.118$\\
1214.4	& $1.364^{+0.016}_{-0.031}$ & $17.41^{+0.14}_{-0.27}$ & $1.325$\\ 
\hline
1214.4	& $3.299^{+0.007}_{-0.014}$ & $5.43^{+0.02}_{-0.04}$ & $3.280$\\
1225.0	& $0.640^{+0.010}_{-0.020}$ & $4.20^{+0.04}_{-0.08}$ & $0.614$\\
1235.1	& $1.410^{+0.006}_{-0.012}$ & $8.18^{+0.04}_{-0.07}$ & $1.394$\\
1255.7	& $0.470^{+0.007}_{-0.013}$ & $12.36^{+0.25}_{-0.59}$ & $0.452$\\
1286.2	& $0.057^{+0.004}_{-0.008}$ & $7.66^{+0.86}_{-1.61}$ & $0.046$\\
1304.0	& $0.167^{+0.007}_{-0.018}$ & $6.59^{+0.29}_{-0.57}$ & $0.139$\\
1335.3	& $0.045^{+0.001}_{-0.003}$ & $4.01^{+0.15}_{-0.29}$ & $0.042$\\
1388.8	& $0.086^{+0.002}_{-0.004}$ & $13.12^{+0.17}_{-0.35}$ & $0.081$\\
1398.5	& $0.227^{+0.001}_{-0.003}$ & $9.85^{+0.07}_{-0.14}$ & $0.224$\\
1422.1	& $0.038^{+0.001}_{-0.002}$ & $6.73^{+0.23}_{-0.45}$ & $0.035$\\
\hline
\end{tabular}
\end{minipage}
\hfil
\begin{minipage}[t]{0.3\linewidth}\centering
\caption{ Parameters of emission lines for spectrum 15 with $\alpha_{\lambda}=-2.07$.}\label{tab-spec-15}
\fontsize{7}{7}\selectfont
\vspace*{-2ex}
 \begin{tabular}{p{0.7cm}p{1.3cm}p{1.2cm}p{0.6cm}}
\hline
$\lambda_{0}$,\,\AA   & a & w,\,\AA & a-3$\sigma$\\
\hline
1055.8	& $0.165^{+0.014}_{-0.022}$ & $17.41^{+3.38}_{-5.31}$ & $0.138$\\
1066.6	& $0.078^{+0.006}_{-0.011}$ & $3.88^{+0.36}_{-0.65}$ & $0.065$\\
1072.5	& $0.077^{+0.012}_{-0.021}$ & $2.38^{+0.29}_{-0.52}$ & $0.051$\\
1083.2	& $0.066^{+0.012}_{-0.021}$ & $3.36^{+0.52}_{-0.96}$ & $0.040$\\
1108.3	& $0.024^{+0.004}_{-0.008}$ & $1.88^{+0.50}_{-0.90}$ & $0.013$\\
1117.3	& $0.063^{+0.003}_{-0.006}$ & $3.33^{+0.20}_{-0.38}$ & $0.055$\\
1125.5	& $0.063^{+0.004}_{-0.007}$ & $4.07^{+0.32}_{-0.56}$ & $0.054$\\
\hline
1213.7	& $3.154^{+0.007}_{-0.015}$ & $5.06^{+0.02}_{-0.041}$ & $3.135$\\
1223.2	& $1.010^{+0.011}_{-0.021}$ & $5.08^{+0.04}_{-0.07}$ & $0.982$\\
1235.5	& $1.364^{+0.005}_{-0.011}$ & $8.15^{+0.03}_{-0.07}$ & $1.350$\\
1256.1	& $0.480^{+0.007}_{-0.012}$ & $12.34^{+0.20}_{-0.38}$ & $0.464$\\
1283.7	& $0.068^{+0.004}_{-0.008}$ & $6.25^{+0.44}_{-0.83}$ & $0.058$\\
1303.2	& $0.171^{+0.006}_{-0.011}$ & $7.78^{+0.32}_{-0.60}$ & $0.156$\\
1335.8	& $0.044^{+0.003}_{-0.005}$ & $4.98^{+0.44}_{-0.80}$ & $0.037$\\
1396.8	& $0.138^{+0.012}_{-0.022}$ & $17.97^{+1.07}_{-1.84}$ & $0.110$\\
1397.3	& $0.157^{+0.014}_{-0.027}$ & $8.27^{+0.33}_{-0.68}$ & $0.122$\\
\hline
\end{tabular}
\end{minipage}
\end{table*}

\begin{table*}
\begin{minipage}[t]{0.3\linewidth}\centering
\caption{ Parameters of emission lines for spectrum 16 with $\alpha_{\lambda}=-2.14$.}\label{tab-spec-16}
\fontsize{7}{7}\selectfont
\vspace*{-2ex}
 \begin{tabular}{p{0.7cm}p{1.3cm}p{1.2cm}p{0.6cm}}
\hline
$\lambda_{0}$,\,\AA   & a & w,\,\AA & a-3$\sigma$\\
\hline
1052.4	& $0.136^{+0.003}_{-0.007}$ & $3.61^{+0.15}_{-0.28}$ & $0.128$\\
1060.6	& $0.137^{+0.005}_{-0.010}$ & $4.15^{+0.12}_{-0.23}$ & $0.124$\\
1071.4	& $0.158^{+0.003}_{-0.005}$ & $4.50^{+0.10}_{-0.19}$ & $0.151$\\
1083.1	& $0.073^{+0.003}_{-0.006}$ & $4.29^{+0.22}_{-0.42}$ & $0.065$\\
1111.3	& $0.057^{+0.003}_{-0.006}$ & $4.42^{+0.22}_{-0.41}$ & $0.049$\\
1123.7	& $0.093^{+0.003}_{-0.006}$ & $6.17^{+0.20}_{-0.38}$ & $0.084$\\
1140.1	& $0.027^{+0.003}_{-0.007}$ & $2.72^{+0.41}_{-0.74}$ & $0.018$\\
1155.5	& $0.064^{+0.007}_{-0.011}$ & $5.71^{+0.69}_{-1.13}$ & $0.050$\\
1165.7	& $0.033^{+0.005}_{-0.010}$ & $2.19^{+0.26}_{-0.51}$ & $0.021$\\
1174.3	& $0.114^{+0.005}_{-0.010}$ & $6.10^{+0.31}_{-0.60}$ & $0.102$\\
1214.1	& $1.351^{+0.019}_{-0.037}$ & $17.12^{+0.20}_{-0.39}$ & $1.303$\\
\hline
1214.1	& $3.092^{+0.009}_{-0.017}$ & $5.21^{+0.03}_{-0.05}$ & $3.070$\\
1224.1	& $0.917^{+0.014}_{-0.028}$ & $4.83^{+0.05}_{-0.09}$ & $0.879$\\
1235.6	& $1.176^{+0.006}_{-0.012}$ & $7.74^{+0.06}_{-0.12}$ & $1.160$\\
1252.4	& $0.516^{+0.007}_{-0.014}$ & $15.01^{+0.33}_{-0.67}$ & $0.497$\\
1285.6	& $0.037^{+0.005}_{-0.010}$ & $6.83^{+1.27}_{-2.13}$ & $0.024$\\
1302.2	& $0.155^{+0.007}_{-0.015}$ & $7.15^{+0.39}_{-0.75}$ & $0.134$\\
1333.7	& $0.046^{+0.002}_{-0.004}$ & $5.10^{+0.31}_{-0.57}$ & $0.041$\\
1343.7	& $0.021^{+0.003}_{-0.005}$ & $8.66^{+1.55}_{-2.56}$ & $0.014$\\
1371.8	& $0.041^{+0.004}_{-0.006}$ & $9.95^{+0.46}_{-0.88}$ & $0.033$\\
1397.4	& $0.300^{+0.004}_{-0.006}$ & $10.71^{+0.07}_{-0.14}$ & $0.293$\\
1416.8	& $0.016^{+0.002}_{-0.004}$ & $3.99^{+0.45}_{-0.84}$ & $0.010$\\
1425.7	& $0.045^{+0.004}_{-0.006}$ & $7.35^{+0.57}_{-1.01}$ & $0.038$\\
\hline
\end{tabular}
\end{minipage}
\hfil
\begin{minipage}[t]{0.65\linewidth}\centering
 \caption{The values of $b$ (or $d$ in case of range 5) characterizing continuum for each range. Here n is the number of spectrum.}\label{tab:b-param}
 \fontsize{8}{8}\selectfont
\vspace*{-2ex}
\begin{tabular}{c|c|c|c|c|c}
\hline
n & range 1 & range 2 & range 3 & range 4 & range 5\\
\hline
1 & $0.940^{+0.011}_{-0.032}$ & $0.953^{+0.003}_{-0.007}$ & $0.973^{+0.007}_{-0.018}$ & $1.052^{+0.005}_{-0.011}$ & $717^{+2}_{-4}$ \\ 
2 & $0.963^{+0.012}_{-0.034}$ & $0.977^{+0.003}_{-0.007}$ & $0.989^{+0.015}_{-0.044}$ & $1.067^{+0.005}_{-0.011}$ & $1136^{+2}_{-4}$ \\ 
3 & $0.971^{+0.007}_{-0.016}$ & $0.995^{+0.003}_{-0.008}$ & $1.014^{+0.010}_{-0.027}$ & $1.068^{+0.006}_{-0.013}$ & $1634^{+8}_{-21}$ \\ 
4 & $0.973^{+0.014}_{-0.028}$ & $0.957^{+0.005}_{-0.009}$ & $1.047^{+0.007}_{-0.017}$ & $1.084^{+0.006}_{-0.013}$ & $1920^{+2}_{-5}$ \\ 
5 & $1.011^{+0.003}_{-0.005}$ & $1.001^{+0.004}_{-0.010}$ & $1.010^{+0.006}_{-0.014}$ & $1.102^{+0.005}_{-0.012}$ & $5660^{+8}_{-16}$ \\ 
6 &--- 			      & $1.000^{+0.002}_{-0.006}$ & $1.072^{+0.008}_{-0.018}$ & $1.113^{+0.005}_{-0.010}$ & $17836^{+65}_{-140}$ \\ 
7 & $1.041^{+0.007}_{-0.020}$ & $1.038^{+0.013}_{-0.043}$ & $1.083^{+0.011}_{-0.024}$ & $1.113^{+0.005}_{-0.010}$ & $29541^{+148}_{-304}$  \\ 
8 & $1.077^{+0.020}_{-0.042}$ & $1.007^{+0.011}_{-0.025}$ & $1.099^{+0.005}_{-0.011}$ & $1.123^{+0.006}_{-0.013}$ & $30490^{+37}_{-88}$ \\ 
9 & $1.066^{+0.003}_{-0.006}$ & $1.066^{+0.003}_{-0.006}$ & $1.099^{+0.012}_{-0.025}$ & $1.128^{+0.007}_{-0.015}$ & ---\\ 
10 & $1.071^{+0.008}_{-0.018}$ & $1.087^{+0.005}_{-0.013}$ & $1.089^{+0.006}_{-0.013}$ & $1.126^{+0.008}_{-0.016}$ & --- \\ 
11 & $1.066^{+0.004}_{-0.008}$ & $1.075^{+0.006}_{-0.014}$ & $1.123^{+0.008}_{-0.017}$ & $1.160^{+0.004}_{-0.009}$ & $292221^{+327}_{-699}$ \\ 
12 & $1.103^{+0.015}_{-0.030}$ & $1.124^{+0.006}_{-0.016}$ & $1.165^{+0.014}_{-0.031}$ & $1.149^{+0.013}_{-0.032}$ & $865732^{+587}_{-1183}$ \\ 
13 & $1.142^{+0.003}_{-0.006}$ & $1.121^{+0.002}_{-0.005}$ & $1.119^{+0.004}_{-0.008}$ & $1.178^{+0.003}_{-0.007}$ & $1136890^{+6492}_{-17148}$ \\ 
14 & ---  & $1.136^{+0.003}_{-0.007}$ & $1.149^{+0.003}_{-0.006}$ & $1.193^{+0.006}_{-0.012}$ & $2452040^{+1525}_{-3088}$  \\ 
15 & $1.143^{+0.015}_{-0.052}$ & $1.155^{+0.004}_{-0.009}$ & --- & $1.187^{+0.006}_{-0.012}$ & $3499960^{+8783}_{-20747}$ \\ 
16 & $1.147^{+0.003}_{-0.005}$ & $1.124^{+0.003}_{-0.007}$ & $1.123^{+0.008}_{-0.021}$ & $1.208^{+0.007}_{-0.014}$ & $5788270^{+21075}_{-47788}$ \\ 
\hline
\end{tabular}
\end{minipage}
\end{table*}

\clearpage

\clearpage

\clearpage
\clearpage \vspace*{1ex} \label{lastpage}

\end{document}

%% file: ivashchenko.sergijenko.torbaniuk.bbl
\let\jnlstyle=\rm\def\jref#1{{\jnlstyle#1}}\def\aj{\jref{AJ}}
  \def\araa{\jref{ARA\&A}} \def\apj{\jref{ApJ}\ } \def\apjl{\jref{ApJ}}
  \def\apjs{\jref{ApJS}} \def\ao{\jref{Appl.~Opt.}} \def\apss{\jref{Ap\&SS}}
  \def\aap{\jref{A\&A}} \def\aapr{\jref{A\&A~Rev.}} \def\aaps{\jref{A\&AS}}
  \def\azh{\jref{AZh}} \def\baas{\jref{BAAS}} \def\jrasc{\jref{JRASC}}
  \def\memras{\jref{MmRAS}} \def\mnras{\jref{MNRAS}}
  \def\pra{\jref{Phys.~Rev.~A}\ } \def\prb{\jref{Phys.~Rev.~B}}
  \def\prc{\jref{Phys.~Rev.~C}\ } \def\prd{\jref{Phys.~Rev.~D}}
  \def\pre{\jref{Phys.~Rev.~E}} \def\prl{\jref{Phys.~Rev.~Lett.}}
  \def\pasp{\jref{PASP}} \def\pasj{\jref{PASJ}} \def\qjras{\jref{QJRAS}}
  \def\skytel{\jref{S\&T}} \def\solphys{\jref{Sol.~Phys.}}
  \def\sovast{\jref{Soviet~Ast.}} \def\ssr{\jref{Space~Sci.~Rev.}}
  \def\zap{\jref{ZAp}} \def\nat{\jref{Nature}\ } \def\iaucirc{\jref{IAU~Circ.}}
  \def\aplett{\jref{Astrophys.~Lett.}}
  \def\apspr{\jref{Astrophys.~Space~Phys.~Res.}}
  \def\bain{\jref{Bull.~Astron.~Inst.~Netherlands}}
  \def\fcp{\jref{Fund.~Cosmic~Phys.}} \def\gca{\jref{Geochim.~Cosmochim.~Acta}}
  \def\grl{\jref{Geophys.~Res.~Lett.}} \def\jcp{\jref{J.~Chem.~Phys.}}
  \def\jgr{\jref{J.~Geophys.~Res.}}
  \def\jqsrt{\jref{J.~Quant.~Spec.~Radiat.~Transf.}}
  \def\memsai{\jref{Mem.~Soc.~Astron.~Italiana}}
  \def\nphysa{\jref{Nucl.~Phys.~A}} \def\physrep{\jref{Phys.~Rep.}}
  \def\physscr{\jref{Phys.~Scr}} \def\planss{\jref{Planet.~Space~Sci.}}
  \def\procspie{\jref{Proc.~SPIE}} \let\astap=\aap \let\apjlett=\apjl
  \let\apjsupp=\apjs \let\applopt=\ao

%% file: ivashchenko.sergijenko.torbaniuk.bbl
\begin{thebibliography}{}

\bibitem[\protect\citeauthoryear{{Abazajian} \& {et al.}}{{Abazajian} et~al.}{2009}]{Abazadjian_2009}
{Abazajian} K.~N. {et al.}, 2009, \apjs, 182, 543

\bibitem[\protect\citeauthoryear{{Baldwin}}{{Baldwin}}{1977}]{baldwin}
{Baldwin} J.~A.,  1977, \apj, 214, 679

\bibitem[\protect\citeauthoryear{{Bernardi} {et al.}}{{Bernardi} et~al.}{2003}]{bernardi+03}
{Bernardi} M. {et al.}, 2003, \aj, 125, 32

\bibitem[\protect\citeauthoryear{{Binette} \& {Krongold}}{{Binette} \&
  {Krongold}}{2008}]{binette+08} {Binette} L.,  {Krongold} Y.,  2008, {\aap}, 477, 413

\bibitem[\protect\citeauthoryear{{Brotherton}, {Tran}, {Becker}, {Gregg},
  {Laurent-Muehleisen} \& {White}}{{Brotherton} et~al.}{2001}]{brotherton+01}
{Brotherton} M.~S.,  {Tran} H.~D.,  {Becker} R.~H.,  {Gregg} M.~D.,
  {Laurent-Muehleisen} S.~A.,    {White} R.~L.,  2001, \apj, 546, 775

\bibitem[\protect\citeauthoryear{{Brotherton}, {Wills}, {Francis} \&
  {Sheidel}}{{Brotherton} et~al.}{1994}]{brotherton+94}
{Brotherton} M.~S.,  {Wills} B.~J.,  {Francis} P.~J.,    {Sheidel} C.~C.,
  1994, \apj, 430, 495

\bibitem[\protect\citeauthoryear{{Busca} {et al.}}{{Busca} et~al.}{2012}]{busca+12}
{Busca} N.~G. {et al.},  2012, ArXiv e-prints (arXiv:1211.2616)

\bibitem[\protect\citeauthoryear{{Carballo}, {Gonz{\'a}lez-Serrano}, {Benn},
  {S{\'a}nchez} \& {Vigotti}}{{Carballo} et~al.}{1999}]{carballo+99}
{Carballo} R.,  {Gonz{\'a}lez-Serrano} J.~I.,  {Benn} C.~R.,  {S{\'a}nchez}
  S.~F.,    {Vigotti} M.,  1999, \mnras, 306, 137

\bibitem[\protect\citeauthoryear{{Croom}, {Boyle}, {Shanks}, {Smith}, {Miller},
  {Outram}, {Loaring}, {Hoyle} \& {da {\^A}ngela}}{{Croom}
  et~al.}{2005}]{Croom+2005}
{Croom} S.~M.,  {Boyle} B.~J.,  {Shanks} T.,  {Smith} R.~J.,  {Miller} L.,
  {Outram} P.~J.,  {Loaring} N.~S.,  {Hoyle} F.,    {da {\^A}ngela} J.,  2005,
  \mnras, 356, 415

\bibitem[\protect\citeauthoryear{{da {\^A}ngela}, {Outram}, {Shanks}, {Boyle},
  {Croom}, {Loaring}, {Miller} \& {Smith}}{{da {\^A}ngela}
  et~al.}{2005}]{daAngela+2005}
{da {\^A}ngela} J.,  {Outram} P.~J.,  {Shanks} T.,  {Boyle} B.~J.,  {Croom}
  S.~M.,  {Loaring} N.~S.,  {Miller} L.,    {Smith} R.~J.,  2005, \mnras, 360,
  1040

\bibitem[\protect\citeauthoryear{{da {\^A}ngela} et al.}{{da {\^A}ngela}
  et~al.}{2008}]{daAngela+2008} {da {\^A}ngela} J. et al.,  2008, \mnras, 383, 565

\bibitem[\protect\citeauthoryear{{Desjacques}, {Nusser} \& {Sheth}}{{Desjacques} et~al.}{2007}]{desjacques_07}
{Desjacques} V.,  {Nusser} A.,    {Sheth} R.~K.,  2007, \mnras, 374, 206

\bibitem[\protect\citeauthoryear{{Francis}, {Hewett}, {Foltz}, {Chaffee},
  {Weymann} \& {Morris}}{{Francis} et~al.}{1991}]{francis+91}
{Francis} P.~J.,  {Hewett} P.~C.,  {Foltz} C.~B.,  {Chaffee} F.~H.,  {Weymann}
  R.~J.,    {Morris} S.~L.,  1991, \apj, 373, 465

\bibitem[\protect\citeauthoryear{{Gaskell}}{{Gaskell}}{1982}]{Gaskell+1982}
{Gaskell} C.~M.,  1982, \apj, 263, 79

\bibitem[\protect\citeauthoryear{{Hewett} \& {Wild}}{{Hewett} \&
  {Wild}}{2010}]{hewett_10}
{Hewett} P.~C.,  {Wild} V.,  2010, \mnras, 405, 2302

\bibitem[\protect\citeauthoryear{{Ivashchenko}, {Zhdanov} \&  {Tugay}}{{Ivashchenko} et~al.}{2010}]{Ivashchenko+2010}
{Ivashchenko} G.,  {Zhdanov} V.~I.,    {Tugay} A.~V.,  2010, \mnras, 409, 1691

\bibitem[\protect\citeauthoryear{{Laor}, {Bahcall}, {Jannuzi}, {Schneider} \&
  {Green}}{{Laor} et~al.}{1995}]{Laor+1995}
{Laor} A.,  {Bahcall} J.~N.,  {Jannuzi} B.~T.,  {Schneider} D.~P.,    {Green}
  R.~F.,  1995, \apjs, 99, 1

\bibitem[\protect\citeauthoryear{{Laor}, {Bancall} \& {Jannuzi}}{{Laor}
  et~al.}{1994}]{Laor+94}
{Laor} A.,  {Bancall} J.~N.,    {Jannuzi} B.~T.,  1994, \apj, 420, 110

\bibitem[\protect\citeauthoryear{{Laor}, {Jannuzi}, {Green} \&
  {Boroson}}{{Laor} et~al.}{1997}]{laor+97}
{Laor} A.,  {Jannuzi} B.~T.,  {Green} R.~F.,    {Boroson} T.~A.,  1997, \apj,
  489, 656

\bibitem[\protect\citeauthoryear{{Leighly}, {Halpern}, {Jenkins} \&
  {Casebeer}}{{Leighly} et~al.}{2007}]{leighly+07}
{Leighly} K.~M.,  {Halpern} J.~P.,  {Jenkins} E.~B.,    {Casebeer} D.,  2007,
  \apjs, 173, 1

\bibitem[\protect\citeauthoryear{{McDonald} et al.}{{McDonald} et~al.}{2006}]{mcdonald+06}
{McDonald} P. et al.,  2006, \apjs, 163, 80

\bibitem[\protect\citeauthoryear{{Mountrichas}, {Sawangwit}, {Shanks}, {Croom},
  {Schneider}, {Myers} \& {Pimbblet}}{{Mountrichas}
  et~al.}{2009}]{Mountrichas+2009}
{Mountrichas} G.,  {Sawangwit} U.,  {Shanks} T.,  {Croom} S.~M.,  {Schneider}
  D.~P.,  {Myers} A.~D.,    {Pimbblet} K.,  2009, \mnras, 394, 2050

\bibitem[\protect\citeauthoryear{{Outram}, {Hoyle}, {Shanks}, {Boyle}, {Croom},
  {Loaring}, {Miller} \& {Smith}}{{Outram} et~al.}{2001}]{Outram+2001}
{Outram} P.~J.,  {Hoyle} F.,  {Shanks} T.,  {Boyle} B.~J.,  {Croom} S.~M.,
  {Loaring} N.~S.,  {Miller} L.,    {Smith} R.~J.,  2001, \mnras, 328, 174

\bibitem[\protect\citeauthoryear{{Pieri}, {Frank}, {Weinberg}, {Mathur} \&
  {York}}{{Pieri} et~al.}{2010}]{pieri+10}
{Pieri} M.~M.,  {Frank} S.,  {Weinberg} D.~H.,  {Mathur} S.,    {York} D.~G.,
  2010, \apjl, 724, L69

\bibitem[\protect\citeauthoryear{{Polinovskyi}}{{Polinovskyi}}{2010}]{greg+10}
{Polinovskyi} G.,  2010, in WDS'06 Proceedings of Contributed Papers: Part III - Physics (eds. J. Safrankova and
  J. Pavlu), Prague, Matfyzpress, 163

\bibitem[\protect\citeauthoryear{{Press}, {Rybicki} \& {Schneider}}{{Press}
  et~al.}{1993}]{press+93} {Press} W.~H.,  {Rybicki} G.~B.,    {Schneider} D.~P.,  1993, \apj, 414, 64

\bibitem[\protect\citeauthoryear{{Reichard} et al.}{{Reichard}
  et~al.}{2003}]{reichard+03} {Reichard} T.~A. et al.,  2003, \aj, 126, 2594

\bibitem[\protect\citeauthoryear{{Richards} et al.}{{Richards} et~al.}{2011}]{richards+11}
{Richards} G.~T. et al.,  2011, \aj, 141, 167

\bibitem[\protect\citeauthoryear{{Richards} et al.}{{Richards} et~al.}{2006}]{richards_06}
{Richards} G.~T. et al.,  2006, \aj, 131, 2766

\bibitem[\protect\citeauthoryear{{Richards}, {Vanden Berk}, {Reichard}, {Hall},
  {Schneider}, {SubbaRao}, {Thakar} \& {York}}{{Richards}
  et~al.}{2002}]{Richards+2002} {Richards} G.~T.,  {Vanden Berk} D.~E.,  {Reichard} T.~A.,  {Hall} P.~B.,
  {Schneider} D.~P.,  {SubbaRao} M.,  {Thakar} A.~R.,    {York} D.~G.,  2002,
  \aj, 124, 1

\bibitem[\protect\citeauthoryear{{Schlegel}, {Finkbeiner} \&
  {Davis}}{{Schlegel} et~al.}{1998}]{schlegel+98}
{Schlegel} D.~J.,  {Finkbeiner} D.~P.,    {Davis} M.,  1998, \apj, 500, 525

\bibitem[\protect\citeauthoryear{{Schneider} et al.}{{Schneider}
  et~al.}{2010}]{Schneider_2010}
{Schneider} D.~P. et al., 2010, \aj, 139, 2360

\bibitem[\protect\citeauthoryear{{Scott}, {Kriss}, {Brotherton}, {Green},
  {Hutchings}, {Shull} \& {Zheng}}{{Scott} et~al.}{2004}]{scott2+04}
{Scott} J.,  {Kriss} G.,  {Brotherton} M.,  {Green} R.,  {Hutchings} J.,
  {Shull} J.,    {Zheng} W.,  2004, in ASP Conf. Ser., 311, AGN
  Physics with the Sloan Digital Sky Survey, ed. {G.~T.~Richards \& P.~B.~Hall},  31

\bibitem[\protect\citeauthoryear{{Shen} et al.}{{Shen} et~al.}{2007}]{Shen+2007}
{Shen} Y. et al., 2007, \aj, 133, 2222

\bibitem[\protect\citeauthoryear{{Songaila}}{{Songaila}}{2004}]{songaila-04}
{Songaila} A.,  2004, \aj, 127, 2598

\bibitem[\protect\citeauthoryear{{Tang}, {Shang}, {Gu}, {Brotherton} \&
  {Runnoe}}{{Tang} et~al.}{2012}]{tang+12}
{Tang} B.,  {Shang} Z.,  {Gu} Q.,  {Brotherton} M.~S.,    {Runnoe} J.~C.,
  2012, \apjs, 201, 38

\bibitem[\protect\citeauthoryear{{Telfer}, {Zheng}, {Kriss} \&
  {Davidsen}}{{Telfer} et~al.}{2002}]{telfer+02}
{Telfer} R.~C.,  {Zheng} W.,  {Kriss} G.~A.,    {Davidsen} A.~F.,  2002, \apj,
  565, 773

\bibitem[\protect\citeauthoryear{{Torbaniuk}, {Ivashchenko} \&
  {Sergijenko}}{{Torbaniuk} et~al.}{2012}]{torbaniuk+12}
{Torbaniuk} O.,  {Ivashchenko} G.,    {Sergijenko} O.,  2012, in WDS'12
  Proceedings of Contributed Papers: Part III - Physics (eds. J. Safrankova and
  J. Pavlu), Prague, Matfyzpress, 123

\bibitem[\protect\citeauthoryear{{Tytler} \& {Fan}}{{Tytler} \&
  {Fan}}{1992}]{Tytler+1992} {Tytler} D.,  {Fan} X.-M.,  1992, \apjs, 79, 1

\bibitem[\protect\citeauthoryear{{Tytler}, {O'Meara}, {Suzuki}, {Kirkman},
  {Lubin} \& {Orin}}{{Tytler} et~al.}{2004}]{tytler+04}
{Tytler} D.,  {O'Meara} J.~M.,  {Suzuki} N.,  {Kirkman} D.,  {Lubin} D.,
  {Orin} A.,  2004, \aj, 128, 1058

\bibitem[\protect\citeauthoryear{{vanden Berk} , {Yip}, {Connolly}, {Jester} \&
  {Stoughton}}{{Vanden Berk} et~al.}{2004}]{vandenberk+04}
{Vanden Berk} D.,  {Yip} C.,  {Connolly} A.,  {Jester} S.,    {Stoughton} C.,
  2004, in ASP Conf. Ser. 311, AGN Physics with the Sloan Digital Sky Survey, ed. {Richards} G.~T.,  {Hall} P.~B., 21

\bibitem[\protect\citeauthoryear{{Vanden Berk} et al.}{{Vanden Berk}
  et~al.}{2001}]{vandenberk+01} {Vanden Berk} D.~E. et al.,  2001, \aj, 122, 549

\bibitem[\protect\citeauthoryear{{Vestergaard} \& {Wilkes}}{{Vestergaard} \&
  {Wilkes}}{2001}]{Vestergaard+01} {Vestergaard} M.,  {Wilkes} B.~J.,  2001, \apjs, 134, 1

\bibitem[\protect\citeauthoryear{{Wild} \& {Hewett}}{{Wild} \&
  {Hewett}}{2005}]{Wild_05} {Wild} V.,  {Hewett} P.~C.,  2005, \mnras, 358, 1083

\bibitem[\protect\citeauthoryear{{Wild} \& {Hewett}}{{Wild} \&
  {Hewett}}{2010}]{wild_10} {Wild} V.,  {Hewett} P.~C.,  2010, ArXiv e-prints (arXiv:1010.2500)

\bibitem[\protect\citeauthoryear{{Wolfe}, {Gawiser} \& {Prochaska}}{{Wolfe}
  et~al.}{2005}]{wolfe_05} {Wolfe} A.~M.,  {Gawiser} E.,    {Prochaska} J.~X.,  2005, \araa, 43, 861

\bibitem[\protect\citeauthoryear{{Yip} et al.}{{Yip} et~al.}{2004}]{yip+04}
{Yip} C.~W. et al.,  2004, \aj, 128, 2603

\bibitem[\protect\citeauthoryear{{Zheng}, {Kriss}, {Telfer}, {Grimes} \&
  {Davidsen}}{{Zheng} et~al.}{1997}]{zheng+97} {Zheng} W.,  {Kriss} G.~A.,  {Telfer} R.~C.,  {Grimes} J.~P.,    {Davidsen}
  A.~F.,  1997, \apj, 475, 469

\end{thebibliography}
